\documentclass[twocolumn]{aastex631}

\shorttitle{AGE-PRO III: Lupus Disks}
\shortauthors{Deng et al.}
\usepackage{placeins} 

\graphicspath{{./}{figures/}} 

\begin{document}

\title{The ALMA Survey of Gas Evolution of PROtoplanetary Disks (AGE-PRO): \\
III. Dust and Gas Disk Properties in the Lupus Star-forming Region}

\author[0000-0003-0777-7392]{Dingshan Deng}
\affiliation{Lunar and Planetary Laboratory, the University of Arizona, Tucson, AZ 85721, USA}

\correspondingauthor{Dingshan Deng}
\email{dingshandeng@arizona.edu}

\author[0000-0002-4147-3846]{Miguel Vioque}
\affiliation{European Southern Observatory, Karl-Schwarzschild-Str. 2, 85748 Garching bei München, Germany}
\affiliation{Joint ALMA Observatory, Alonso de Córdova 3107, Vitacura, Santiago 763-0355, Chile}

\author[0000-0001-7962-1683]{Ilaria Pascucci}
\affiliation{Lunar and Planetary Laboratory, the University of Arizona, Tucson, AZ 85721, USA}

\author[0000-0002-1199-9564]{Laura M. P\'erez}
\affiliation{Departamento de Astronom\'ia, Universidad de Chile, Camino El Observatorio 1515, Las Condes, Santiago, Chile}

\author[0000-0002-0661-7517]{Ke Zhang}
\affiliation{Department of Astronomy, University of Wisconsin-Madison, 475 N Charter St, Madison, WI 53706, USA}

\author[0000-0002-2358-4796]{Nicol\'as T. Kurtovic}
\affiliation{Max-Planck-Institut f\"ur Extraterrestrische Physik, Giessenbachstrasse 1, D-85748 Garching, Germany}
\affiliation{Max-Planck-Institut f\"ur Astronomie (MPIA), Konigstuhl 17, 69117 Heidelberg, Germany}

\author[0000-0002-8623-9703]{Leon Trapman}
\affiliation{Department of Astronomy, University of Wisconsin-Madison, 475 N Charter St, Madison, WI 53706, USA}

\author[0000-0001-9961-8203]{Estephani E. TorresVillanueva}
\affiliation{Department of Astronomy, University of Wisconsin-Madison, 475 N Charter St, Madison, WI 53706, USA}

\author[0000-0002-7238-2306]{Carolina Agurto-Gangas}
\affiliation{Departamento de Astronom\'ia, Universidad de Chile, Camino El Observatorio 1515, Las Condes, Santiago, Chile}

\author[0000-0003-2251-0602]{John Carpenter}
\affiliation{Instituto de Estudios Astrof\'isicos, Universidad Diego Portales, Av. Ejercito 441, Santiago, Chile}
\affiliation{Millennium Nucleus on Young Exoplanets and their Moons (YEMS), Chile}
\affiliation{Center for Interdisciplinary Research in Astrophysics and Space Exploration (CIRAS), Universidad de Santiago de Chile, Chile}

\author[0000-0001-8764-1780]{Paola Pinilla}
\affiliation{Mullard Space Science Laboratory, University College London, Holmbury St Mary, Dorking, Surrey RH5 6NT, UK}

\author[0000-0002-3311-5918]{Uma Gorti}
\affiliation{NASA Ames Research Center, Moffett Field, CA 94035, USA}
\affiliation{Carl Sagan Center, SETI Institute, Mountain View, CA 94043, USA}

\author[0000-0002-1103-3225]{Benoît Tabone}
\affiliation{Université Paris-Saclay, CNRS, Institut d'Astrophysique Spatiale, 91405 Orsay, France}

\author[0000-0002-5991-8073]{Anibal Sierra}
\affiliation{Departamento de Astronom\'ia, Universidad de Chile, Camino El Observatorio 1515, Las Condes, Santiago, Chile}
\affiliation{Mullard Space Science Laboratory, University College London, Holmbury St Mary, Dorking, Surrey RH5 6NT, UK}

\author[0000-0003-4853-5736]{Giovanni P. Rosotti}
\affiliation{Dipartimento di Fisica, Università degli Studi di Milano, Via Celoria 16, I-20133 Milano, Italy}

\author[0000-0002-2828-1153]{Lucas A. Cieza}
\affiliation{Instituto de Estudios Astrofísicos, Universidad Diego Portales, Av. Ejercito 441, Santiago, Chile}
\affiliation{Millennium Nucleus on Young Exoplanets and their Moons (YEMS), Chile}

\author[0009-0004-8091-5055]{Rossella Anania}
\affiliation{Dipartimento di Fisica, Università degli Studi di Milano, Via Celoria 16, I-20133 Milano, Italy}

\author[0000-0003-4907-189X]{Camilo Gonz\'alez-Ruilova}
\affiliation{Instituto de Estudios Astrofísicos, Universidad Diego Portales, Av. Ejercito 441, Santiago, Chile}
\affiliation{Millennium Nucleus on Young Exoplanets and their Moons (YEMS), Chile}
\affiliation{Center for Interdisciplinary Research in Astrophysics and Space Exploration (CIRAS), Universidad de Santiago de Chile, Chile}

\author[0000-0001-5217-537X]{Michiel R. Hogerheijde}
\affiliation{Leiden Observatory, Leiden University, PO Box 9513, 2300 RA Leiden, the Netherlands}
\affiliation{Anton Pannekoek Institute for Astronomy, University of Amsterdam, the Netherlands}

\author[0000-0002-1575-680X]{James Miley}
\affiliation{Departamento de Física, Universidad de Santiago de Chile, Av. Victor Jara 3659, Santiago, Chile}
\affiliation{Millennium Nucleus on Young Exoplanets and their Moons (YEMS), Chile}
\affiliation{Center for Interdisciplinary Research in Astrophysics and Space Exploration (CIRAS), Universidad de Santiago de Chile, Chile}

\author[0000-0003-3573-8163]{Dary A. Ruiz-Rodriguez}
\affiliation{National Radio Astronomy Observatory; 520 Edgemont Rd., Charlottesville, VA 22903, USA}
\affiliation{Joint ALMA Observatory, Avenida Alonso de Córdova 3107, Vitacura, Santiago, Chile}

\author[0000-0003-0522-5789]{Maxime Ruaud}
\affiliation{Carl Sagan Center, SETI Institute, Mountain View, CA 94043, USA}

\author[0000-0002-6429-9457]{Kamber Schwarz}
\affiliation{Max-Planck-Institut f\"ur Astronomie (MPIA), Konigstuhl 17, 69117 Heidelberg, Germany}


\begin{abstract}

We present Band~6 and Band~7 observations of 10 Lupus disks around M3-K6 stars from the ALMA survey of Gas Evolution in PROtoplanetary disks (AGE-PRO) Large Program. In addition to continuum emission in both bands, our Band~6 setup covers the $\mathrm{{}^{12}CO}$, $\mathrm{{}^{13}CO}$ and $\mathrm{C^{18}O}$\,$J$=2-1 lines, while our Band~7 setup covers the $\mathrm{N_2H^+}$\,$J$=3-2 line. All of our sources are detected in $\mathrm{{}^{12}CO}$ and $\mathrm{{}^{13}CO}$, 7 out of 10 are detected in $\mathrm{C^{18}O}$, and 3 are detected in $\mathrm{N_2H^+}$. We find strong correlations between the CO isotopologue line fluxes and the continuum flux densities. With the exception of one disk, we also identify a strong correlation between the $\mathrm{C^{18}O}$\,$J$=2-1 and $\mathrm{N_2H^+}$\,$J$=3-2 fluxes, indicating similar CO abundances across this sample. For the two sources with well-resolved continuum and $\mathrm{{}^{12}CO}$\,$J$=2-1 images, we find that their gas-to-dust size ratio is consistent with the median value of $\sim 2$ inferred from a larger sample of Lupus disks. We derive dust disk masses from continuum flux densities. We estimate gas disk masses by comparing $\mathrm{C^{18}O}$\,$J$=2-1 line fluxes with those predicted by the limited grid of self-consistent disk models of \citet{ruaud_C18O_2022}. A comparison of these mass estimates with those derived by \citet{Trapman_AGEPRO_V_gas_masses}, using a combination of CO isotopologue and $\mathrm{N_2H^+}$ line emission, shows that the masses are consistent with each other. Some discrepancies appear for small and faint disks, but they are still within the uncertainties. Both methods find gas disk masses increase with dust disk masses, and gas-to-dust mass ratios are between $10-100$ in the AGE-PRO Lupus sample.

\end{abstract}

\section{Introduction}
\label{sec:intro}

Disks around young stars are the sites of planet formation, hence the name protoplanetary disks.
Among their different properties, disk masses and sizes are among the most critical ones to test disk evolution and planet formation models \citep[see][for recent reviews]{Drazkowska_PPVII_2023,Manara_2023_PPVII}.
For example, the initial dust and gas disk masses determine what type of planetary systems could form \citep[e.g.,][]{Mordasini_Planet_Formation_Synthesis_2012}.
Disk sizes as a function of time can be used to distinguish disk evolutionary models. 
In the classic disk viscous accretion models \citep[e.g.,][]{Pringle_disk_1981}, the angular momentum is redistributed outward which leads to larger gas disk sizes over time. 
In the case of the magneto-hydrodynamic (MHD) wind-driven accretion scenario \citep[e.g.,][for a recent review]{Pascucci_2023_PPVII}, the angular momentum is extracted from the disk through the wind \citep[e.g.,][]{Suzuki_diskwind_2016, Tabone_MHDwind_2022MNRAS} so that gas disk sizes are not expected to increase with time.
Dust disk sizes cannot directly probe gas disk evolution, but, when coupled with gas disk sizes, they can indicate the extent of radial drift of millimeter-size dust particles, often called pebbles \citep[e.g.,][]{trapman_disk_size_comp_2019, trapman_radial_drift_2020, Toci_sizes_2021}.

In the past decade, the characterization of disk masses and sizes for both the gas and dust components has been significantly improved by the Atacama Large Millimeter/Submillimeter Array (ALMA; see \citealt{Miotello_PPVII_2023} for a recent review).
Initial surveys of nearby star-forming regions detected emission from pebble-sized grains in most disks around a large range of stellar masses (from $\sim$0.1 to 1 $M_\odot$; e.g., \citealt{Ansdell_2016ApJ_Lupus, Barenfeld_USco_survey_2016, Pascucci2016_massrelation}). 
These surveys revealed that the millimeter flux, a proxy for the dust disk mass, scales with the mass of the central star with a relation that steepens with time. 
They also enabled measuring dust disk sizes and showed that they are smaller in older regions \citep{Hendler_disk_sizes_2020}. 
Several ALMA surveys have also probed the gaseous component, particularly in carbon monoxide (CO) and its isotopologues.
Spatially-resolved $\mathrm{{}^{12}CO}$ observations were used to measure gas disk sizes and the optically thin line emission from rarer CO isotopologues (e.g., $\mathrm{{}^{13}CO}$, $\mathrm{C^{18}O}$) was used to derive gas disk masses when combined with thermo-chemical disk models \citep[e.g.,][]{miotello_lupus_2017, woitke_models_2019, ruaud_C18O_2022, deng_diskmint_2023}.
However, the detected CO and its rarer isotopologues have been found to be weaker than what were expected from early models \citep[e.g.,][]{Ansdell_2016ApJ_Lupus, Ansdell_Lupus_2018ApJ, Barenfeld_USco_survey_2016, long_alma_chamI_2017}.
As such, gas disk mass estimates are only available for the bright disks in each star-forming region where their $\mathrm{{}^{13}CO}$ and/or $\mathrm{C^{18}O}$ lines were detected, corresponding to $\sim 50\%$ of the Lupus disks in the literature \citep[e.g.,][]{miotello_lupus_2017, zhang_maps_2021, schwarz_MAPS_2021, deng_diskmint_2023, pascucci_largedisks_2023}.

The ALMA survey of Gas Evolution in PROtoplanetary disks (AGE-PRO; Project ID: 2021.1.00128.L, PI: K. Zhang) is an ALMA Cycle~8 Large Program. 
It builds on the initial lower-sensitivity surveys but goes much deeper to detect gas emission from the rarer CO isotopologues and $\mathrm{N_2H^+}$.
The survey is designed to study  disks at different ages in a consistent observational set up, aiming to systematically trace the evolution of gas masses and sizes throughout the lifetime of protoplanetary disks \citep{Zhang_AGEPRO_I_overview}.
The disks in the Lupus star-forming region have been selected to represent disks at the age of $\sim 1-3\,\mathrm{Myrs}$ because of the richness of literature data and well-characterized stellar and disk parameters.

The Lupus molecular cloud complex is one of the closest low-mass star-forming regions, and it is part of the Scorpius-Centaurus OB association \citep{Comeron_2008_Lupus_book}.
It consists of four main clouds with active star formation (Lupus~I-IV) located at slightly different distances of $\sim 150 - 160\,\mathrm{pc}$. 
The mean distance to the Lupus region is $\sim 158\,\mathrm{pc}$ \citep{Luhman_n_Esplin_2020AJ, Galli_Lupus_DANCe_2020, Gaia_DR3_2022yCat.1355....0G}.
The average age of the region is $\sim 1-3\,\mathrm{Myrs}$: the youngest region is Lupus~I (median age $\sim1.2\,{\rm Myrs}$), the middle age is Lupus~III-IV ($\sim2.4-2.5\,{\rm Myrs}$), and older targets are off cloud ($\sim3.2\,{\rm Myrs}$, \citealt{Galli_Lupus_DANCe_2020}).

There are many initial and near-complete surveys characterizing the protoplanetary disks and their stars in the Lupus star-forming region.
Observations in the near- and mid- infrared with \emph{Spitzer} \citep{Merin_Spitzer_2008, Evans_Spitzer_2009, Dunham_Spitzer_2015} and far-infrared with \textit{Herschel} \citep{Bustamante_Herschel_2015} found a disk fraction of $\sim70-80\%$.
Later, \citet{Ansdell_2016ApJ_Lupus} carried out the first ALMA survey of disks in Lupus, followed by other extended surveys \citep[e.g.,][]{Ansdell_Lupus_2018ApJ, Miotello_Compat_Disks_2021A&A...651A..48M} as well as several deep observations focusing on specific targets and lines \citep[e.g.,][]{Cleeves_IMLup_2016, Tsukagoshi_Sz91_2019}.
Moreover, \citet{alcala_x-shooter_2014, alcala_x-shooter_2017, alcala_hst_2019} conducted a spectroscopic survey using VLT/X-Shooter, which derived stellar properties (e.g., mass, luminosity, effective temperature, and age) and mass accretion rates for these star-disk systems in Lupus.
Therefore, a near-complete catalog including both stellar and disk properties for the Lupus disks has been compiled \citep[see][for a recent summary]{Manara_2023_PPVII}.

This comprehensive stellar and disk characterization allow us to select 10 sources to form a sample of disks around M3-K6 type stars.
The new AGE-PRO ALMA observations of CO isotopologues in Band~6 and $\mathrm{N_2H^+}$ emission lines in Band~7 enable a detailed and in-depth study of the gas disk masses and sizes.
In this paper, we introduce the ALMA observations and basic analysis of the AGE-PRO Large Program Lupus sample, while more detailed analysis of the AGE-PRO datasets and comparisons across the regions are presented in accompanying papers (\citealt{Zhang_AGEPRO_I_overview} provides an overview, which is also provided on the AGE-PRO website\footnote{The AGE-PRO website: \url{https://agepro.das.uchile.cl}}).
The Lupus target selection and observations are introduced in Section~\ref{sec:targetnobservation}. The data reduction, including calibration and imaging procedures, is discussed in Section~\ref{sec:DataProcessing}, while the analysis and main results are presented in Section~\ref{sec:Results}.  Section~\ref{sec:discussions} discusses the main results while Section~\ref{sec:Summary} summarizes the key findings.

\section{Targets and Observations}
\label{sec:targetnobservation}

\subsection{Targets in Lupus Star-forming Region}
\label{subsec:targets_details}

The AGE-PRO disks in the Lupus star-forming region were selected based on the following criteria:

\begin{itemize}
\item Stars were chosen to have spectral types (SpT) between M3-K6, roughly corresponding to stellar masses  ($M_\star$) $\sim 0.3 - 0.8\,M_\odot$ according to evolutionary tracks \citep[e.g.,][]{Feiden_YSO_tracks_2016}.
\item Stars identified as close-binaries in previous literature were excluded.
\item Class II disks were selected based on the their Spectral Energy Distributions (SEDs) and their infrared spectral indices $-1.6 < \alpha_{\rm IR} < -0.3$ \citep{Williams_n_Cieza_2011ARAA}.
\item Sources were chosen to have ALMA Band~7 detections in the continuum and $\mathrm{{}^{13}CO}$ J = 3-2 line \citep{Ansdell_2016ApJ_Lupus, Ansdell_Lupus_2018ApJ}.
\item Known highly-inclined (inclination angle $>75$\arcdeg) and edge-on disks were excluded because the properties of their central stars cannot be well constrained.
\end{itemize}

By using these criteria, we first identified 24 sources in the Lupus star-forming region. 
From this sample, we selected 10 disks to cover the range of millimeter luminosities in the Lupus star-forming region. 
The target list with their stellar properties is compiled in Table~\ref{Tab:targets}.

\begin{splitdeluxetable*}{lcccccBcccccccc}
\tablecaption{Stellar properties \label{Tab:targets}}
\tablewidth{0.45\textwidth}
\tablehead{
\colhead{ID} &
\colhead{Source} & 
\colhead{R.A.} & \colhead{Decl.} & \colhead{Distance} &
\colhead{Cloud} &
\colhead{SpT} & \colhead{$T_\mathrm{eff}$ } & 
\colhead{$L_\mathrm{\star}$} & 
\colhead{$A_{V}$} & 
\colhead{$\log_{10}(\dot{M}_{\rm acc})$} & 
\colhead{$M_\mathrm{\star}$} & 
\colhead{Age} \\ 
\colhead{} & 
\colhead{} & 
\colhead{} & \colhead{} & \colhead{(pc)} &
\colhead{} &
\colhead{} & \colhead{(K)} & 
\colhead{($L_\odot$)} & 
\colhead{(mag)} &
\colhead{$(M_\odot \mathrm{yr}^{-1})$} &
\colhead{($M_\odot$)} &
\colhead{(Myr)} 
}
\startdata
1 &              Sz65 & 15:39:27.753 & -34:46:17.577 &      153.00 &     I &     K7 &    4060 &         0.869 &   0.6 &         -9.480 & $0.68^{+0.20}_{-0.14}$ & $1.17^{+1.06}_{-0.60}$ \\
2 &              Sz71 & 15:46:44.709 & -34:30:36.054 &      154.07 &     I &   M1.5 &    3632 &         0.327 &   0.5 &         -9.033 & $0.42^{+0.12}_{-0.10}$ & $1.74^{+1.98}_{-0.94}$ \\
3 & J16124373-3815031 & 16:12:43.736 & -38:15:03.471 &      158.65 &   off &     M1 &    3705 &         0.390 &   0.8 &         -9.084 & $0.49^{+0.14}_{-0.11}$ & $1.74^{+2.83}_{-1.08}$ \\
4 &              Sz72 & 15:47:50.608 & -35:28:35.779 &      155.49 &     I &     M2 &    3560 &         0.272 &  0.75 &         -8.601 & $0.39^{+0.12}_{-0.09}$ & $1.78^{+2.02}_{-0.98}$ \\
5 &              Sz77 & 15:51:46.941 & -35:56:44.531 &      154.82 &   off &     K7 &    4060 &         0.593 &     0 &         -8.690 & $0.73^{+0.19}_{-0.15}$ & $2.29^{+2.50}_{-1.22}$ \\
6 & J16085324-3914401 & 16:08:53.227 & -39:14:40.553 &      161.48 &   III &     M3 &    3415 &         0.198 &   1.9 &        -10.030 & $0.31^{+0.05}_{-0.05}$ & $1.78^{+3.01}_{-1.19}$ \\
7 &             Sz131 & 16:00:49.414 & -41:30:04.263 &      159.14 &    IV &     M3 &    3415 &         0.150 &   1.3 &         -9.171 & $0.31^{+0.05}_{-0.05}$ & $2.75^{+2.49}_{-1.37}$ \\
8 &              Sz66 & 15:39:28.264 & -34:46:18.450 &      154.41 &     I &     M3 &    3415 &         0.216 &     1 &         -8.508 & $0.30^{+0.05}_{-0.04}$ & $1.58^{+1.37}_{-0.77}$ \\
9 &              Sz95 & 16:07:52.293 & -38:58:06.446 &      159.19 &   III &     M3 &    3415 &         0.267 &   0.8 &         -9.381 & $0.30^{+0.05}_{-0.04}$ & $1.20^{+1.49}_{-0.69}$ \\
10 &          V1094Sco & 16:08:36.160 & -39:23:02.879 &      154.78 &   III &     K6 &    4205 &         1.210 &   1.7 &         -7.880 & $0.82^{+0.21}_{-0.15}$ & $0.89^{+1.01}_{-0.38}$ \\
\enddata
\tablecomments{effective temperature ($T_{\rm eff}$), stellar luminosity ($L_\star$), extinction ($A_V$), and mass accretion rate ($\dot{M}_{\rm acc}$) are from \citet{Manara_2023_PPVII}, summarizing \citet{alcala_x-shooter_2014, alcala_x-shooter_2017, alcala_hst_2019, Herczeg_n_Hillenbrand_TTauri_SpT}, and rescaled to the new \citet{Gaia_DR3_2022yCat.1355....0G} distances.
Geometric distances are from \emph{Gaia/r}-band \citep{Bailer-Jones_2021AJ_Gaia_geodistance}.
Cloud information is based on the dynamical analysis by \citet{Galli_Lupus_DANCe_2020}.
We adopt a Bayesian framework as in \citet{Pascucci2016_massrelation} to estimate stellar masses ($M_{\star}$) and ages from stellar evolutionary tracks in the HR-diagram (Figure~\ref{fig:HRD_Lupus} and Section~\ref{subsec:targets_details}). 
Uncertainties for $M_{\star}$ and ages represent a 68\% confidence interval.
}
\end{splitdeluxetable*}

\begin{figure*}
   \gridline{\fig{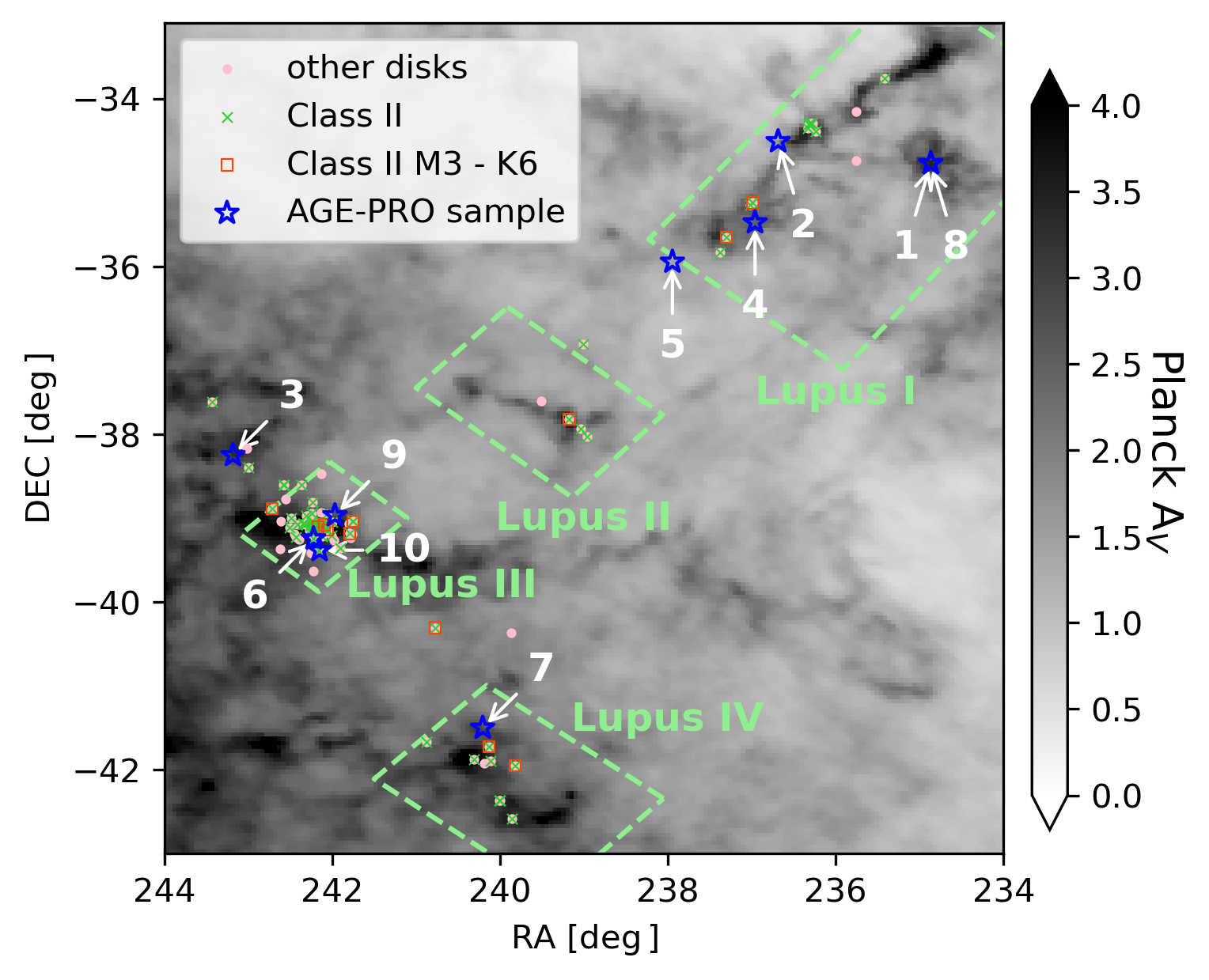}{0.9\textwidth}{}
            }
  \caption{
  Lupus disks (pink; as selected in the \citealt{Manara_2023_PPVII} table) and the selected AGE-PRO sample (blue stars with assigned IDs) on the top of Planck dust map \citep{Planck_2014_dust_emission_2014}.
  Green crosses on top of the pink circles show the Class~II sources, and orange squares mark the disks around stars with SpT M3-K6.
  The four green dashed squares mark the boundaries of the four different sub-clouds in the Lupus star-forming region based on \citet{Galli_Lupus_DANCe_2020}.
  }
  \label{fig:disks_on_dustmap}
\end{figure*}

Figure~\ref{fig:disks_on_dustmap} presents the spatial distributions of our sample with the Planck extinction map \citep{Planck_2014_dust_emission_2014} as background \citep[compiled from the \texttt{Python} package \texttt{dustmap}\footnote{\url{https://github.com/gregreen/dustmaps}}, ][]{Green_2018JOSS....3..695M}.
Comparing with the recent dynamical analysis by \citet{Galli_Lupus_DANCe_2020}, we confirm that all selected targets are from the Lupus star-forming region and  cover subgroups with slightly different ages.
We have four targets from Cloud~I, four from Clouds~III-IV, and two off-cloud targets, see Table~\ref{Tab:targets} for the specific target names.

\begin{figure}
   \gridline{\fig{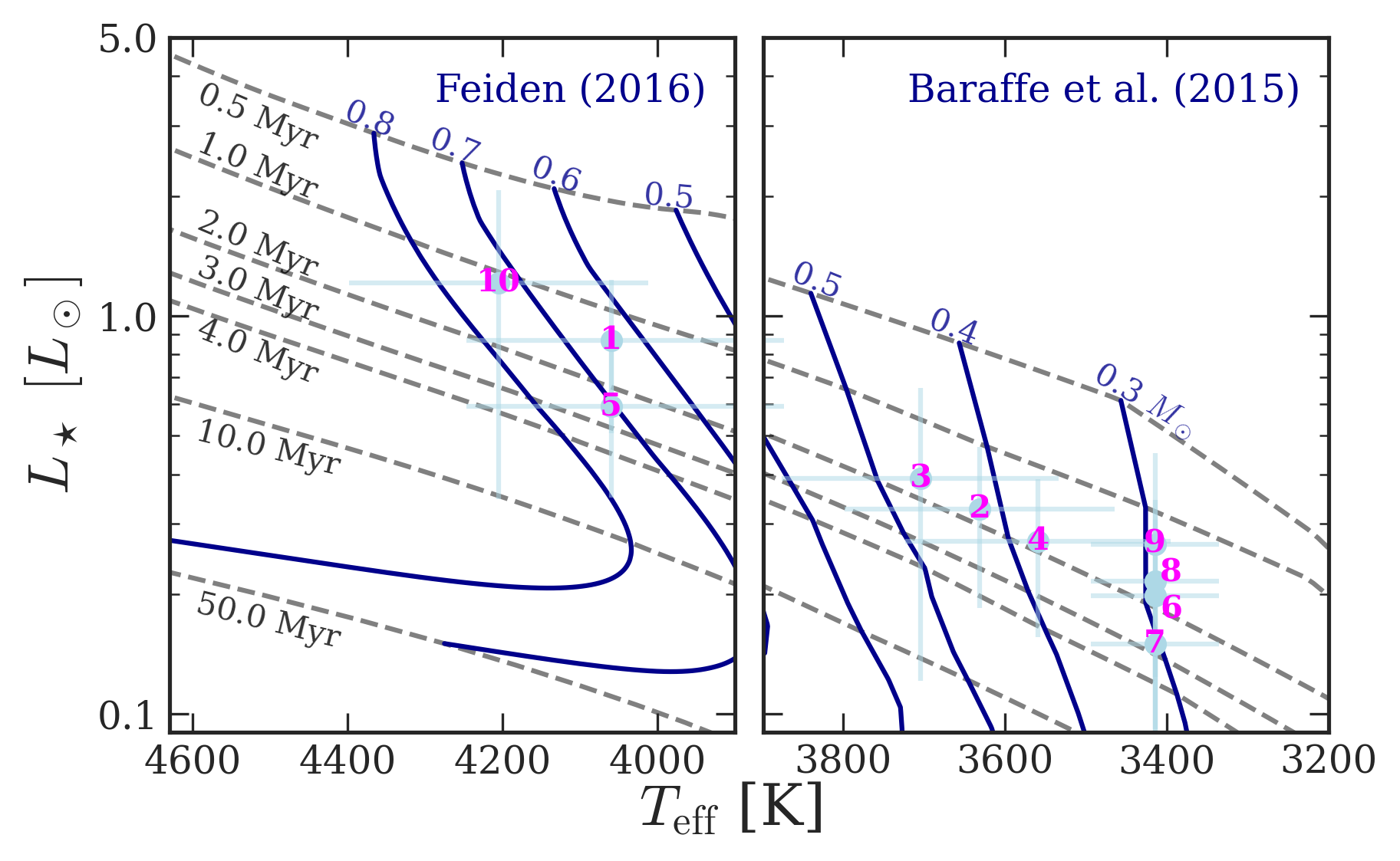}{0.47\textwidth}{}
            }
  \caption{
  Hertzsprung–Russell diagram for the Lupus targets, marked with blue points and magenta labels for their IDs (see Table~\ref{Tab:targets} for IDs and stellar properties). 
  Isochrones and pre-main sequence stellar evolutionary tracks are presented as blue solid and gray dashed lines, respectively.
  Following \cite{Pascucci2016_massrelation}, the tracks from \citet{Feiden_YSO_tracks_2016} are adopted for the stars with $T_{\rm eff} > 3,900\,{\rm K}$ and presented in the left panel, while the \citet{baraffe_stellar_model_2015} tracks are used for cooler stars and presented in the right panel.
  }
  \label{fig:HRD_Lupus}
\end{figure}

Since stellar ages are critical for the AGE-PRO survey but are not available in the literature, we use a new \texttt{Python} package \texttt{ysoisochrone}\footnote{\url{https://github.com/DingshanDeng/ysoisochrone}} \citep{Deng_2025_ysoisochrone} to estimate stellar masses and ages from stellar evolutionary tracks in a consistent way. 
This Python package is based on the \texttt{IDL} code developed by \cite{Pascucci2016_massrelation}, which applies a Bayesian inference approach to estimate uncertainties in the inferred masses and ages \citep[e.g.,][]{Jorgensen_2005_Bayesian_isochrone, Gennaro_2012_Bayesian_isochrone, Andrews_2013_Bayesian_isochrone}.
As in \citet{Pascucci2016_massrelation}, we use non-magnetic tracks: from \citet{baraffe_stellar_model_2015} for targets with $T_{\rm eff} < 3900\,{\rm K}$ and from \citet{Feiden_YSO_tracks_2016} for hotter stars, see Figure~\ref{fig:HRD_Lupus} for the position of the Lupus sources in the Hertzsprung–Russell diagram.
The conditional likelihood function assumes uniform priors on the model properties, and we use the uncertainties on $T_{\rm eff}$ and $L_{\star}$ reported in \citet{alcala_x-shooter_2014, alcala_x-shooter_2017, alcala_hst_2019}. 
The estimated stellar masses and ages, along with their 68\% confidence intervals, are summarized in Table~\ref{Tab:targets}.
The $M_{\star}$ estimated here are consistent with the ones reported in the literature \citep{alcala_hst_2019, Manara_2023_PPVII}.
The method is applied to all the disks in the AGE-PRO sample, including the other disks in Ophiuchus and Upper~Sco star-forming regions \citep{Ruiz-Rodriguez_AGEPRO_II_Ophiuchus, Agurto-Gangas_AGEPRO_IV_UpperSco}.

\begin{figure*}
   \gridline{\fig{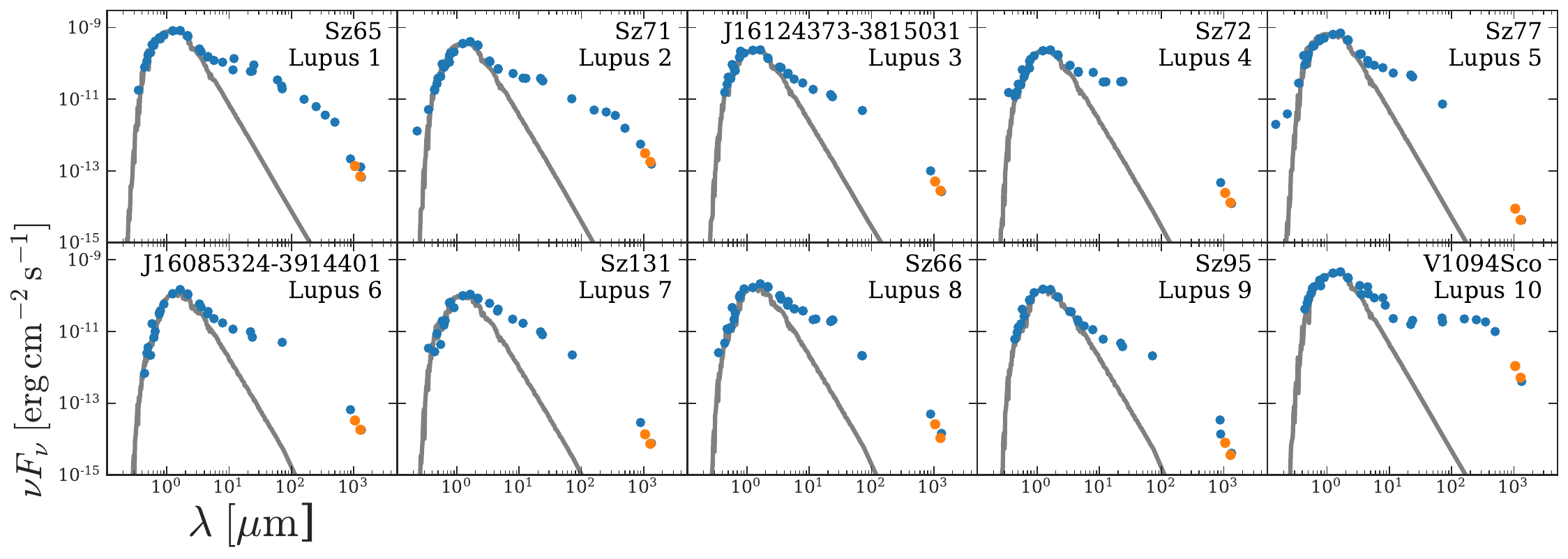}{0.95\textwidth}{}
            }
  \caption{
  Spectral Energy Distributions for the AGE-PRO Lupus disks. 
  The gray lines are the stellar photospheric spectra from \texttt{PHOENIX} models \citep{Husser_PHOENIX_model_2013} corresponding to their $T_{\rm eff}$. 
  We also apply the extinction (Table~\ref{Tab:targets}) to the models and then scale it to the distance of each star. 
  The SEDs evidence the accuracy of the stellar parameters, and the presence of the IR~excess suggesting they are Class~II sources.
  The orange points are the continuum flux densities measured in our observations (see Section~\ref{sec:Results} for more details), and the blue points are the observations collected from literature: XMM-Newton \citep{Page_XMM-Newton_2012MNRAS}, GALEX \citep{Bianchi_GALEX_2017ApJS}, GSC~II \citep{Lasker_GSCII_2008AJ}, SDSS \citep{Ahumada_SDSS_2020ApJS}, APASS DR9 \citep{Henden_2015_APASS}, Gaia DR3 \citep{Gaia_DR3_2022yCat.1355....0G}, 2MASS \citep{Cutri_2MASS_2003yCat}, Spitzer \citep{Evans_Spitzer_2003PASP}, WISE \citep{Cutri_WISE_2014yCat}, ROSAT \citep{Voges_ROSAT_1999}, Herschel \citep{Benedettini_Herschel_Lupus_2018}, ALMA \citep{Ansdell_2016ApJ_Lupus, Ansdell_Lupus_2018ApJ}.
  }
  \label{fig:SEDs}
\end{figure*}

We also confirm that the selected Lupus targets are Class~II sources by compiling their de-extincted broadband SEDs (see Figure~\ref{fig:SEDs} which also superimposes the \texttt{Phoenix} stellar photospheric models from \citealt{Husser_PHOENIX_model_2013}).
All sources have excess emission from a few micron out to millimeter wavelengths, and their spectral indices in the infrared ($\sim 2-24\micron$) are in the range of $-1.5 \lesssim \alpha_{\rm IR} \lesssim 0$, as expected from Class~II sources. 

\begin{figure*}
   \gridline{\fig{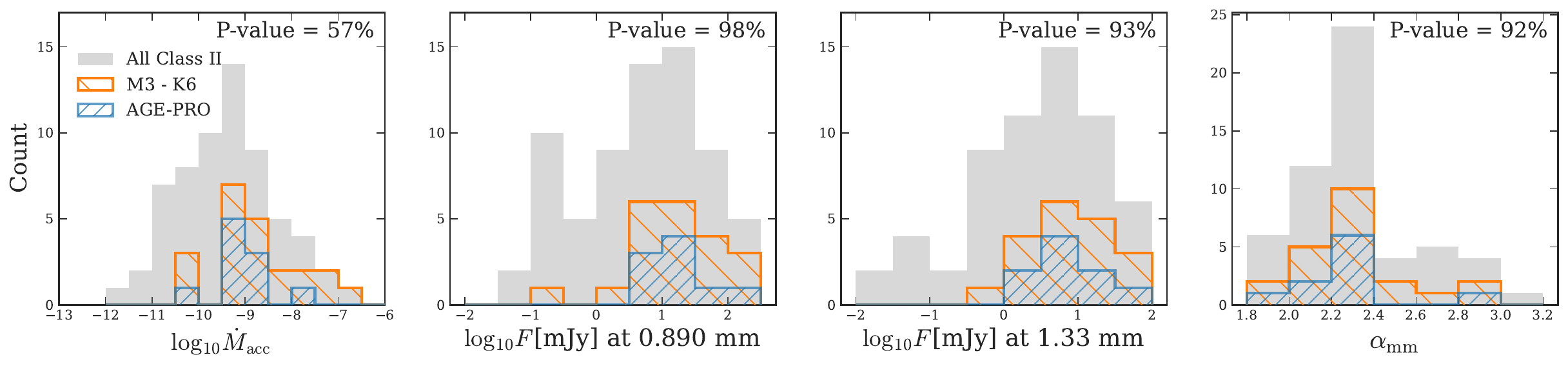}{0.97\textwidth}{}
            }
  \caption{
  The AGE-PRO Lupus sample vs. the Lupus Class~II sample that summarized in \citet{Manara_2023_PPVII}. 
  Mass accretion rates are from \citet{alcala_x-shooter_2014, alcala_x-shooter_2017, alcala_hst_2019}, mm fluxes from \citet{Ansdell_2016ApJ_Lupus, Ansdell_Lupus_2018ApJ, Tsukagoshi_Sz91_2019, Sanchis_conti_obs_2020}. 
The spectral indices ($\alpha_{\rm mm}$) between these two continuum flux densities are computed in this work (Section~\ref{subsec:targets_details}).
  The P-values from the Kolmogorov-Smirnov tests comparing the selected Lupus sample in AGE-PRO (blue histograms) with the same spectral type range Class~II sample (orange histograms) are all much larger than 5\%, suggesting that the AGE-PRO sample is representative for the M3-K6 Class~II sources in Lupus.
  }
  \label{fig:age-pro-sample-in-lupus}
\end{figure*}

Our aim was to select a representative disk sample from the Lupus star-forming region, covering a wide range of properties.
To evidence this, Figure~\ref{fig:age-pro-sample-in-lupus} shows the histograms of the distribution of mass accretion rates ($\dot{M}_\mathrm{acc}$), millimeter fluxes ($F_{\rm mm}$), and spectral indices ($\alpha_{\rm mm}$) for the AGE-PRO sample compared with the Class~II disk population in the Lupus star-forming region and the subset with the same spectral type range (M3-K6). 
While the AGE-PRO disks have larger $\dot{M}_\mathrm{acc}$ and $F_{\rm mm}$ compared to the whole sample, the distribution restricted to the same spectral type range is similar.

We carry out a two-sample Kolmogorov-Smirnov test (K-S test) to compare the two datasets and determine if the null hypothesis that they come from the same parent distribution can be rejected, which requires a probability $P-{\rm value} < 5\%$. 
Indeed, the K-S tests applied to the AGE-PRO and the SpT-restricted sample give very high probabilities ($\gg 5$\,\%), suggesting that the AGE-PRO sample is representative of the Lupus disks in the M3-K6 SpT range.

\subsection{Observations}

Our targets were observed between December 2021 and June 2022 with the ALMA 12\,m array in Band~6 (234\,GHz, or 1.28\,mm) using the C43-2 (15-314\,m) and C43-5 (15-1400\,m) array configurations, which we refer to as `short' and `long' baselines, respectively.
Targets were also observed in Band~7 (285\,GHz, or 1.05\,mm) in April 2022 using only the C43-3 (15-500\,m) configuration. 
We also complemented our data with ALMA archival data: Band 6 long baseline data for Sz~65 were obtained in August 2018 (2017.1.00569.S, PI: Hsi-Wei Yen) while Band 7 data for Sz~65 and J16085324-3914401 were obtained between October and November 2019 (2019.1.01135.S, PI: Dana Anderson).
The observational log for AGE-PRO observations and the archival data are summarized in Table~\ref{table:obs_log} in Appendix~\ref{appendix:observation_log}.

\section{Data Reduction}
\label{sec:DataProcessing}

In AGE-PRO, we carry out a similar data reduction strategy for disks across different star-forming regions to ensure consistency in the products. 
The data reduction procedure is described in detail in the overview paper \citep[][Section~3]{Zhang_AGEPRO_I_overview}, and the data reduction scripts are released on the AGE-PRO website.
Here we summarize main procedures and the special steps that are taken for the 10 disks in the Lupus star-forming region.

\subsection{Calibration Strategy}

The execution blocks were initially calibrated individually by the ALMA staff using the CASA pipeline version \texttt{6.2.1-7-pipeline} for both short and long baseline Band~6 and Band~7 data, and \texttt{5.6.1-8-pipeline} for archival Band~7 data.

Self-calibration of the continuum visibilities is performed for all the sources in the AGE-PRO Lupus sample. 
The continuum visibilities are first created by flagging any possible line emission in each spectral window and averaging into a maximum 125 MHz channel width.
We find that the systemic local standard of rest (LSR) velocities ($v_{\rm sys}$) of the Lupus targets are $\sim$ 4 km\,s$^{-1}$, thus we flag channels with velocities between -19.0 and 21.0 km\,s$^{-1}$ as spectral lines, and these channels are removed from the continuum analysis.

Then we perform astrometric alignment and flux-scale alignment among different executions. 
We first self-calibrate the short baseline data (for the Band~7 setting this is where we stop because we only have short baseline data), and we concatenate the self-calibrated short-baseline data with the long-baseline data.
The combined visibilities are then self-calibrated together. 
First, iterations of phase-only self-calibration are performed.
We start with a large solution interval (\texttt{solint=`300s'} or \texttt{`inf'}), and after each iteration of self-calibration, the resulting peak intensity and image SNR are used to determine if we continue with another iteration.
As the last step, one amplitude self-calibration is attempted, usually with a solution interval equal to the scan length.
Self-calibration improves the SNR for the disks in AGE-PRO Lupus sample by $\sim 50\%$ within a few iterations.

These alignments and calibration tables are also applied to the original visibilities that do not have any spectral averaging and include the spectral lines.
Then we subtract the continuum from these aligned and self-calibrated visibilities to provide a continuum-subtracted measurement set for each source for further analysis on different spectral lines.

The data products, continuum+line measurement
sets and data reduction scripts are available for download from the AGE-PRO website.

\subsection{Imaging Strategy}

The continuum images in both Bands~6 (234\,GHz) and~7 (285\,GHz) are generated while the self-calibration is conducted. For the final continuum image product after self-calibration, we use the \texttt{tclean} task with an elliptical mask, and a Briggs robust parameter of 0.5 to achieve an angular resolution of $\sim 0.35\,\arcsec$ for Band~6 and $\sim 0.70\,\arcsec$ for Band~7.

Before the line imaging, we split the data into individual measurement sets for each target line.
We also use the constraint of \texttt{uvrange = `>50klambda'} for Band~6 data to remove the cloud contamination in large-scale structures $\ge 4\arcsec$ that heavily affect $\mathrm{{}^{12}CO}\,(2-1)$ line emission, and sometimes in the $\mathrm{{}^{13}CO}\,(2-1)$ line. 
The choice of the 4\arcsec criteria is based on the initial images, where we see the $\mathrm{{}^{12}CO}\,(2-1)$ line emission size of the disk is $< 4\arcsec$, but cloud contamination structures can be more extended.
We do not apply such constraint for V1094~Sco (Lupus 10), because it has a large disk (see Section~\ref{sec:Results}).
Then the split measurement sets are re-sampled onto a common velocity grid using the \texttt{cvel2} task.
A velocity spacing of $0.1\,\mathrm{km\,s^{-1}}$ is used for Band~6, $\mathrm{{}^{12}CO}\,(2-1)$ line, and $0.2\,\mathrm{km\,s^{-1}}$ is used for all other lines in both Band~6 and~7, unless otherwise stated.
Only a few targets have a previously measured system velocity ($v_\mathrm{sys}$), thus we set a large range of channels from $-19.0\,\mathrm{km\,s^{-1}}$ to $+21\,\mathrm{km\,s^{-1}}$ to capture the entire disk emission.

The line images are then produced with the \texttt{tclean} task.
We start by using the \texttt{tclean} tasks with a Briggs robust parameter of 0.5 for all line images to achieve the same high spatial resolution of $\sim 0.35\,\arcsec$ as the continuum images, which also gives high image fidelity for the $\mathrm{{}^{12}CO}$.
However, for other weaker lines including $\mathrm{{}^{13}CO}$, $\mathrm{C^{18}O}$ and $\mathrm{N_2H^+}$, the robust value of 0.5 resulted in a low SNR.
Therefore, to boost the SNR and provide more accurate total fluxes, a robust value of 1.0 is chosen to generate images for other lines.
This choice results in a larger beam size of $\sim 0.50\,\arcsec$ (Band~6) and $\sim 0.80\,\arcsec$ (Band~7).

The \texttt{CLEAN} masks for the \texttt{tclean} tasks are adjusted manually in \texttt{interactive} mode and checked channel by channel by visual inspection starting from the assumption of Keplerian rotation for the disk gas.
In this step, we apply the following strategy: 
(a) The disk's systemic velocity ($v_\mathrm{sys}$) and its geometric parameters (such as inclinations and position angles) are iteratively adjusted based on the channel maps of the $\mathrm{{}^{12}CO}\,(2-1)$ line emission. 
Note that the geometric parameters for the gas disk slightly deviate from those derived from continuum images, as detailed in Table~\ref{Tab:Keplerian_Masks}. 
(b) To account for non-Keplerian or more extended emissions, we use a large mask in each channel by adopting larger stellar masses than those in Table~\ref{Tab:targets}, along with a large maximum radius $R_{\rm max}$ and large heights of the emission layers.
Motivated by deep spatially resolved observations of a few disks \citep[e.g.,][]{law_MAPS_emittingsurface_2021}, we set the emitting height at 1\arcsec\ at $z_0/r_1\sim 0.23$ for $\mathrm{{}^{12}CO}\,(2-1)$, $z_0/r_1\sim 0.15$ for $\mathrm{{}^{13}CO}\,(2-1)$, $z_0/r_1\sim 0.10$ for other lines.
We remind the reader that the mask has several poorly constrained parameters -- e.g., stellar mass, inclination, and the emitting height are all degenerate when describing the extension of the disk emission -- and the values adopted here are to ensure we adopt a conservative mask that includes the entire disk to start self-calibration.

In accompanying works, \citet{Trapman_AGEPRO_XI_gas_disk_sizes} and \citet{Vioque_AGEPRO_X_dust_disks} carry out detailed analyses on deconvolved $\mathrm{{}^{12}CO}$\,$J$=2-1 images and continuum visibilities to estimate the gas and dust disk geometry and morphology, respectively. 
The estimated disk inclination and position angles from the two works are summarized in \citet{Vioque_AGEPRO_X_dust_disks} Table~1. 
Although their disk inclinations and position angles are in agreement with those adopted in the conservative Keplerian masks during \texttt{tclean} (see Table~\ref{Tab:Keplerian_Masks}), for the following steps, which include the generation of images shown in this paper and the measurements of fluxes and radii, we adopt the values from \citet{Vioque_AGEPRO_X_dust_disks} for consistency among the AGE-PRO papers.

\begin{deluxetable*}{c|cc|ccc|ccc}
\tablecaption{Measured disk geometries and systemic velocities from images and Keplerian mask parameters \label{Tab:Keplerian_Masks}}
\tablewidth{0.95\textwidth}
\tablehead{
\colhead{ID} &  \multicolumn{2}{c}{cont. image} &  \multicolumn{3}{c}{$\mathrm{{}^{12}CO}$ (2-1) image} & \multicolumn{3}{c}{other mask parameters}  \\
  \colhead{} &  \colhead{incl} &  \colhead{PA} &  \colhead{incl} &  \colhead{PA} &  \colhead{$v_\mathrm{sys}$} &  \colhead{$M_\star$} &  \colhead{$R_{\rm max}$} &  \colhead{$z_0/r_1$} \\
  \colhead{} &  \colhead{[deg]} &  \colhead{[deg]} &  \colhead{[deg]} &  \colhead{[deg]} &  \colhead{[${\rm km\,s^{-1}}$]} &  \colhead{[$M_\odot$]} &  \colhead{[arcsec]} &  \colhead{}
}
\startdata
1  & 61.61 & 289.58 & 62    & 290    & 4.40 & 0.84 & 1.6 & 0.30 \\
2  & 37.52 & 36.40  & 35    & 35     & 3.60 & 0.46 & 2.5 & 0.20 \\
3  & 45.00 & 196.59  & 57    & 195    & 4.60 & 0.52 & 1.5 & 0.23 \\
4  & 32.07 & 175.04  & 30    & 225    & 3.60 & 0.42 & 0.5 & 0.23 \\
5  & 17.00 & 86.98  & 45    & 43     & 3.80 & 0.87 & 0.5 & 0.20 \\
6  & 52.79 & 256.47  & 53    & 284    & 4.20 & 0.32 & 0.5 & 0.20 \\
7  & 42.90 & 342.14 & 43    & 342    & 3.40 & 0.33 & 0.7 & 0.20 \\
8  & 70.15 & 347.46 & 45    & 270    & 4.30 & 0.32 & 0.7 & 0.23 \\
9  & -     & -      & -     & -      & 3.10 & -    & 1.2\tablenotemark{a} & -   \\
10 & 49.91 & 106.11 & 53    & 108    & 5.40 & 1.21 & 8.0 & 0.20 \\
\enddata
\tablenotetext{a}{For Lupus~9, we are unable to measure its disk geometry from the continuum or $\mathrm{{}^{12}CO}$ (2-1) images due to low SNR. Hence, we do not apply a Keplerian mask during \texttt{tclean}; instead, we use a circularized mask with a radius of $1.2\arcsec$ for both cont. and gas line images.}
\end{deluxetable*}

We use the same set of inclination and position angles for all lines in Band 6 and 7 images, and the targets are shifted to the center of the image before applying the mask. The only exceptions are for Lupus 1 (Sz~65) and 8 (Sz~66) in Band~7, where new and archival data is first shifted to the center of the AGE-PRO execution and then mosaicked together in the imaging step.


We also re-image the Band~6 continuum and all CO isotopologue lines with a circularized beam $-$ these images are used to measure the dust and gas disk radii (see Section~\ref{subsec:measure_flux_n_radii}).
The circularization is carried out first with \texttt{uvtap} in the UV-plane to reach a circularization more than 90\%, and then we apply \texttt{imsmooth} to make a perfectly circularized image.
The emission is \texttt{CLEANed} down to $1 \times \mathrm{RMS}$ (root-mean-square of the background noise outside of the mask, which is referenced as $\sigma$ in this work), estimated in a line-free region.
More details on why we choose this \texttt{CLEAN} criterion is discussed in the accompanying AGE-PRO overview paper \citep{Zhang_AGEPRO_I_overview}.

The data products are published on the AGE-PRO website, and they include: (a) both Band 6 and 7 continuum images with typical noise level of $\sigma \sim 0.1 \,\mathrm{mJy}$ generated with robust = 0.5; (b) $\mathrm{{}^{12}CO}$ line datacubes with noise level of $\sigma \sim 7.0 \,\mathrm{mJy\,km\,s^{-1}}$ generated with robust = 0.5 and circularized beams; (c) $\mathrm{{}^{13}CO}$ ($\sigma \sim 4.0 \,\mathrm{mJy\,km\,s^{-1}}$) and $\mathrm{C^{18}O}$ ($\sigma \sim 4.0 \,\mathrm{mJy\,km\,s^{-1}}$) line datacubes with robust = 1.0 and circularized beams; (d) line datacubes with robust = 1.0 for all other lines including $\mathrm{N_2H^+}$ ($\sigma \sim 3.0 \,\mathrm{mJy\,km\,s^{-1}}$).

\section{Results}
\label{sec:Results}

\subsection{Band 6 Images}
\label{subsec:band6observation}

The AGE-PRO Band 6 set up includes the continuum image at 234\,GHz, three CO isotopologues $\mathrm{{}^{12}CO}$\,$J$=2-1 (230.5\,GHz), $\mathrm{{}^{13}CO}$\,$J$=2-1 (220.3\,GHz), and $\mathrm{C^{18}O}$\,$J$=2-1 (219.6\,GHz), and other lines in this wavelength coverage: $\mathrm{H_{2}CO}$ $J$=$\mathrm{3_{03}}$-$\mathrm{2_{02}}$ (218.2\,GHz), $\mathrm{DCN}$\,$J$=3-2 (217.2\,GHz), and $\mathrm{N_2D^+}$\,$J$=3-2 (231.3\,GHz). 

All sources are detected in the continuum (with SNR $\gtrsim 20$), $\mathrm{{}^{12}CO}$\,$J$=2-1 (SNR $\gtrsim 10$), and $\mathrm{{}^{13}CO}$\,$J$=2-1 (SNR $\gtrsim 5$ except Lupus 9 with SNR $\sim 3$).
$\mathrm{C^{18}O}$\,$J$=2-1 emission is detected with ${\rm SNR} > 3 \sigma$ for Lupus~1, 2, 4, 6, 7, 8 and 10, and with ${\rm SNR} \sim 1-3 \sigma$ for Lupus~3, 5 and 9.

\begin{figure*}
   \gridline{\fig{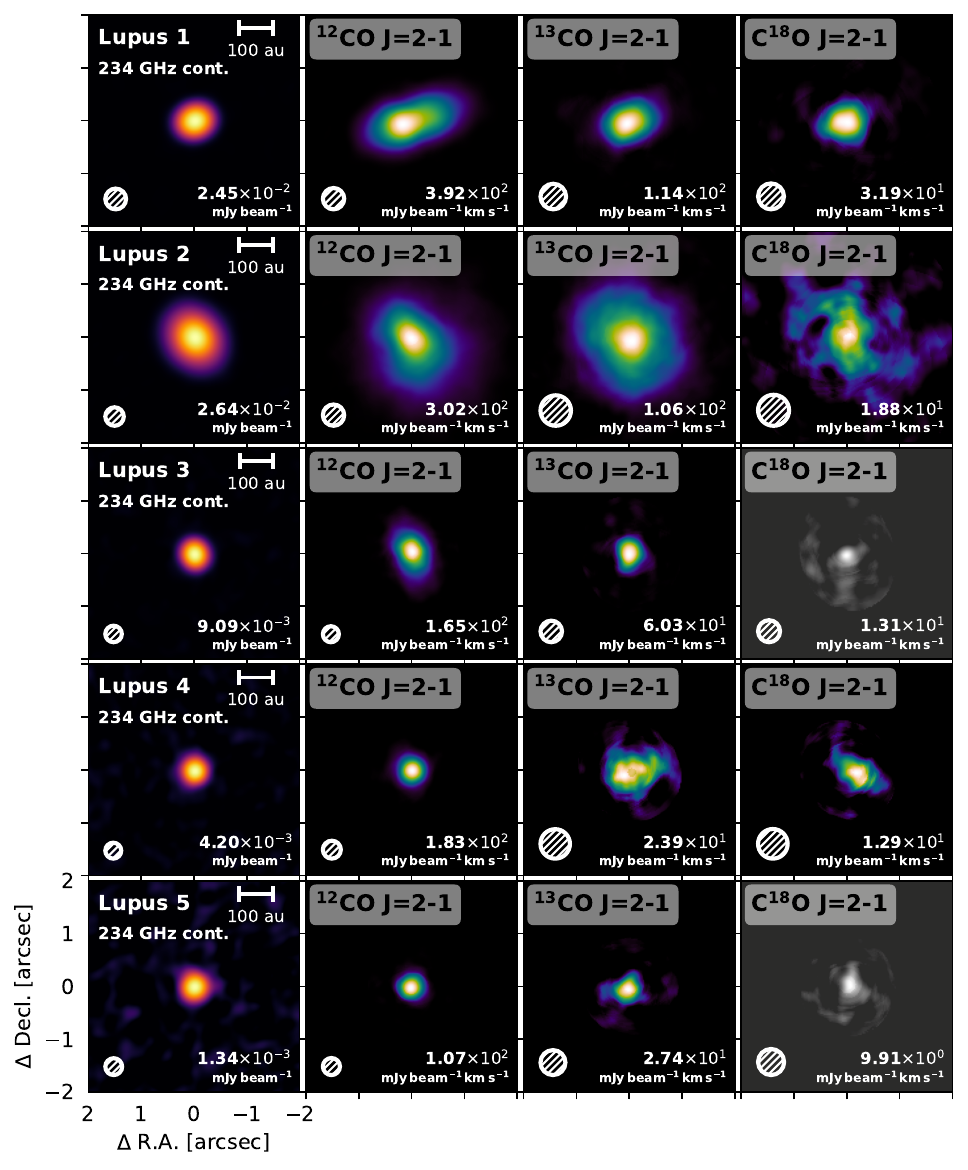}{0.37\textheight}{}
   \fig{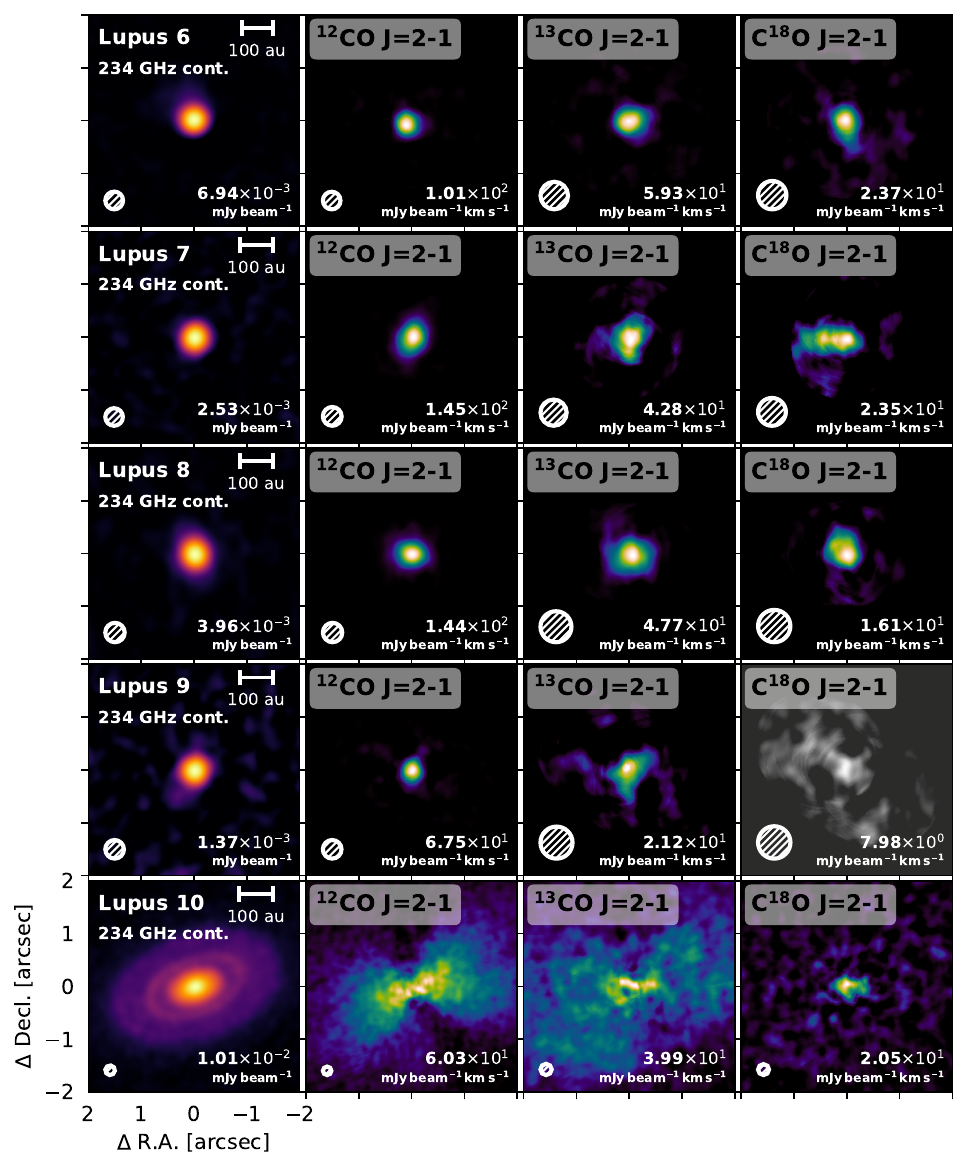}{0.37\textheight}{}
            }
  \caption{
  Continuum and CO isotopologue gas line images for the AGE-PRO Lupus targets. 
  The $\mathrm{{}^{12}CO\,(2-1)}$ images use robust 0.5 with circularized beam, while the $\mathrm{{}^{13}CO\,(2-1)}$ and $\mathrm{C^{18}O\,(2-1)}$ images use robust 1.0 with circularized beam.
  The beams are shown in the lower-left corner and the peak flux is in the lower-right corner of each panel. For the three disks without 3$\sigma$ detections on $\mathrm{C^{18}O\,(2-1)}$ line emission, their plots are shown in gray. 
  }
  \label{fig:gasline-gallery-1}
\end{figure*}

\begin{figure}[hbp]
   \gridline{\fig{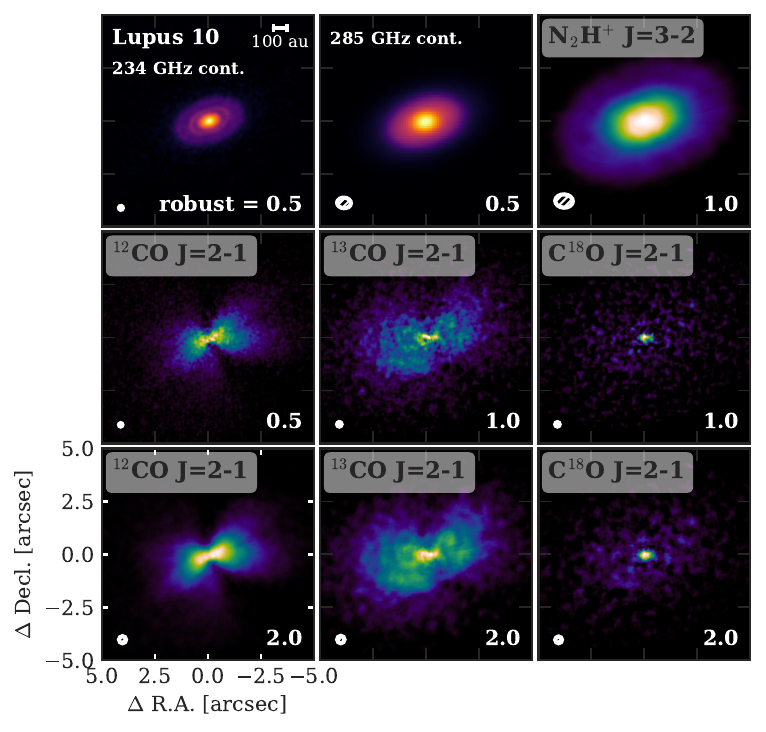}{0.47\textwidth}{}
            }
  \caption{
  Image gallery for Lupus~10, including continuum images in both bands, $\mathrm{N_2H^+}$\,$J$=3-2 and CO isotopologues with larger image size to show the extent of the full  disk. 
  The robust values are shown at the lower-right corner of each panel.
  We also use larger robust values for CO isotopologues in the bottom row during the imaging to increase the signal-to-noise ratio in each beam. 
  All other notations follow Figure~\ref{fig:gasline-gallery-1}.
  }
  \label{fig:gasline-gallery-3}
\end{figure}

Figure~\ref{fig:gasline-gallery-1} shows the Band 6 continuum images and the zeroth-moment (integrated flux) maps of CO isotopologue lines.
For Lupus 10, a zoom-out image with size of $\pm 5\arcsec$ is shown in Figure~\ref{fig:gasline-gallery-3} to present its full gas disk.

\begin{figure*}
   \gridline{\fig{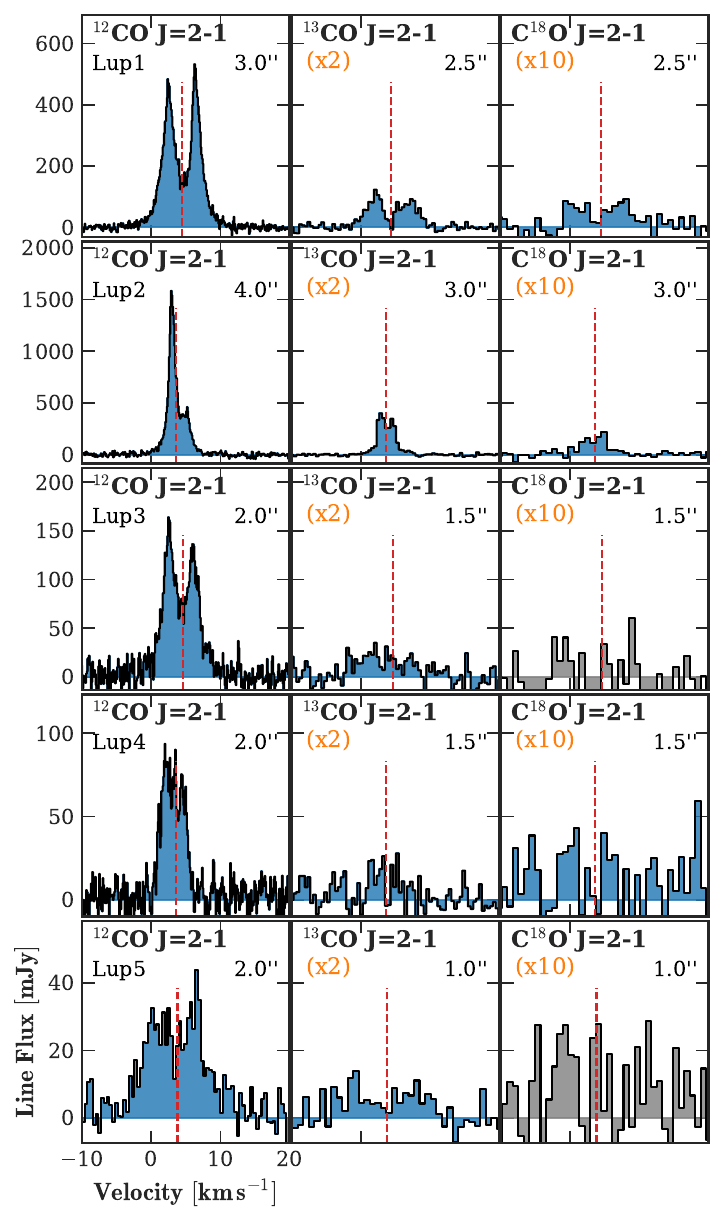}{0.37\textheight}{}
   \fig{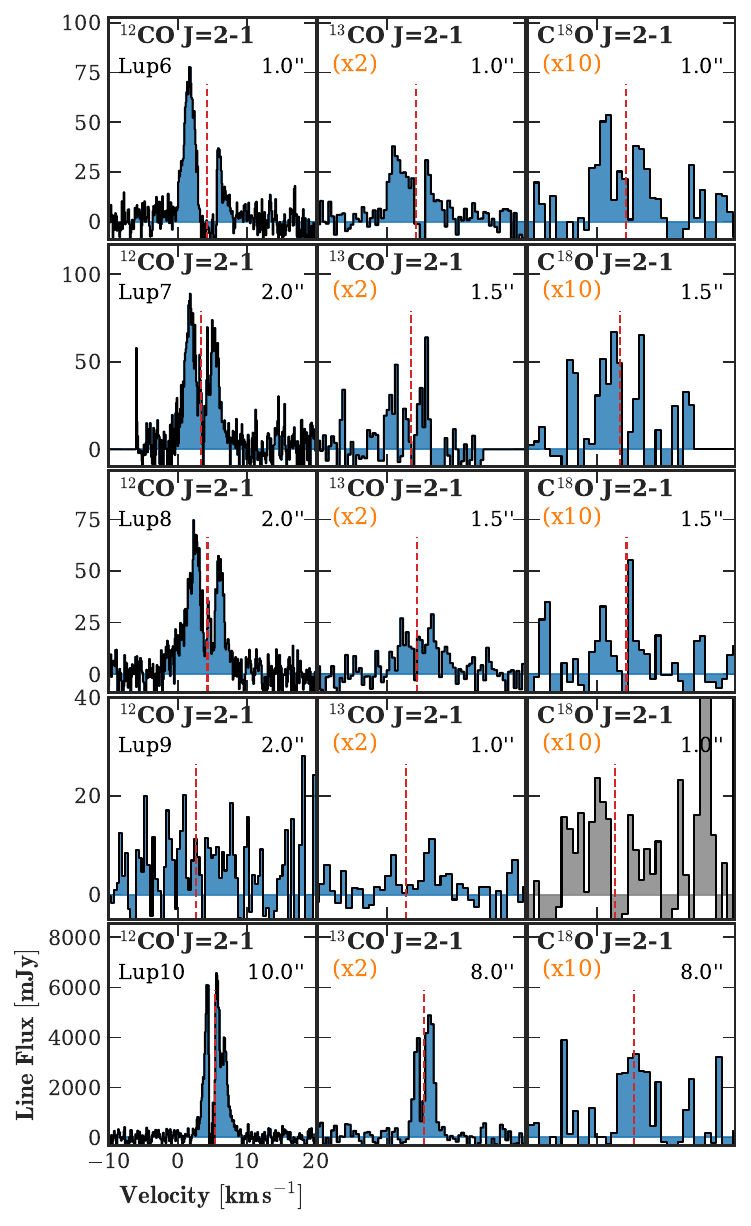}{0.385\textheight}{}
            }
  \caption{
Disk-integrated line spectra of CO isotopologues for AGE-PRO Lupus targets.
To boost the SNR in spectra, the channel width of $\mathrm{{}^{13}CO\,(2-1)}$ is re-binned to $0.4\,\mathrm{km\,s^{-1}}$, and that of $\mathrm{C^{18}O\,(2-1)}$ is re-binned to $0.8\,\mathrm{km\,s^{-1}}$.
  We use the aperture corresponding to the gas disk radii defined in $\mathrm{{}^{12}CO\,(2-1)}$, and smaller apertures for $\mathrm{{}^{13}CO\,(2-1)}$ and $\mathrm{C^{18}O\,(2-1)}$, and the adopted values are shown on the upper-right corner in each panel.
  The fluxes of $\mathrm{{}^{13}CO}\,(2-1)$ and $\mathrm{C^{18}O}\,(2-1)$ are scaled up by a factor of $\times 2$ and $\times 10$ from observed fluxes for better visualization, respectively.
  For the three disks without $3\,\sigma$ detections on $\mathrm{C^{18}O}\,(2-1)$ line emission, their plot are shown in gray color.
  The disk systemic velocities identified from $\mathrm{{}^{12}CO}\,(2-1)$ lines are shown as red dashed lines.
  }
  \label{fig:gasline-lineprofile-1}
\end{figure*}

Figure~\ref{fig:gasline-lineprofile-1} shows the spectra of each target.
We use the aperture corresponding to the gas disk radii defined in $\mathrm{{}^{12}CO}$ (see Section~\ref{subsec:measure_flux_n_radii}), and smaller apertures for $\mathrm{{}^{13}CO}$ and $\mathrm{C^{18}O}$, and the adopted apertures are shown on the upper-right corner on each panel.
To make these spectra, we re-bin the channel width of $\mathrm{{}^{13}CO}$ and $\mathrm{C^{18}O}$ to 0.4 and 0.8\,$\mathrm{km\,s^{-1}}$ respectively to boost the SNR in each channel.

In the zeroth-moment (moment~0) maps and the line spectrum of $\mathrm{{}^{12}CO}$ images, the emission is asymmetric for Lupus 1, and 2. 
This is due to cloud contamination, which absorbs the emission from the disk in certain velocity channels. 
For example, a clear cloud contamination can be seen in the channels around 5$\,\mathrm{km\,s^{-1}}$ for the Lupus 2 $\mathrm{{}^{12}CO}$ (Figure~\ref{fig:gasline-lineprofile-1}), which leads to a clear absorption of the $\mathrm{{}^{12}CO}$ emission, and thus creating an asymmetric line profile.
An asymmetry is also presented in the first-moment (moment~1) maps (Figure~\ref{fig:gasline-gallery-CO_M1maps} in Appendix~\ref{appendix:m1maps}).
Although we removed the smallest baselines to reduce contamination from large scale structures, absorption from the nearby cloud is still an issue and strongly affects part of the channels in the $\mathrm{{}^{12}CO}$ images.
We further discuss how to correct for cloud contamination in Section~\ref{subsec:measure_flux_n_radii} when measuring the disk emission and size from the $\mathrm{{}^{12}CO}$  images.
The $\mathrm{{}^{13}CO}$ emission experiences a similar cloud contamination, though the extent is significantly less pronounced -- with more symmetric line spectra, moment~0 and moment~1 maps -- and therefore we do not attempt to remove the contamination there.

While $\mathrm{C^{18}O}$ does not show evidence of cloud contamination in the images or spectra, there are some structures present in the moment~0 maps: a `spiral-like' structure in Lupus~2, and `tail-like' structure in Lupus~4 and 7.
These structures only appear in $\mathrm{C^{18}O}$ but not in the other two CO isotopologues, and could be signatures of possible interaction with the cloud \citep[e.g.,][]{pineda_bubbles_ISM_PPVII_2023, winter_spatially_Lupus_ISM_2024}.
However, the flux level of these structures is just $\sim 1-2$ times of the $\mathrm{RMS}$.
Therefore, these structures may not be real kinematic structures, and are most likely due to the low sensitivity in those regions.
The confirmation of such substructures needs deep follow-up observations.

\begin{figure*}
   \gridline{\fig{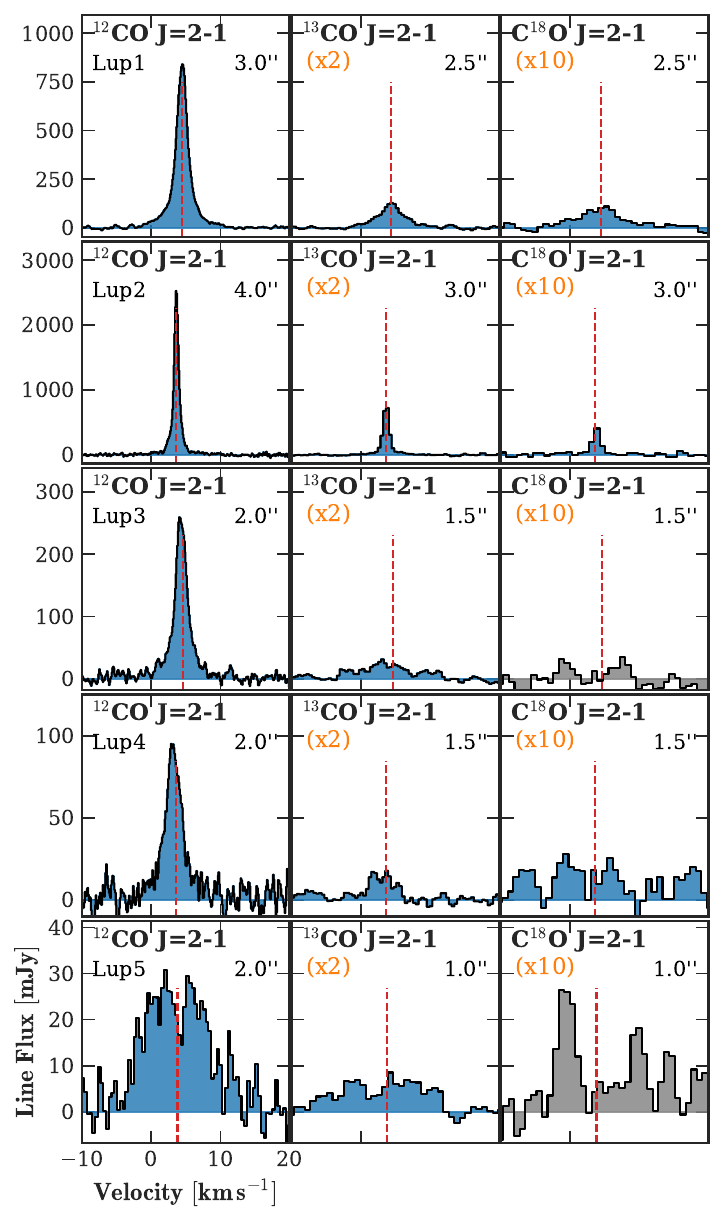}{0.44\textwidth}{}
   \fig{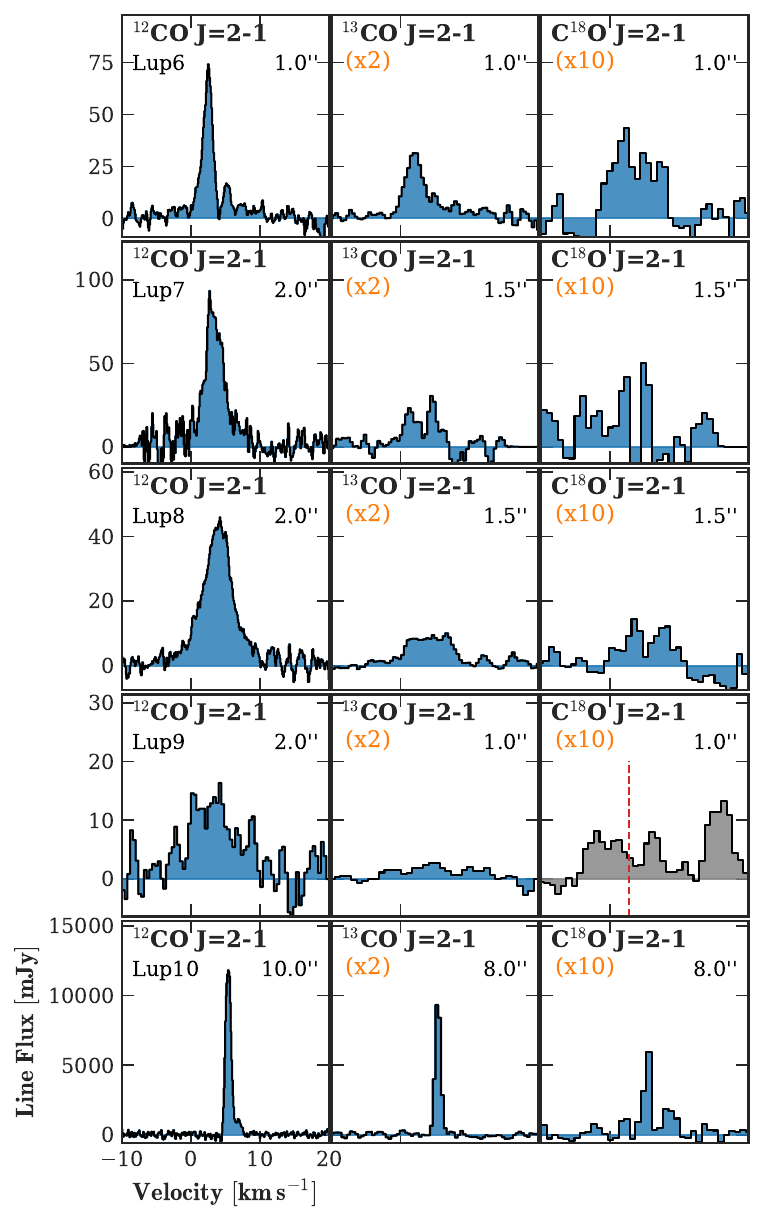}{0.465\textwidth}{}
            }
  \caption{
  The gas line velocity stacked spectra (created with \texttt{GoFish}) of CO isotopologues for AGE-PRO Lupus targets.
  We adopt the stellar parameters in Table~\ref{Tab:targets}, and the disk geometric parameters derived from $\mathrm{{}^{12}CO}\,(2-1)$ images in Table~\ref{Tab:Keplerian_Masks}.
  This figure follows the same notation as Figure~\ref{fig:gasline-lineprofile-1}.
  }
  \label{fig:gasline-lineprofile-vstacked-1}
\end{figure*}

\begin{figure*}
   \gridline{\fig{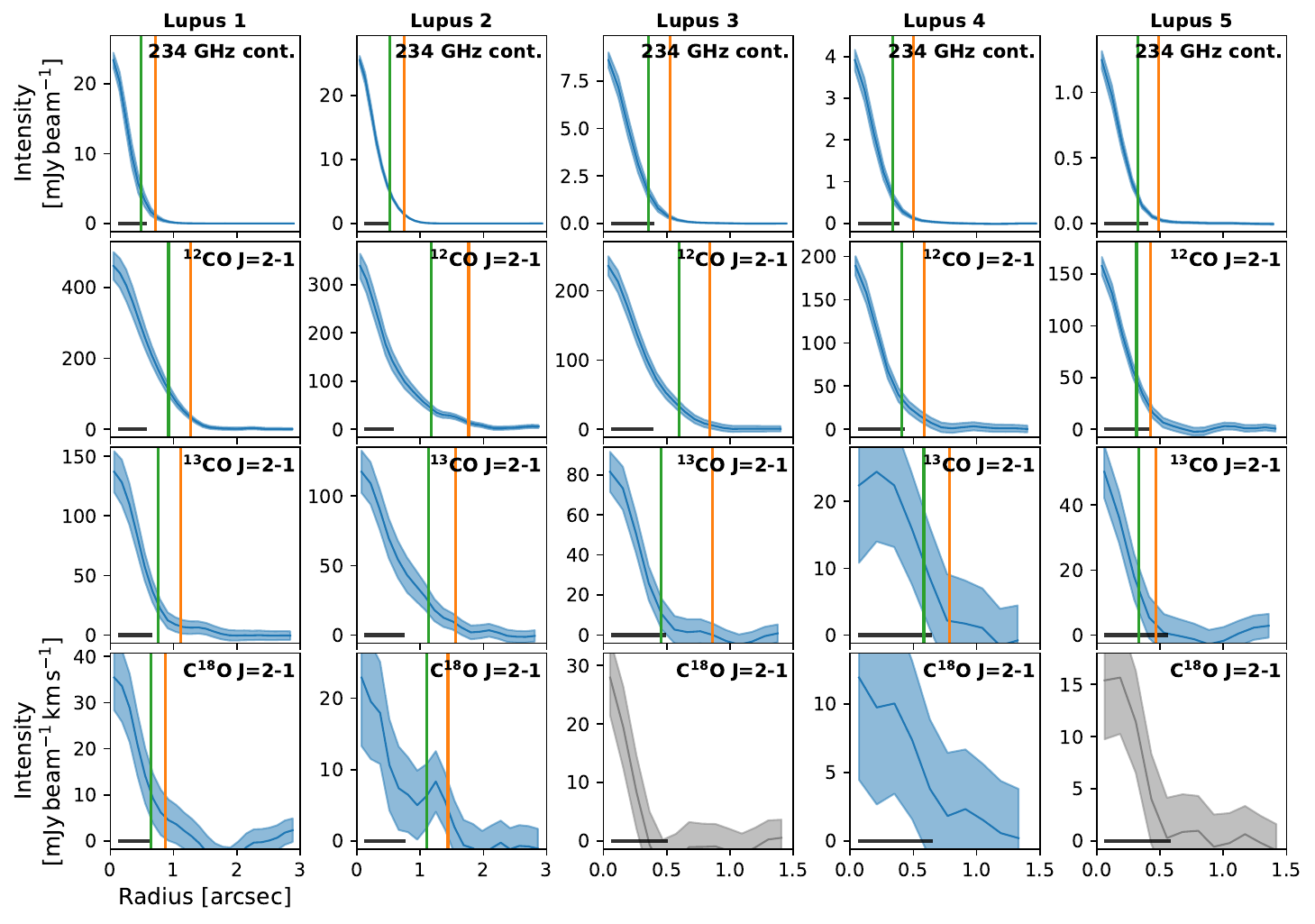}{0.80\textwidth}{}
            }
    \gridline{\fig{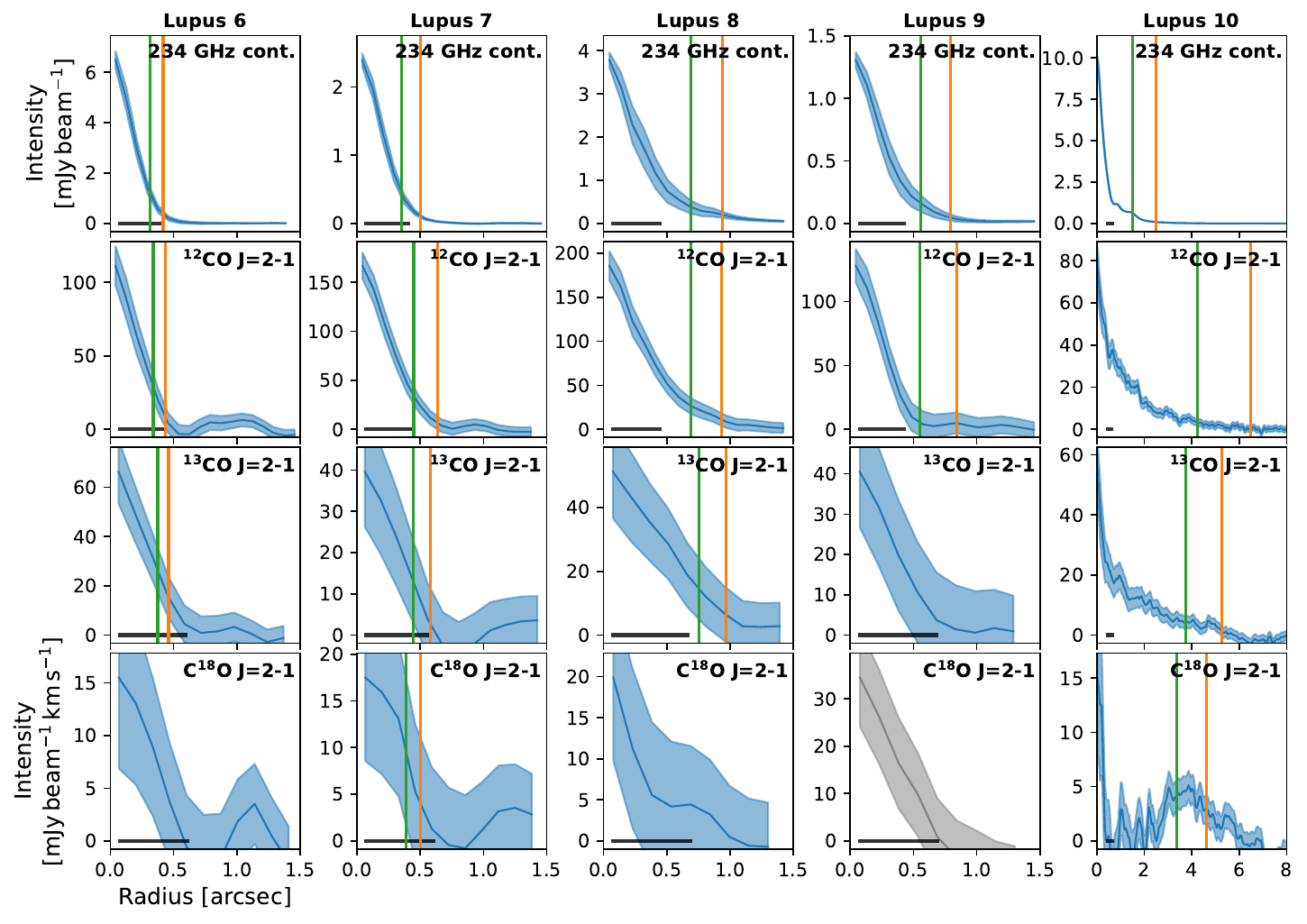}{0.80\textwidth}{}}
  \caption{
  Gas line radial profiles of CO isotopologues for the Lupus targets extracted by \texttt{GoFish}.
  These images are de-projected using the stellar parameters in Table~\ref{Tab:targets}, and the disk geometric parameters derived from $\mathrm{{}^{12}CO}\,(2-1)$ images in Table~\ref{Tab:Keplerian_Masks} using the inclination and position angles in latter columns from \citet{Vioque_AGEPRO_X_dust_disks}.
  The $R_{68}$ and $R_{90}$ are shown as green and orange solid lines, respectively.
  The beam sizes are presented as black lines in the lower-left corner of each panel.
  The line emissions without $3\sigma$ detections are shown in gray color.
  }
  \label{fig:gasline-radialprofile-1}
\end{figure*}

We also use \texttt{GoFish} code \footnote{\url{https://github.com/richteague/gofish}} \citep{GoFish} to generate: i) the velocity-stacked spectra (Figure~\ref{fig:gasline-lineprofile-vstacked-1}) -- where the line spectra are de-projected according to the Keplerian velocity field, using the method first introduced in \citet{Yen_Vstack_spectra2016} -- and ii) the radial profile of the CO isotopologue line images (Figure~\ref{fig:gasline-radialprofile-1}, where the continuum radial profile is also presented) to help confirm the line detections.
These two plots are de-projected with the disk geometry (inclination angle and position angle from visibility fitting summarized in \citealt{Vioque_AGEPRO_X_dust_disks}) to boost the SNR of the line images.

Some weaker CO isotopologue lines present as single-peak velocity-stacked spectra (Figure~\ref{fig:gasline-lineprofile-vstacked-1}), and also clearly centered in the radial profile, where their emission peaks at the image center and drops outwards (Figure~\ref{fig:gasline-radialprofile-1}).
Combining moment~0 maps, line spectra, and radial profiles, we can conclude that we have detections of $\mathrm{{}^{13}CO\,(2-1)}$ for Lupus~4, 5, and 9, and $\mathrm{C^{18}O\,(2-1)}$ for Lupus 7 and 8. 
However, the $\mathrm{C^{18}O\,(2-1)}$ for Lupus~3, 5 and 9 do not have a single-peaked line profile in these velocity-stacked (de-projected) spectra, so that whether they have a real detection is not clear.
Therefore, we conclude that we have clear detections on $\mathrm{{}^{12}CO\,(2-1)}$ and $\mathrm{{}^{13}CO\,(2-1)}$ for all the targets, but might not have good detections on $\mathrm{C^{18}O\,(2-1)}$ for Lupus~3, 5 and 9.

We also have detections of: $\mathrm{H_2CO}\,J\mathrm{=3_{03}-2_{02}}$ for Lupus~2, 3 and 10; $\mathrm{DCN}$\,$J$=3-2 for Lupus 2 and 3; and $\mathrm{N_2D^+}$\,$J$=3-2 for Lupus~2.
The image galleries for other molecular lines in Band~6 are presented in Figures~\ref{fig:allinone_M0_gallery} and ~\ref{fig:allinone_M0_gallery_Lupus10}, in  Appendix~\ref{appendix:other_weak_lines}.
We also find potential companions around Lupus~6 which is discussed in Appendix~\ref{appendix:lupus6companions}.

\subsection{Band 7 Images}
\label{subsec:band7observation}

The AGE-PRO Band~7 set-up includes the continuum at 285\,GHz, the $\mathrm{N_2H^+}$\,$J$=3-2 (279.5\,GHz) line, and other molecular lines including $\mathrm{DCO^+}$\,$J$=4-3 (288.1\,GHz), $\mathrm{C^{34}S}$\,$J$=6-5 (289.2\,GHz), $\mathrm{DCN}$\,$J$=4-3 (289.6\,GHz), and $\mathrm{H_2CO}$\,$J$=$\mathrm{4_{04}}$-$\mathrm{3_{03}}$ (290.6\,GHz).
We note that there are two exceptions to the velocity spacing in the Band~7 data, where we do not use the value of $0.2\,\mathrm{km\,s^{-1}}$: (a) We adopt $1.2\,\mathrm{km\,s^{-1}}$ for the  $\mathrm{H_2CO}\,J=4_{04}-3_{03}$ line because it is in a continuum window that has low spectral resolution; (b) For the archival data, we use $0.3\,\mathrm{km\,s^{-1}}$ for all the lines except $\mathrm{N_2H^+\,(3-2)}$, because their native spectral resolution is only $252\,\mathrm{m\,s^{-1}}$.

$\mathrm{N_2H^+}\,(3-2)$ emission is detected in 3 disks (Lupus~2, 3 and 10) out of 10 targets with ${\rm SNR} > 5\,\sigma$, 1 target (Lupus~1) with ${\rm SNR} \sim 3$, and there is centrally-peaked emission in other 4 disks (Lupus~4, 6, 8, 9) but with $< 3\,\sigma$. There is no central emission in the two remaining disks (Lupus~5 and 7).

\begin{figure*}
   \gridline{\fig{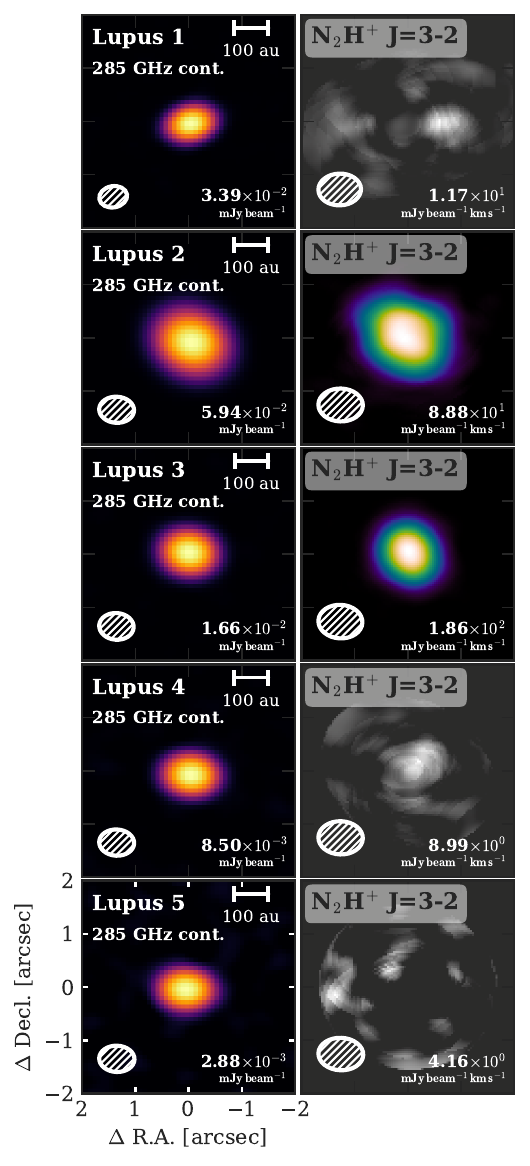}{0.35\textwidth}{}
   \fig{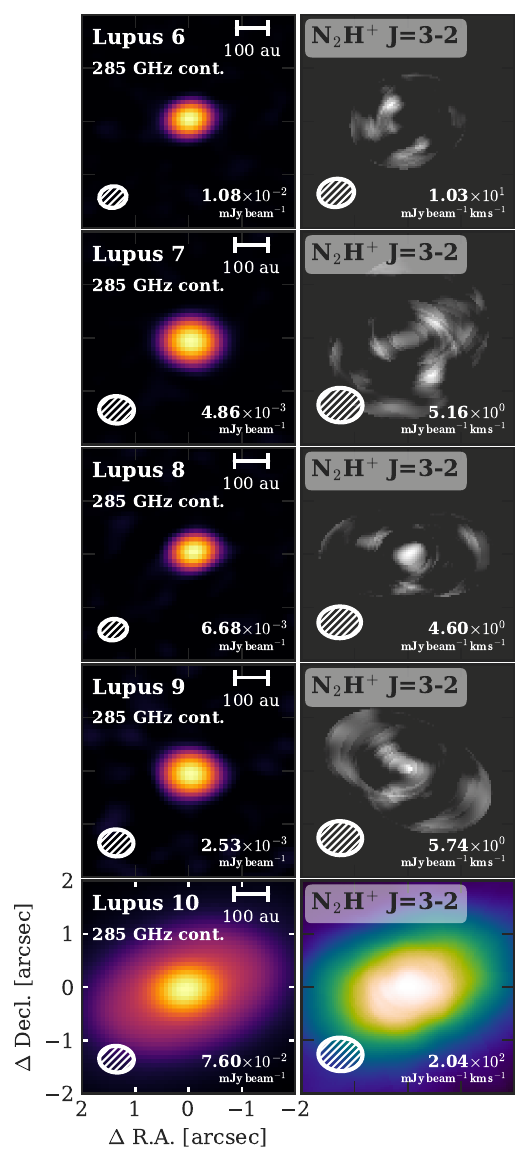}{0.35\textwidth}{}
            }
    \caption{
    The Band~7 continuum and $\mathrm{N_2H^+}\,(3-2)$ zeroth-moment maps gallery of Band 7. 
    The disks with clear $\mathrm{N_2H^+}\,(3-2)$ detections are in \texttt{rainforest} color and non-detections in gray color.
    This Figure follows the same notations as Figure~\ref{fig:gasline-gallery-1}.
    The extended images of Band~7 continuum and $\mathrm{N_2H^+}\,(3-2)$ zeroth-moment maps for Lupus~10 is also shown in Figure~\ref{fig:gasline-gallery-3}.
    }
  \label{fig:gasline-gallery-B7-0}
\end{figure*}

Figures~\ref{fig:gasline-gallery-B7-0} shows the Band 7 continuum images and the $\mathrm{N_2H^+}\,(3-2)$ zeroth-moment maps, where only the ones with clear detections are color-coded.
The $\mathrm{N_2H^+}\,(3-2)$ emission for Lupus 2 and 3 are barely resolved.
But for the largest disk, Lupus 10, the $\mathrm{N_2H^+}\,(3-2)$ emission is resolved and its emission extends to $\sim 4 \arcsec$ ($\sim 600\,\mathrm{AU}$).

We also have detections of $\mathrm{DCO^+}$\,$J$=4-3, $\mathrm{DCN}$\,$J$=4-3, $\mathrm{H_2CO}\,J=4_{04}-3_{03}$ in Lupus~2, 3, and 10; and $\mathrm{C^{34}S}$\,$J$=6-5 in Lupus 10.
The image galleries for these lines are also presented in Figures~\ref{fig:allinone_M0_gallery} and ~\ref{fig:allinone_M0_gallery_Lupus10}, together with their measured fluxes in Appendix~\ref{appendix:other_weak_lines}.

\subsection{Fluxes and Disk Radii}
\label{subsec:measure_flux_n_radii}

The total flux and the radius of the continuum and gas emission lines are measured from the curve-of-growth method following the same method adopted in the literature \citep[e.g.,][]{Ansdell_2016ApJ_Lupus, Ansdell_Lupus_2018ApJ, Andrews_DSHARP_2018ApJ...869L..41A, Sanchis_Lupus_radius_2021A&A...649A..19S}.

For gas emission lines, the moment~0 maps are made with a Keplerian mask using the parameters given in Table~\ref{Tab:Keplerian_Masks} with disk inclination and position angle from \citet{Vioque_AGEPRO_X_dust_disks}, but with $1.5\times$ larger $R_\mathrm{max}$ to make sure all the emission is included. 
Then, we create a series of elliptical apertures upon the moment~0 maps. 
The apertures correspond to the circular aperture projected to the corresponding disk inclination and position angle. 
The apertures are set up with different radii centered at the stellar location, and the flux inside each aperture is measured to compile a curve-of-growth of flux vs. radius $r$. 
The disk total flux is measured when the curve-of-growth reaches the first peak of the oscillating plateau modulated by noise, and the variation of total flux after the peak does not exceed the noise.
Flux uncertainties are estimated as the square root of the number of beams that are inside the aperture for measuring the total flux, and multiplied by the per-beam noise measured from an emission free part of the image outside of that aperture.
To validate these uncertainties estimated from root-mean-square (RMS), we also carry out a bootstrapping method on three CO isotopologue lines (described in Appendix~\ref{appendix:bootstrap}). 
We find the uncertainties from both methods are in agree with each other, except for the large disk around Lupus~10 (Figure~\ref{fig:flux_comp_bootstrap} and Appendix~\ref{appendix:bootstrap} for details).
We report the $R\mathrm{_{68}}$ and $R\mathrm{_{90}}$ at the radius where the curve-of-growth flux is 68\% and 90\% of the total flux.
Uncertainties in these radii are estimated by calculating the range of radii corresponding to the uncertainties of the curve-of-growth flux at the 68\% and 90\% flux levels. 

For continuum images, the same curve-of-growth is generated simply on top of the images, and their total flux densities and radii are measured in the same way.
More details on curve-of-growth method are described in \citet{Zhang_AGEPRO_I_overview} Section~4.2.

The $\mathrm{{}^{12}CO\,(2-1)}$ images for Lupus~2, 3, and 6 suffer significantly from cloud contamination, making the flux and radius measurements obtained inaccurate. 
As a first-order correction, we compare the two halves of the image based on their position angles and line spectra, and employ the curve-of-growth method on the half of the image that is comparably less affected by cloud contamination to determine the flux and radius.
For Lupus~2, 3, and 6, we measure the total flux from this uncontaminated half using a velocity range of $v \leq v_{\rm sys}$, denoting this as $\frac{1}{2}F^{\rm cor}_\mathrm{{}^{12}CO\,(2-1)}$. 
We then multiply it by $\times 2$ to obtain the corrected flux $F^{\rm cor}_\mathrm{{}^{12}CO\,(2-1)}$. 

The measured flux from the less-contaminated half could still be underestimated because of some minor contamination not accounted for in our approach.
Contamination could be better treated by utilizing only the `clean' channels and derive the disk properties through visibility modeling, but the estimated fluxes could still be associated with large uncertainties because of the missing information behind contamination \citep[e.g.,][]{kurtovic_n_Pinilla_2024_recover_gasproperties_MHO6}.
The disk sizes measured from less-contaminated halves are not sensitive to these minor contamination, and we find they are consistent with the beam-deconvolved sizes estimated through models, where the contaminated channels are removed (Figure~\ref{fig:comp-with-literature-radius} in Appendix~\ref{appendix:compare_literature}).

The measured fluxes for the continuum, CO isotopologues, and $\mathrm{N_2H^+\,(3-2)}$ from the curve-of-growth are presented in Table~\ref{Tab:COG-flux}, along with the corrected $\mathrm{{}^{12}CO\,(2-1)}$ fluxes for Lupus~2, 3, and 6.
We report the fluxes for the lines without $> 3\sigma$ detections, but we recommend to use 3$\times$ the uncertainties as a conservative upper limits.
The disk radii measured from dust continuum and $\mathrm{{}^{12}CO}$ line images are summarized in Table~\ref{Tab:COG-radius}.

In Table~\ref{Tab:COG-flux}, we also report the global millimeter spectral index $\alpha_{\rm mm}$ for each target in our sample using the continuum fluxes from Band 6 (234\,GHz) and 7 (285\,GHz) by:
\begin{equation}
    \alpha_{\rm mm} = \frac{\ln{F_{\rm 234\,GHz}}-\ln{F_{\rm 285\,GHz}}}{\ln{\rm 234\,GHz}-\ln{\rm 285\,GHz}}.
\end{equation} 
The errors are estimated with the same procedure described in \citet{Chiang_spectralindex_2012}, which includes the systematic calibration uncertainty.
The spectral indices are useful indicators of the dust evolution in protoplanetary disks, including grain growth and possible presence of dust traps \citep[e.g.,][]{draine_submillimeter_dust_2006, ricci_dust_2010, testi_dust_2014, Pinilla2020_dust_traps, Stadler2022_dynamic_bumps, delussu2024_disk_substructure}.
These numbers are compared with dust evolutionary models and further discussed in the accompanying paper by \citet{Kurtovic_AGEPRO_VI_DustEvolution}.
We also identify that the Band~7 archival continuum image for Lupus~1 (Sz~65, Project ID: 2019.1.01135.S, PI: D. Anderson) has a flux mis-calibration based on the surprisingly low initial spectral indices of $\sim 1.10$.
We find its flux density is off by a factor of 1.22 based on the analysis on Lupus~8 (Sz~66), which was covered in the field-of-view of the archival data and is targeted as part of the new AGE-PRO observation.
Therefore, we correct the flux density of Lupus~1 and find a spectral index of $\sim 2.0$, which is consistent with the spectral indices of $\sim 2.3 \pm 0.3$ derived between the Band~6 and archival Band~8 ($\sim$340\,GHz) data (Project ID: 2019.2.00054.S, PI: M. Ansdell).
The corrected flux density and spectral index are reported in Table~\ref{Tab:COG-flux}. 

We also derive dust disk masses ($M_{\rm dust}$) from Band~6 continuum flux densities for each disk.
We adopt the same approach as in the literature \citep[e.g.,][]{Andrews_2013_Bayesian_isochrone, Miotello_PPVII_2023, Manara_2023_PPVII, Ruiz-Rodriguez_AGEPRO_II_Ophiuchus, Agurto-Gangas_AGEPRO_IV_UpperSco}, assuming the dust continuum fluxes are optically thin and derive the dust disk masses from the equation: 
\begin{equation}
    M_{\rm dust} = \frac{d^2}{B_\nu(T_{\rm dust}) \kappa_\nu} F_{\nu},
\end{equation}
where $d$ is the distance to the object, $B_\nu$ is the Plank spectrum at the average dust temperature $T_{\rm dust}$, and $\kappa_\nu$ is the dust opacity.
We adopt the average dust temperature of 20\,K, and the dust opacity as $\kappa_\nu = 2.3\,(\nu/{\rm 230\,GHz})\,{\rm cm^2\,g^{-1}}$.
The derived dust disk masses are presented in Table~\ref{Tab:COG-flux}.

For other lines besides CO isotopologues and $\mathrm{N_2H^+}$, we measure the total flux with the same curve-of-growth method for the resolved lines, and they are presented in Appendix~\ref{appendix:other_weak_lines}.
We do not measure their radii because most of them are weak and unresolved.

\begin{deluxetable*}{lrrrrrrrrr}
\tablecaption{Measured Fluxes from curve-of-growth method}\label{Tab:COG-flux}
\tablehead{
  \colhead{ID} &  \colhead{$F_{\rm 234\,GHz}$} &  \colhead{$F_{\rm 285\,GHz}$} &  \colhead{$\alpha_{\rm mm}$} &  \colhead{$M_{\rm dust}$} &  
  \colhead{$F_\mathrm{{}^{12}CO\,(2-1)}$} & \colhead{$F^{\rm cor}_\mathrm{{}^{12}CO\,(2-1)}$}\tablenotemark{a} &  \colhead{$F_\mathrm{{}^{13}CO\,(2-1)}$} &  \colhead{$F_\mathrm{C^{18}O\,(2-1)}$} & \colhead{$F_\mathrm{N_2H^+\,(3-2)}$} \\
  \colhead{} &  \colhead{[mJy]} &  \colhead{[mJy]} & \colhead{} & \colhead{[$M_\oplus$]} &  \colhead{[${\rm mJy\,km\,s^{-1}}$]} & \colhead{[${\rm mJy\,km\,s^{-1}}$]} &  \colhead{[${\rm mJy\,km\,s^{-1}}$]} &  \colhead{[${\rm mJy\,km\,s^{-1}}$]} & \colhead{[${\rm mJy\,km\,s^{-1}}$]} 
}
\startdata
    Lupus 1 &             30.3$\pm$0.1 &             48.2$\pm$0.2 &       2.35$\pm$0.02 &                19.41 &         1979.9$\pm$6.1 &               -- &          259.1$\pm$5.7 &           55.3$\pm$3.8 &           13.1$\pm$4.1 \\
    Lupus 2 &             76.4$\pm$0.1 &            109.4$\pm$0.1 &       1.83$\pm$0.01	 &                49.96 &        2602.3$\pm$10.4 &            3359.2$\pm$9.0 &          574.4$\pm$9.1 &           77.0$\pm$6.1 &          288.2$\pm$6.4 \\
    Lupus 3 &             12.0$\pm$0.1 &             17.9$\pm$0.1 &       2.00$\pm$0.04 &                 8.36 &         643.8$\pm$10.2 &             746.4$\pm$7.0 &           81.5$\pm$7.7 &            \textcolor{gray}{5.3$\pm$2.3\tablenotemark{b}} &          321.5$\pm$4.7 \\
    Lupus 4 &              5.6$\pm$0.1 &              8.6$\pm$0.1 &       2.15$\pm$0.10 &                 3.74 &          305.1$\pm$7.1 &               -- &           36.9$\pm$5.5 &            9.7$\pm$2.7 &            \textcolor{gray}{7.9$\pm$3.4\tablenotemark{b}} \\
    Lupus 5 &              1.8$\pm$0.1 &              3.1$\pm$0.1 &       2.66$\pm$0.29 &                 1.21 &          155.7$\pm$6.2 &               -- &           20.8$\pm$5.8 &            \textcolor{gray}{5.2$\pm$2.4\tablenotemark{b}} &            \textcolor{gray}{0.0$\pm$4.6\tablenotemark{bc}} \\
    Lupus 6 &              7.8$\pm$0.1 &             11.6$\pm$0.2 &       1.99$\pm$0.08 &                 5.65 &          131.7$\pm$5.4 &             205.4$\pm$5.4 &           46.7$\pm$3.8 &           16.3$\pm$2.5 &            \textcolor{gray}{8.8$\pm$5.6\tablenotemark{b}} \\
    Lupus 7 &              3.2$\pm$0.1 &              4.8$\pm$0.1 &       2.08$\pm$0.14 &                 2.22 &          278.6$\pm$9.7 &               -- &           50.1$\pm$6.5 &           15.2$\pm$3.9 &            \textcolor{gray}{0.0$\pm$1.29\tablenotemark{bc}} \\
    Lupus 8 &              4.6$\pm$0.1 &              9.1$\pm$0.1 &       3.48$\pm$0.08 &                 3.03 &          213.9$\pm$3.0 &               -- &           36.7$\pm$2.1 &            9.0$\pm$1.4 &            \textcolor{gray}{6.1$\pm$3.7\tablenotemark{b}} \\
    Lupus 9 &              1.5$\pm$0.1 &              2.8$\pm$0.1 &       2.94$\pm$0.21 &                 1.08 &           72.2$\pm$4.2 &               -- &           10.9$\pm$2.7 &            \textcolor{gray}{5.5$\pm$1.8\tablenotemark{b}} &            \textcolor{gray}{1.3$\pm$1.4\tablenotemark{b}} \\
    Lupus 10 &            220.2$\pm$0.5 &            382.7$\pm$0.1 &       2.80$\pm$0.01 &               149.32 &       12273.4$\pm$47.3\tablenotemark{d} &               -- &        6574.0$\pm$41.7\tablenotemark{d} &         970.0$\pm$21.5\tablenotemark{d} &        4017.6$\pm$13.8\tablenotemark{d} \\
\enddata
\tablenotetext{a}{$F^{\rm cor}_\mathrm{{}^{12}CO\,(2-1)}$: The corrected flux for $\mathrm{^{12}CO}\,(2-1)$ line measured from the velocity channels where the cloud contamination is not significant.}
\tablenotetext{b}{The measured flux is below $3\,\sigma$ detection limits. We recommend using the measured flux plus $3\times$ the uncertainties as upper limits.}
\tablenotetext{c}{The measured flux from the curve-of-growth is giving a negative flux (the total flux drops as the radius) because of low SNR and here we assign the median values manually to 0.0 and show their RMS from the moment~0 maps as the uncertainties.}
\tablenotetext{d}{The RMS uncertainties measured for Lupus~10 line images could be underestimated, the more conservative uncertainty estimates from the bootstrapping method are about a factor of $\sim 10$ larger for CO isotopologues (see Appendix~\ref{appendix:bootstrap} for details).}
\end{deluxetable*}

\begin{deluxetable*}{c|c|cr|cr|c|cr|cr}
\tablecaption{Measured radius from curve-of-growth method from images \label{Tab:COG-radius}}
\tablehead{
\multicolumn{1}{c|}{} & \multicolumn{5}{c|}{Band~6 ${\rm 234\,GHz\, cont.}$} & \multicolumn{5}{c}{$\mathrm{{}^{12}CO\,(2-1)}$}
\\
\multicolumn{1}{c|}{ID} &    \colhead{Beam} &  \multicolumn{2}{c}{$R\mathrm{_{68}}$} &  \multicolumn{2}{c|}{$R\mathrm{_{90}}$} & \colhead{Beam} &  \multicolumn{2}{c}{$R\mathrm{_{68}}$} & \multicolumn{2}{c}{$R\mathrm{_{90}}$} \\
\multicolumn{1}{c|}{} &     \colhead{[arcsec]} &  \colhead{[arcsec]} & \colhead{[AU]} &  \colhead{[arcsec]} &  \multicolumn{1}{c|}{[AU]} & \colhead{[arcsec]} &  \colhead{[arcsec]} &  \colhead{[AU]} &  \colhead{[arcsec]} &  \colhead{[AU]}
}
\startdata
    Lupus 1 & 0.393 $\times$ 0.393 & 0.49$\mathrm{^{+0.001}_{-0.001}}$ &  75.4$\mathrm{^{+0.2}_{-0.2}}$ & 0.71$\mathrm{^{+0.003}_{-0.003}}$ & 109.8$\mathrm{^{+0.5}_{-0.5}}$ & 0.404 $\times$ 0.404 & 0.92$\mathrm{^{+0.002}_{-0.002}}$ & 142.8$\mathrm{^{+0.4}_{-0.4}}$ & 1.27$\mathrm{^{+0.008}_{-0.008}}$ &   196.8$\mathrm{^{+1.2}_{-1.2}}$ \\
    Lupus 2 & 0.350 $\times$ 0.350 & 0.53$\mathrm{^{+0.001}_{-0.001}}$ &  82.0$\mathrm{^{+0.1}_{-0.1}}$ & 0.75$\mathrm{^{+0.002}_{-0.002}}$ & 116.8$\mathrm{^{+0.3}_{-0.2}}$ & 0.403 $\times$ 0.403 & 1.18$\mathrm{^{+0.005}_{-0.005}}$ & 184.1$\mathrm{^{+0.7}_{-0.7}}$ & 1.77$\mathrm{^{+0.015}_{-0.015}}$ &   275.6$\mathrm{^{+2.4}_{-2.4}}$ \\
    Lupus 3 & 0.313 $\times$ 0.313 & 0.36$\mathrm{^{+0.002}_{-0.002}}$ &  57.1$\mathrm{^{+0.3}_{-0.3}}$ & 0.52$\mathrm{^{+0.005}_{-0.005}}$ &  83.8$\mathrm{^{+0.8}_{-0.8}}$ & 0.303 $\times$ 0.303 & 0.60$\mathrm{^{+0.009}_{-0.009}}$ &  95.5$\mathrm{^{+1.4}_{-1.4}}$ & 0.84$\mathrm{^{+0.024}_{-0.020}}$ &   134.0$\mathrm{^{+3.8}_{-3.2}}$ \\
    Lupus 4 & 0.302 $\times$ 0.302 & 0.34$\mathrm{^{+0.004}_{-0.004}}$ &  52.7$\mathrm{^{+0.6}_{-0.6}}$ & 0.50$\mathrm{^{+0.018}_{-0.018}}$ &  78.4$\mathrm{^{+2.8}_{-2.8}}$ & 0.339 $\times$ 0.339 & 0.41$\mathrm{^{+0.008}_{-0.008}}$ &  63.4$\mathrm{^{+1.3}_{-1.3}}$ & 0.58$\mathrm{^{+0.031}_{-0.031}}$ &    91.0$\mathrm{^{+4.8}_{-4.8}}$ \\
    Lupus 5 & 0.319 $\times$ 0.319 & 0.32$\mathrm{^{+0.011}_{-0.011}}$ &  50.3$\mathrm{^{+1.7}_{-1.7}}$ & 0.49$\mathrm{^{+0.074}_{-0.049}}$ & 75.3$\mathrm{^{+11.5}_{-7.6}}$ & 0.323 $\times$ 0.323 & 0.31$\mathrm{^{+0.009}_{-0.009}}$ &  48.7$\mathrm{^{+1.4}_{-1.4}}$ & 0.43$\mathrm{^{+0.024}_{-0.024}}$ &    66.0$\mathrm{^{+3.7}_{-3.7}}$ \\
    Lupus 6 & 0.336 $\times$ 0.336 & 0.31$\mathrm{^{+0.002}_{-0.002}}$ &  52.6$\mathrm{^{+0.3}_{-0.3}}$ & 0.42$\mathrm{^{+0.003}_{-0.003}}$ &  70.1$\mathrm{^{+0.6}_{-0.6}}$ & 0.333 $\times$ 0.333 & 0.34$\mathrm{^{+0.010}_{-0.010}}$ &  56.9$\mathrm{^{+1.7}_{-1.7}}$ & 0.44$\mathrm{^{+0.023}_{-0.019}}$ &    73.4$\mathrm{^{+3.8}_{-3.1}}$ \\
    Lupus 7 & 0.333 $\times$ 0.333 & 0.35$\mathrm{^{+0.006}_{-0.004}}$ &  56.1$\mathrm{^{+0.9}_{-0.6}}$ & 0.51$\mathrm{^{+0.016}_{-0.016}}$ &  81.1$\mathrm{^{+2.5}_{-2.5}}$ & 0.355 $\times$ 0.355 & 0.45$\mathrm{^{+0.016}_{-0.012}}$ &  72.1$\mathrm{^{+2.6}_{-1.9}}$ & 0.64$\mathrm{^{+0.053}_{-0.036}}$ &   102.4$\mathrm{^{+8.6}_{-5.7}}$ \\
    Lupus 8 & 0.367 $\times$ 0.367 & 0.69$\mathrm{^{+0.006}_{-0.006}}$ & 109.0$\mathrm{^{+0.9}_{-0.9}}$ & 0.94$\mathrm{^{+0.012}_{-0.012}}$ & 148.0$\mathrm{^{+1.9}_{-1.9}}$ & 0.367 $\times$ 0.367 & 0.69$\mathrm{^{+0.010}_{-0.010}}$ & 108.6$\mathrm{^{+1.6}_{-1.6}}$ & 0.93$\mathrm{^{+0.024}_{-0.024}}$ &   146.3$\mathrm{^{+3.8}_{-3.7}}$ \\
    Lupus 9 & 0.353 $\times$ 0.353 & 0.56$\mathrm{^{+0.019}_{-0.017}}$ &  88.5$\mathrm{^{+3.1}_{-2.7}}$ & 0.80$\mathrm{^{+0.043}_{-0.043}}$ & 125.8$\mathrm{^{+6.8}_{-6.8}}$ & 0.352 $\times$ 0.352 & 0.55$\mathrm{^{+0.044}_{-0.035}}$ &  86.8$\mathrm{^{+6.9}_{-5.6}}$ & 0.84$\mathrm{^{+0.112}_{-0.084}}$ & 133.0$\mathrm{^{+17.7}_{-13.3}}$ \\
    Lupus 10 & 0.182 $\times$ 0.182 & 1.52$\mathrm{^{+0.003}_{-0.003}}$ & 233.4$\mathrm{^{+0.5}_{-0.5}}$ & 2.49$\mathrm{^{+0.019}_{-0.019}}$ & 381.8$\mathrm{^{+3.0}_{-3.0}}$ & 0.161 $\times$ 0.161 & 4.25$\mathrm{^{+0.023}_{-0.023}}$ & 652.6$\mathrm{^{+3.5}_{-3.5}}$ & 6.49$\mathrm{^{+0.038}_{-0.038}}$ &   997.4$\mathrm{^{+5.9}_{-5.9}}$ \\
\enddata
\end{deluxetable*}

\section{Discussion}
\label{sec:discussions}

While our selection criteria for the Lupus AGE-PRO disks were based on the shallow and moderate resolution Band~7 survey by \citet{Ansdell_2016ApJ_Lupus}, the Lupus region was also observed in Band~6 in the continuum and main CO isotopologue lines \citep{Andrews_DSHARP_2018ApJ...869L..41A, Ansdell_Lupus_2018ApJ}.
When comparing our results with those obtained from the shallower Band~6 survey, we find that AGE-PRO increases the detection rates of the rare isotopologue $\mathrm{{}^{13}CO}$ and $\mathrm{C^{18}O}$ from 10\% to 100\% and from 0\% to 90\%, respectively. 
When sources are detected in both surveys, their fluxes agree to within 20\%, with the exceptions of Lupus~5, 6, and 10, where differences are up to a factor of $\sim$2.
These large differences are due to the low sensitivity per beam of previous observations (see Appendix~\ref{appendix:compare_literature} for details).
We also compare the measured dust and gas radii with the ones reported in \citet{Sanchis_Lupus_radius_2021A&A...649A..19S}, where we also find good agreement for the targets that are well-resolved in the images: Lupus~2 and 10 for both dust and gas radii, and Lupus~1 and 7 for gas radii only.
Unlike \citet{Sanchis_Lupus_radius_2021A&A...649A..19S} who carried out visibility fitting in the UV-plane to measure dust radii and did model-based fitting to derive gas radii, our radii are calculated in the image plane using the curve-of-growth method, and they are not de-convolved with the beams. 
Therefore, the AGE-PRO radii summarized here should be considered upper limits for unresolved disks $-$ AGE-PRO beam de-convolved disk sizes are measured and presented in accompanying work by \citet{Trapman_AGEPRO_XI_gas_disk_sizes}.
More information and details of comparing our measurements with the literature as well as the beam-corrected disk sizes from \citet{Trapman_AGEPRO_XI_gas_disk_sizes} are presented in Appendix~\ref{appendix:compare_literature} with Figures~\ref{fig:comp-with-literature-flux} and~\ref{fig:comp-with-literature-radius} comparing fluxes and radii, respectively.

In this section, we discuss general trends of the inferred properties of our Lupus targets.
Section~\ref{subsec:compare_f_n_f} clarifies which fluxes and flux densities are correlated, Section~\ref{subsec:compare_r_n_r} compares the dust and gas disk radii, and Section~\ref{subsec:compare_w_model} compares different gas disk mass estimates.

\subsection{Correlations between the dust emission and gas line fluxes}
\label{subsec:compare_f_n_f}

There have been many shallow surveys of nearby star-forming regions where the continuum flux is often detected while gas lines are more rarely detected, especially for the rarer CO isotopologues \citep[see][as reviews]{Miotello_PPVII_2023, Manara_2023_PPVII}.
With the higher sensitivity of AGE-PRO, we have many detections of CO isotopologues, which enable us to test whether gas line fluxes are correlated with the continuum emission.
If these correlations exist, they could aid in planning future gas observations in other star-forming regions.
Besides CO isotopologues, AGE-PRO also provides more ${\rm N_2H^+}\,(3-2)$ line detections, which is an ionization tracer in disk mid-planes, and can be used to evaluate the CO abundance in the disk together with $\mathrm{C^{18}O\,(2-1)}$ \citep[e.g., ][]{trapman_N2Hp_DALI_2022, Trapman_AGEPRO_V_gas_masses}.
Thus, we also compare the ${\rm N_2H^+\,(3-2)}$ fluxes with the $\mathrm{C^{18}O\,(2-1)}$ fluxes.

\begin{figure*}
    \gridline{\fig{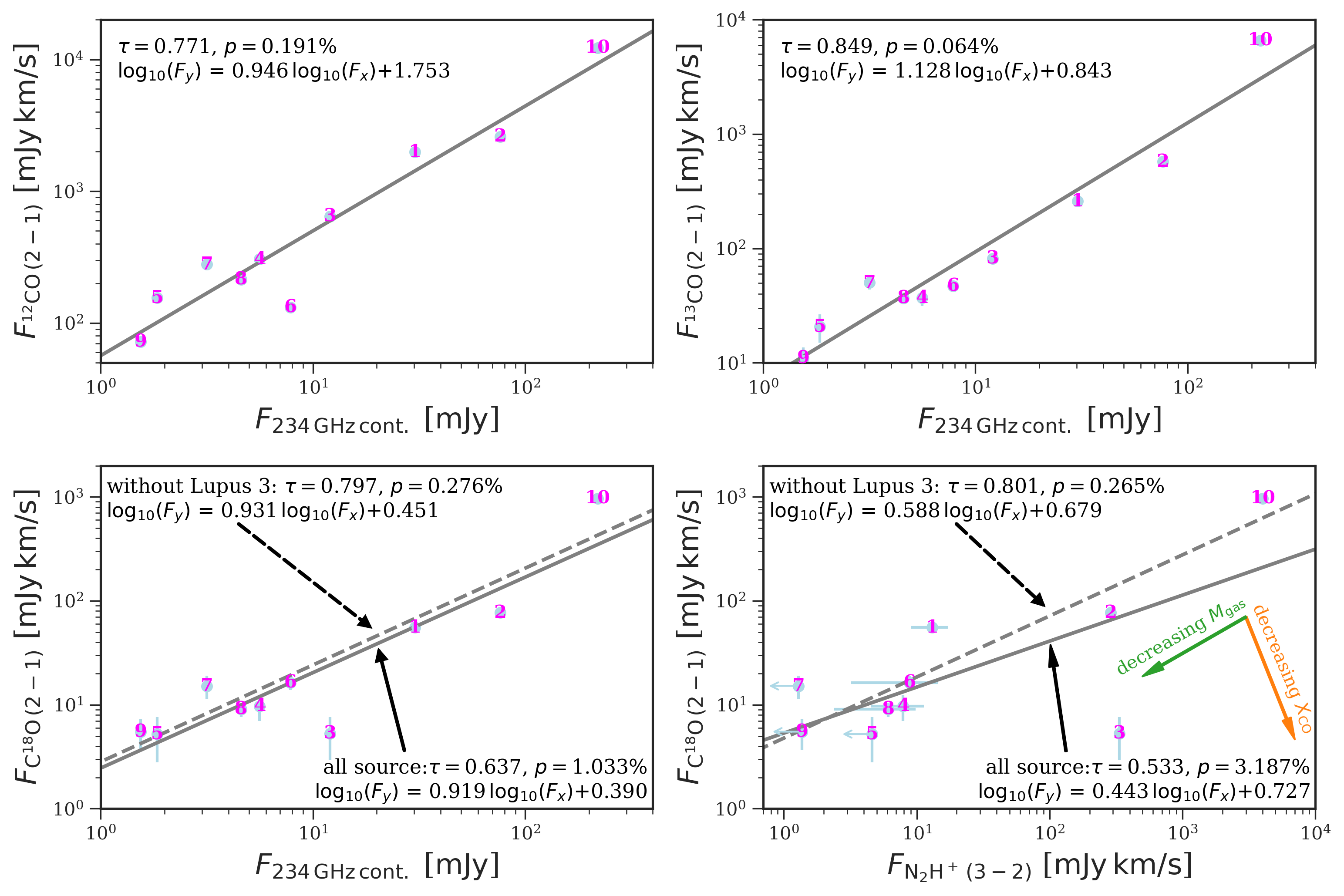}{0.9\textwidth}{}
            }
  \caption{
  Comparisons between different fluxes.
  The results of \texttt{pymccorrelation} Kendall's $\tau$ tests for the Lupus sample are reported in each panel: There are positive correlations between the CO isotopologue fluxes with the Band~6 continuum flux densities.
  Removing the outlier Lupus~3 improves the correlations of the $F_{\mathrm{C^{18}O\,(2-1)}}$ with the $F_{\rm 234\,GHz\,cont.}$ and with the $F_{\mathrm{N_2H^+\,(3-2)}}$.
  The best-fit results from linear-regression in log-log scale are plotted in black solid lines (for all sources) and dashed lines (without Lupus~3) together with the functions at the corners of each panel.
  For $F_{\mathrm{C^{18}O\,(2-1)}}$ versus $F_{\mathrm{N_2H^+\,(3-2)}}$, we also draw two arrows on the right representing two trends of decreasing CO abundance (orange) and decreasing gas disk masses ($M_{\rm gas}$), respectively, predicted by \citet{trapman_N2Hp_DALI_2022}.
  }
  \label{fig:FF_comp}
\end{figure*}

Figure~\ref{fig:FF_comp} shows line fluxes from three CO isotoplogues vs the Band~6 continuum flux density ($F_{\rm 234\,GHz\,cont.}$) and the ${\rm N_2H^+}\,(3-2)$ fluxes.
The figure suggests that all CO isotopologue fluxes are correlated with the $F_{\rm 234\,GHz\,cont.}$, while the $F_\mathrm{C^{18}O\,(2-1)}$ appears to be less well correlated with the $F_{\rm N_2H^+\,(3-2)}$.

To test whether correlations are present we follow \citet{pascucci_largedisks_2023} and use the \texttt{pymccorrelation} routine v0.2.5\footnote{\url{https://github.com/privong/pymccorrelation}} which carries out the non-parametric Kendall's $\tau$ test with censored data \citep{Isobe_Kendall_Tau_1986ApJ...306..490I}.
This routine is capable of including uncertainties and upper limits.
The resulting median values of the Kendall's $\tau$ and the percent probability $p$ are presented in each panel:
$\tau$ gives the direction of the correlation (positive correlation for values $> 0$) while $p$ is the percent probability that the quantities are uncorrelated.
We find that all three CO isotopologue fluxes are positively correlated with the continuum fluxes (positive $\tau$ and p$< 1$\%), especially the $F_{\mathrm{{}^{12}CO\,(2-1)}}$ and $F_{\mathrm{{}^{13}CO\,(2-1)}}$. 
Hence we also provide here the best fit relations shown in the upper-left corner of each panel.
We also find that the $F_{\mathrm{C^{18}O\,(2-1)}}$ is positively correlated with $F_{\rm 234\,GHz\,cont.}$ with $p \sim 2.47\%$.
Here, Lupus~3 appears to be an outlier with very small $F_{\mathrm{C^{18}O\,(2-1)}}$ but relatively large $F_{\rm 234\,GHz\,cont.}$ compared to disks with similar $F_{\mathrm{C^{18}O\,(2-1)}}$ (Lupus~5 and 9).
If we exclude Lupus~3 from the analysis, the correlation is much stronger with $p \sim 0.43\%$.

The dust continuum emission ($F_{\rm cont.}$) from disks is expected to be optically thin, thus the positive correlation between $F_{\rm 234\,GHz\,cont.}$ and the optically thin gas disk tracer $F_{\mathrm{C^{18}O\,(2-1)}}$ could suggest the gas-to-dust mass ratio is similar in our sample expect for Lupus~3.
However, some recent works \citep[e.g.][]{Carasco-Gonzalez_HLTau_2016, Zhu_dust_scattering_2019, xin_measuringMdust_2023, rilinger_determiningMdust_2023} found that the $F_{\rm cont.}$ from disks are not always optically thin even at longer wavelengths ($\gtrsim 3\,{\rm mm}$).
If this is true for the disks in our sample, their correlation between the CO isotopologue line emission and $F_{\rm 234\,GHz\,cont.}$ are more likely due to their similar gas-to-dust size ratio. 
This is also supported by the fact that the optically thick $F_{\mathrm{{}^{12}CO\,(2-1)}}$ and $F_{\mathrm{{}^{13}CO\,(2-1)}}$ lines are more strongly correlated with the $F_{\rm 234\,GHz\,cont.}$, while the $F_{\mathrm{C^{18}O\,(2-1)}}$ is relatively more scattered.

We also find that the $F_{\mathrm{C^{18}O\,(2-1)}}$ and $F_{\mathrm{N_2H^+\,(3-2)}}$ are not well correlated, where the $\tau$ is positive but less than 0.5 and $p > 5\%$. 
Lupus~3 is again an outlier: it has the second strongest $F_{\mathrm{N_2H^+\,(3-2)}}$ and one of the smallest $F_{\mathrm{C^{18}O\,(2-1)}}$.
By excluding Lupus~3 from the analysis, we find a positive correlation between $F_{\mathrm{C^{18}O\,(2-1)}}$ and $F_{\mathrm{N_2H^+\,(3-2)}}$ with $p<1\%$.
It has been suggested that $F_{\mathrm{C^{18}O\,(2-1)}}$ and $F_{\mathrm{N_2H^+\,(3-2)}}$ could be used in combination to break the degeneracy between CO abundance ($X_{\rm CO}$) and the gas disk masses ($M_{\rm gas}$) since CO destroys $\mathrm{N_2H^+}$ \citep[for more details, see][]{trapman_N2Hp_DALI_2022, anderson_N2Hp_2022}.
\citet{trapman_N2Hp_DALI_2022} found that the trends of $F_{\mathrm{N_2H^+\,(3-2)}}$ and $F_{\mathrm{C^{18}O\,(2-1)}}$ would vary depending on the abundance of CO gas in the warm molecular layer where emission arises.
If the CO molecule is preferentially depleted (relative to other species and when compared to typical interstellar medium abundances) then $\mathrm{N_2H^+}$ can survive and we would observe a negative trend between the two fluxes.
If on the other hand, there was no CO depletion but an increase in gas mass, we would see a positive trend. 
These two scenarios are indicated as orange and green arrows, respectively, in Figure~\ref{fig:FF_comp}.
The AGE-PRO Lupus data, with exception of Lupus~3, suggest that the observed $F_{\mathrm{N_2H^+\,(3-2)}}$ and $F_{\mathrm{C^{18}O\,(2-1)}}$ correlation is driven by differences in $M_{\rm gas}$ and they Lupus disks have roughly similar CO abundance.
This is also indicated by the $F_{\rm 234\,GHz\,cont.}$ and $F_{\mathrm{C^{18}O\,(2-1)}}$ correlation, assuming a similar gas-to-dust mass ratio across the sample.

Lupus~3 is a notable outlier, and its disk may have significant CO depletion as indicated by the \citet{trapman_N2Hp_DALI_2022} models.
Not only the $F_{\mathrm{N_2H^+\,(3-2)}}$ is higher relative to the $F_{\mathrm{C^{18}O\,(2-1)}}$, but the $F_{\mathrm{C^{18}O\,(2-1)}}$ is low relative to the dust continuum flux density $F_{\rm 234\,GHz\,cont.}$.
Both trends support the notion of a disk with depleted $X_{\rm CO}$ compared to other disks in our sample.
We note that $\mathrm{N_2H^+}$ emission may also increase with enhanced ionization levels in the disk, as pointed out by \citet{anderson_N2Hp_2019, anderson_N2Hp_2022}.
Interestingly, Lupus~3 is one of the brightest X-ray sources in this sample with X-ray luminosity of $L_X \sim 2.5\times10^{30}\,\mathrm{erg\,s^{-1}}$, a factor of $\sim 2.5$ higher than the average T~Tauri value \citep{Krautter_Xray_1997}.
A more in-depth analysis of Lupus~3 would be valuable to pin down the extent of the CO depletion in this disk.

We also carry out similar tests between the three CO isotopologue line fluxes and the continuum flux densities from Band~7.
These tests give the same results and are thus presented only in the Appendix (see Appendix~\ref{appendix:compare_f_n_f_n_r}).

\subsection{Gas-dust disk size and implications for radial drift}
\label{subsec:compare_r_n_r}

\begin{figure*}
   \gridline{\fig{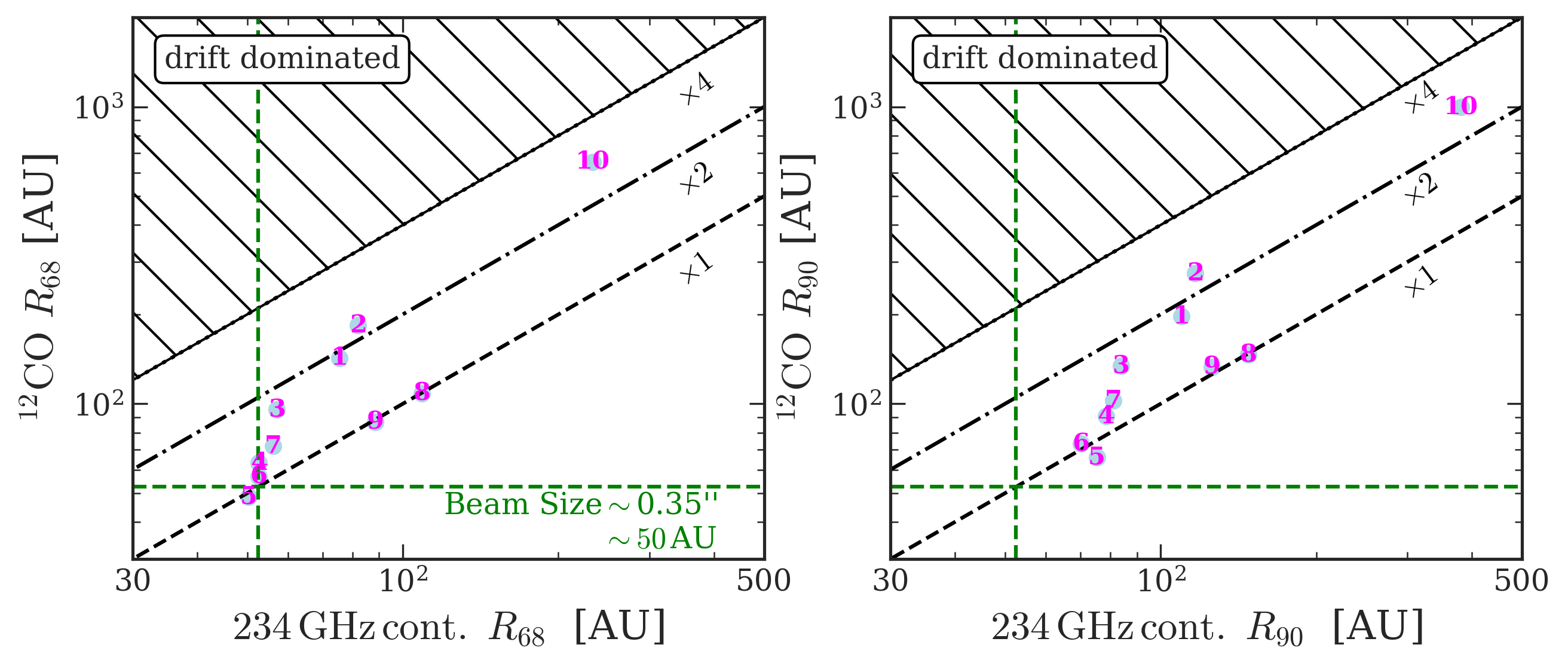}{0.9\textwidth}{}
            }
  \caption{
  $R\mathrm{_{68}}$ (left) and $R\mathrm{_{90}}$ (right) comparisons between CO and Band~6 continuum in the units of AU.
  Typical beam sizes are shown in dashed vertical and horizontal lines.
  The three dashed lines show where the $\mathrm{{}^{12}CO}$ radius is $\times 1$, $\times 2$ and $\times 4$ of the dust disk radius.
  The shaded region above $\times 4$ is for drift-dominated disks \citep{trapman_disk_size_comp_2019, trapman_radial_drift_2020}.
  For the two targets (Lupus~2 and 10) that are well resolved in both CO and Band~6 continuum, the $\mathrm{{}^{12}CO}$ radii are about $\times 2$ of the dust disk radii measured in Band~6.
 Lupus~1 and 3 are not well resolved in the continuum images; and all other smaller disks are not resolved in both continuum and $\mathrm{{}^{12}CO}$ line images, hence their grouping close to the beam sizes.
  The radii reported here are not de-convolved with the beams, and the de-concolved radii are reported in accompanying publication \citep{Vioque_AGEPRO_X_dust_disks,Trapman_AGEPRO_XI_gas_disk_sizes} and briefly compared in Appendix~\ref{appendix:compare_literature}.
  }
  \label{fig:R68_n_R90_comp}
\end{figure*}

The ratio between gas and dust disk radii can be used to identify disks with significant dust evolution and radial drift \citep[e.g.,][]{trapman_disk_size_comp_2019, trapman_radial_drift_2020, Toci_sizes_2021}.
Figure~\ref{fig:R68_n_R90_comp} compares the $\mathrm{{}^{12}CO}$ radii and the Band 6 continuum radii at 234\,GHz, where they represent gas disk radii ($R_{\rm gas}$) and dust disk radii ($R_{\rm dust}$), respectively.
The beam sizes for both tracers are $\sim 0.35\,\arcsec$ and they are represented by vertical and horizontal green dashed lines in each panel of Figure~\ref{fig:R68_n_R90_comp}.

For the two targets that are well resolved in the images of both $\mathrm{{}^{12}CO}\,(2-1)$ and Band~6 continuum -- Lupus~2 and 10 -- their CO gas radii are about two times their dust radii (this is true for both $R_{68}$ and $R_{90}$).
This result is consistent with the trend reported in the literature for Lupus disks as well as disks in other star-forming regions \citep[e.g.,][]{Ansdell_Lupus_2018ApJ, Kurtovic_Sizes_2021, Sanchis_Lupus_radius_2021A&A...649A..19S, Long_Sizes_2022ApJ}, and it is below the region where dust evolution and radial drift dominates ($R_{\rm gas}/R_{\rm dust} > 4$, see \citealt{trapman_disk_size_comp_2019}).

However, the majority of the Lupus disks have dust disk radii close to the beam size.
This means that we can only place an upper limit of $\sim$100\,AU to their dust disk sizes and, from the image plane, we cannot conclude whether or not the gas is significantly more extended than the dust.

In an accompanying publication, \citet{Vioque_AGEPRO_X_dust_disks} carried out visibility fittings of the Band~6 continuum data to obtain dust disk radii and \citet{Trapman_AGEPRO_XI_gas_disk_sizes} estimated beam-corrected deconvolved gas disk radii.
More discussions on the gas-dust disk size differences are presented in \citet{Trapman_AGEPRO_XI_gas_disk_sizes}.
To give a general overview, the gas and dust radii from their works are compared with those measured here in the image plane along with literature values in Appendix~\ref{appendix:compare_literature}. 

Here we conclude that there is no evidence of significant radial drift for the large disks in our sample. 
In an accompanying paper, \citet{Kurtovic_AGEPRO_VI_DustEvolution} demonstrates that the radial drift is not very significant even in compact disks.

\subsection{AGE-PRO data indicate higher gas disk masses compared to literature values}
\label{subsec:compare_w_model}

\begin{figure*}
   \gridline{\fig{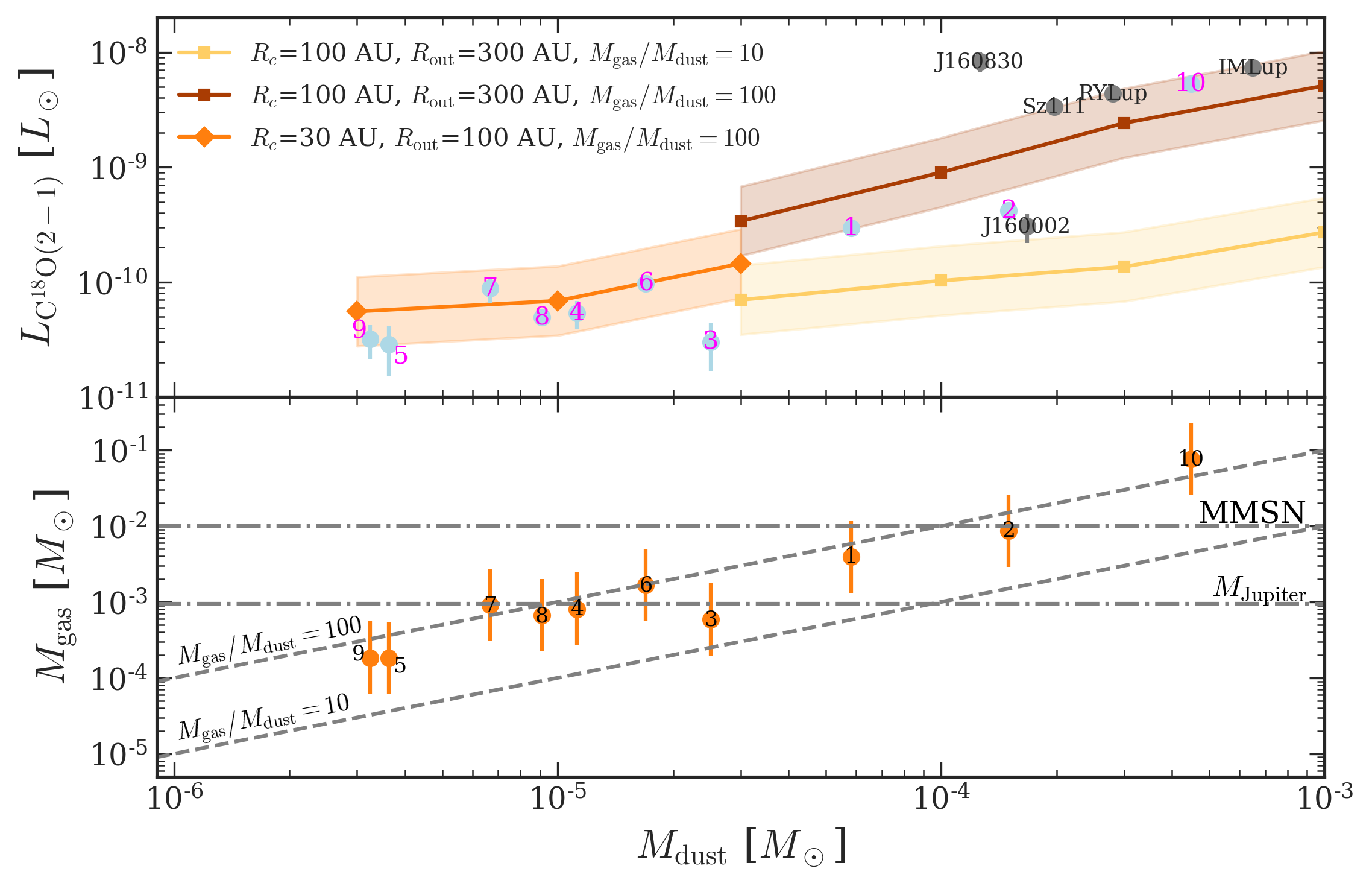}{0.90\textwidth}{}
            }
  \caption{
  $\mathrm{C^{18}O}$ luminosities compared with model predictions (top) and  gas masses estimated from these models (bottom) versus the dust masses
  In the top panel, blue points are the AGE-PRO Lupus sample, and the gray points are the large Lupus disks discussed in \citet{pascucci_largedisks_2023} with data from \citet{Ansdell_Lupus_2018ApJ}.
  Solid lines are  model predictions by \citet{ruaud_C18O_2022} (in brown, orange, and yellow for three different setups as described in Section~\ref{subsec:compare_w_model}) with the shaded area representing the area between the median value $\times\,0.5$ and $\times\,2.0$.
  In the bottom panel, the orange points represent the mass estimates for each target based on the tracks with the closest disk radius from \citet{ruaud_C18O_2022}, and their uncertainties are manually set to be an order of magnitude (the lower and upper limits are median values $\times\,1/3$ , and $\times\,3.0$).
  Two horizontal dash-dotted lines show the minimum mass solar nebula (MMSN) and the Jupiter mass ($M_{\rm Jupiter}$).
  The other two dashed lines show the gas-to-dust mass ratio of 10 and 100.
  }
  \label{fig:derive_Mgas_ruaud}
\end{figure*}

In this section, we compare gas mass estimates from different thermochemical models to the AGE-PRO sample.
We start with the models of \citet{ruaud_C18O_2022} that compute self-consistent disk structures iterating between density, temperature and chemistry and further include a detailed treatment of grain surface chemistry.
Figure~\ref{fig:derive_Mgas_ruaud} shows the AGE-PRO $\mathrm{C^{18}O}\,(2-1)$ line luminosities ($L_\mathrm{C^{18}O\,(2-1)}$) with model predictions from \citet{ruaud_C18O_2022} against the dust disk masses $M_{\rm dust}$ (see Section~\ref{subsec:measure_flux_n_radii} and Table~\ref{Tab:COG-flux}).
In addition, we also include $L_\mathrm{C^{18}O\,(2-1)}$ for other large Lupus disks that are not included in AGE-PRO (gray points) discussed in \citet[][with $\mathrm{C^{18}O}$ fluxes from \citealt{Ansdell_Lupus_2018ApJ}]{pascucci_largedisks_2023}: IM~Lup, RY~Lup, Sz111, J16000236-4222145, and J16083070-3828268.
This demonstrates that although Lupus~10 appears to be an outlier in the AGE-PRO sample, it is in fact representative of the large disks in Lupus.

The \citet{ruaud_C18O_2022} models use standard interstellar medium element abundances in their chemistry and self-consistently compute the distribution of CO in the disk by taking into account the effect of selective photodissociation, freeze-out and subsequent conversion of CO ice into more refractory ice compounds (in particular the ${\rm CO_2}$ ice) at the surfaces of dust grains.
Only a few disk parameters were varied, including the gas and dust disk masses and the radial surface density distribution, which is set as $\Sigma(r) = \Sigma_0 (R/R_c)^{-1} \exp{(-R/R_c)}$ with the critical radius $R_c$ to define the tapering off of the disk mass.
The two sets of models correspond to different disk sizes, one set with outer radius $R_{\rm out} = 100\,{\rm AU}$ and critical radius $R_c = 30\,{\rm AU}$, and a second set with $R_{\rm out} = 300\,{\rm AU}$ and $R_c = 100\,{\rm AU}$.
Two lines with square points in Figure~\ref{fig:derive_Mgas_ruaud} top panel share the same $R_{\rm out} = 300\,{\rm AU}$ but have different gas-to-dust mass ratio of $M_{\rm gas}/M_{\rm dust} = 10$ and $100$.
We crudely estimate the gas disk masses from these sparse tracks as follows.
First, we divide disks into two groups based on radii: the $R_{c} = 100\,{\rm AU}$ model is used for the three larger disks (Lupus~1, 2, and 10) and the $R_{c} = 30\,{\rm AU}$ model is adopted for others. 
Next, we linearly scale the model track that has the same dust disk mass to match each source to derive the gas disk masses, using the following equation:
\begin{equation}
    M_{\rm gas, derived} = M_{\rm gas, model} \frac{L_{\rm C^{18}O, observed}}{L_{\rm C^{18}O, model}}.
\end{equation}
These gas disk mass estimates are shown as orange points in the bottom panel of Figure~\ref{fig:derive_Mgas_ruaud}.
The uncertainties of these estimated masses are manually assigned to cover nearly an order of magnitude, with the upper and lower limits of $M_{\rm gas, derived} \times 3$ and $\times 1/3$, this is to account for both the modeling uncertainties, and the potential differences in stellar parameters and disk sizes between the model assumptions and the observed quantities.

\begin{figure*}
   \gridline{\fig{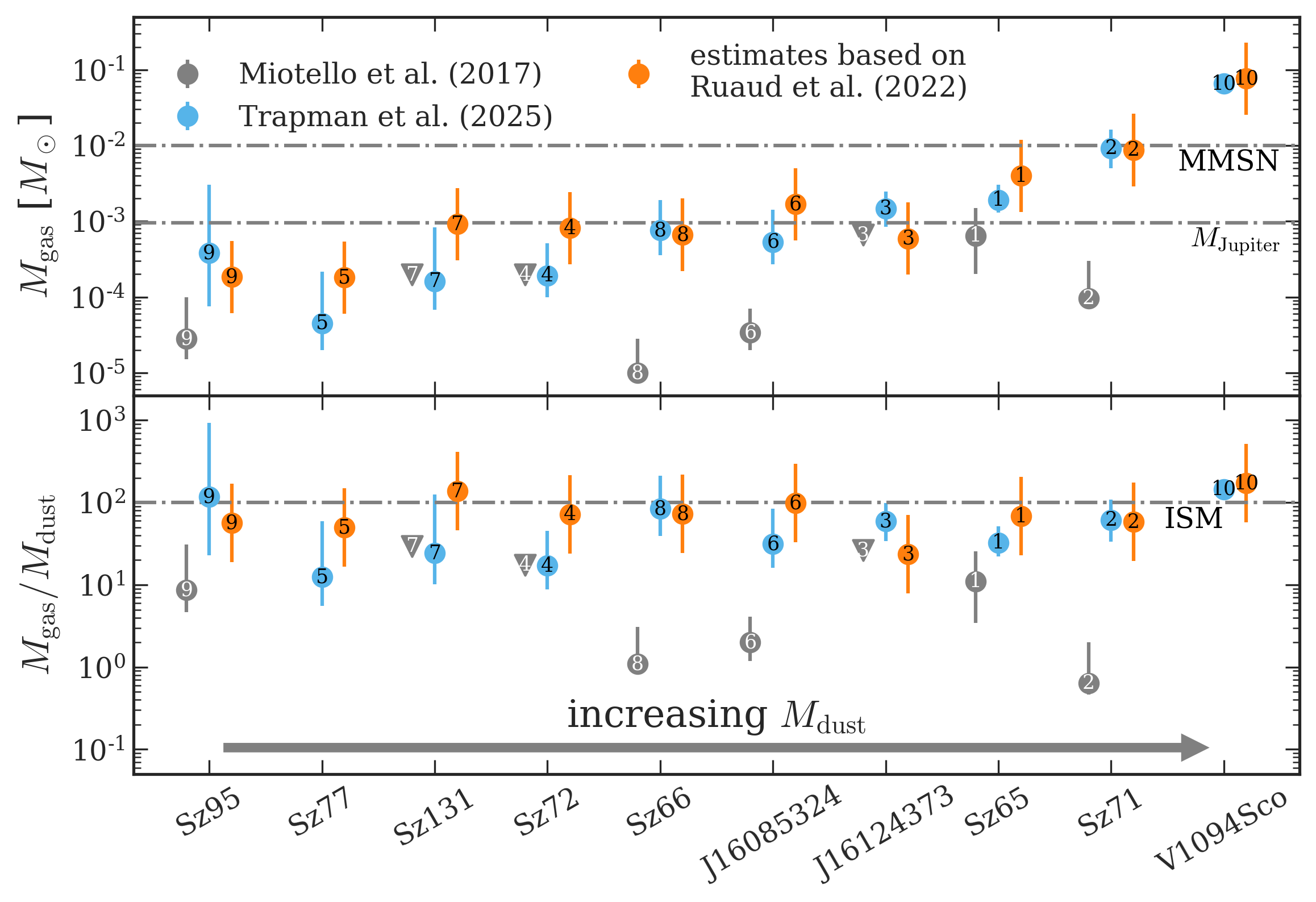}{0.90\textwidth}{}
            }
  \caption{
  Estimated gas masses (top) and gas-to-dust mass ratios (bottom) compared with literature values in a sequence of increasing dust masses ($M_{\rm dust}$) from left to right.
  The orange points are the same as shown in Figure~\ref{fig:derive_Mgas_ruaud}.
  The blue points show the derived gas disk masses and gas-to-dust mass ratio in AGE-PRO using \texttt{DALI} models \citep{Trapman_AGEPRO_V_gas_masses}, and gray points are the ones reported in \citet{miotello_lupus_2017} also using \texttt{DALI} models.
  We adopt the same  dust disk masses derived from the AGE-PRO continuum fluxes (Table~\ref{Tab:COG-flux}) to calculate the gas-to-dust mass ratio for all three model estimates.
  For better visualization, we manually shift the gray and orange points to left and right, respectively, and we place the AGE-PRO ID on the top of each target.
  }
  \label{fig:comp-with-model-threemodels}
\end{figure*}

We then compare these gas masses to two other sets of gas mass estimates, both based on \texttt{DALI} (\texttt{DustAndLInes}) thermochemical models \citep{Bruderer2012, Bruderer2013} in Figure~\ref{fig:comp-with-model-threemodels}.
The first set is from \citet{miotello_lupus_2017} (gray points), who fit previous lower sensitivity $\mathrm{{}^{13}CO}$ and $\mathrm{C^{18}O}$ data, mostly $\mathrm{{}^{13}CO}$ detections in the assumption of those being optically thin, using a modest-size grid of DALI models.
These gas mass estimates were the earliest published for the disks in Lupus.
The second set is from an accompanying AGE-PRO paper, \citet{Trapman_AGEPRO_V_gas_masses} (blue points), who derived gas masses by fitting $\mathrm{{}^{13}CO}$, $\mathrm{C^{18}O}$, $\mathrm{N_2H^+}$ line, and 1.3~mm continuum luminosities, as well as $\mathrm{{}^{12}CO}$ gas disk sizes, using a larger grid of \texttt{DALI} models.
Both \texttt{DALI} and \citet{ruaud_C18O_2022} models consider isotope-selective photodissociation, but the \texttt{DALI} models parameterize the density structure and do not include any grain surface chemistry related to CO. 
The omission of the latter is partly taken into account in our accompanying work by \citet{Trapman_AGEPRO_V_gas_masses}, where the CO abundance in the CO emitting layer is included as a free parameter to account for the CO that would have been converted on grain surfaces.
This reduced model complexity does allow for a much more extensive exploration of parameter space, including disk masses and sizes, resulting in larger and denser grids of models than \citet{ruaud_C18O_2022}, where the CO abundance is derived only from the chemistry and is not a free parameter.
Unlike \citet{miotello_lupus_2017}, who primarily estimated disk masses from $\mathrm{{}^{13}CO}$ fluxes, which may not be optically thin, and attribute lower than expected CO isotopologue emission to lower gas masses; additional constraints including multiple CO isotopologue line emission are accounted in \citet{Trapman_AGEPRO_V_gas_masses}. 
In addition, a major difference is that \citet{Trapman_AGEPRO_V_gas_masses} treat bulk CO abundance as a free parameter, instead of keeping it fixed, and infer it from the CO isotopologues and $\mathrm{N_2H^+}$ line emission as described in \citet{trapman_N2Hp_DALI_2022}.
As a result, \citet{Trapman_AGEPRO_V_gas_masses} find larger disk masses and smaller bulk CO abundances compared to \citet{miotello_lupus_2017}.

We note that, unlike the past \texttt{DALI} estimates reported in \citet{miotello_lupus_2017}, there is good agreement for the mass estimates in \citet{Trapman_AGEPRO_V_gas_masses} and the values inferred from the \citet{ruaud_C18O_2022} tracks (blue and orange points in Figure~\ref{fig:comp-with-model-threemodels}).
The agreement between the thermochemical modeling approaches is better in the case of the four more massive disks (Lupus~1, 2, 3, and 10) with $\mathrm{N_2H^+}$ emission detections, which clearly improves the accuracy of the mass estimates in the DALI models. 
This conclusion is supported by previous work. 
Specifically, \citet{ruaud_C18O_2022} and \citet{pascucci_largedisks_2023} employed the same \citet{ruaud_C18O_2022} grid with the older CO fluxes from \citet{Ansdell_2016ApJ_Lupus} and already found larger gas masses compared to those in \citet{miotello_lupus_2017} for the brightest/more massive disks, which agree with the new ones obtained here and those in \citet{Trapman_AGEPRO_V_gas_masses}.
At the low mass end, we still find reasonable concurrence for the disk masses between the \citet{ruaud_C18O_2022} and the \citet{Trapman_AGEPRO_V_gas_masses} models, with differences smaller than a factor of $\sim 5$ and within the assigned uncertainties.

The gas-to-dust mass ratio (Figure~\ref{fig:comp-with-model-threemodels} lower panel), which is an important indicator of the disk evolutionary stage, seems to be close to (or sometimes even larger than) 100, i.e., the interstellar medium value for the high mass disks.
This has also previously been shown to be the case for disks of different ages and in different regions from a variety of methods, including the mass estimates from the detected $\mathrm{HD}$ line emissions \citep[e.g.,][]{bergin_HD_2013, mcclure_HD_mass_2016}, more detailed disk modelings relying on multiple lines \citep[e.g.,][]{zhang_maps_2021, schwarz_MAPS_2021}, modelings with more comprehensive thermochemical processes \citep[e.g.,][]{powell_COdepletion_2022, pascucci_largedisks_2023, deng_diskmint_2023}, and analyzing disk dynamical masses on self-gravitating disks \citep[e.g.,][]{lodato_dynamical_mass_2022, Martire_dynamical_mass_2024}.
Interestingly, the \citet{ruaud_C18O_2022} models suggest that the ratio remains the same even at the low mass end (with the exception of Lupus~3), while the \citet{Trapman_AGEPRO_V_gas_masses} results suggest that the gas-to-dust mass ratio is lower for three low mass disks (Lupus~4, 5, and 7). 
To better understand the discrepancies and obtain more reliable disk mass estimates, future benchmarking efforts are necessary, along with more detailed source-specific modeling of these small disks \citep[e.g.,][Deng et al. in prep]{deng_diskmint_2023}. 

\section{Summary}
\label{sec:Summary}

In this work, we present ALMA Band~6 and 7 observations and analysis for 10 disks in the Lupus star-forming region.
These disks are part of the ALMA large program AGE-PRO, which aims to study gas disk evolution with the Lupus sample tracing a population of  $\sim 1-3\,\mathrm{Myr}$-old disks.
These disks are selected based on the spectral type of their host stars (M3-K6), their broad-band SED (indicating Class-II systems), and literature observations which suggest they are not close-binaries, not edge-on disks, and also had previous $\mathrm{{}^{13}CO}$ detections.
The high-sensitivity AGE-PRO observations enable us to measure their disk masses and sizes with great precision. 
Combining these values with results from the younger Ophiuchus \citep{Ruiz-Rodriguez_AGEPRO_II_Ophiuchus} and the older Upper~Sco region \citep{Agurto-Gangas_AGEPRO_IV_UpperSco} gives us the opportunity to study the evolution of dust and gas disks  through time \citep{Trapman_AGEPRO_V_gas_masses, Kurtovic_AGEPRO_VI_DustEvolution, Tabone_AGEPRO_VII_GasEvolution}.
Our main conclusions for the AGE-PRO Lupus disks are as follows: 

\begin{enumerate}
    \item We find that the Lupus AGE-PRO sample is representative for all disks around M3-K6 stars in the Lupus region with respect to mass accretion rates, millimeter fluxes, and spectral indices.
    \item In ALMA Band~6 observations, all disks are detected in the 234\,GHz continuum, $\mathrm{^{12}CO}\,(2-1)$ and $\mathrm{^{13}CO}\,(2-1)$ with SNR $\gtrsim 3$. $\mathrm{C^{18}O}\,(2-1)$ emission is detected in 4 targets with SNR $>3\sigma$, in other 5 with SNR $\sim 2-3\sigma$, and for 1 target with SNR $\sim 1\sigma$. We also have $>3\,\sigma$ detections of $\mathrm{H_2CO}\,J\mathrm{=3_{03}-2_{02}}$ in three targets, $\mathrm{DCN}$\,$J$=3-2 in two targets, and $\mathrm{N_2D^+}$\,$J$=3-2 in one target.
    \item In ALMA Band~7 observations, all disks are detected in the 285\,GHz continuum. $\mathrm{N_2H^+}\,(3-2)$ emission is detected in 3 disks with SNR $> 5 \sigma$, 1 disk with SNR$\sim 3\sigma$, and in other 4 disks with SNR $< 3 \sigma$. No any $\mathrm{N_2H^+}\,(3-2)$ emission is found from the other 2 disks. We also have $>3\,\sigma$ detections of $\mathrm{DCO^+}$\,$J$=4-3, $\mathrm{DCN}$\,$J$=4-3, $\mathrm{H_2CO}\,J=4_{04}-3_{03}$ in three targets, and $\mathrm{C^{34}S}$\,$J$=6-5 in one target.
    \item We measure the total flux (densities) from the curve-of-growth method for dust continuum in both ALMA Bands~6 and 7, CO isotopologue line emission, and $\mathrm{N_2H^+}\,(3-2)$ emission. We find that there are positive correlations between all CO isotopologue line emission and the dust continuum flux density, and also a positive correlation between $\mathrm{C^{18}O}\,(2-1)$ and $\mathrm{N_2H^+}\,(3-2)$, with the exception of Lupus~3. While the sample on the whole appears to have similar CO abundances, Lupus~3 shows enhanced $\mathrm{N_2H^+}$ emission and may be depleted in CO.
    \item We also measure the 68\% and 90\% radii for Band~6 dust continuum and $\mathrm{^{12}CO\,(2-1)}$ emission with curve-of-growth method from the image plane. We find that only the two largest disks are well-resolved in the image, and they follow the relationship of $R_\mathrm{gas} \sim 2\times R_\mathrm{dust}$, indicating no significant radial drift. For other targets that are not well-resolved we refer to \citet{Trapman_AGEPRO_XI_gas_disk_sizes} and \citet{Vioque_AGEPRO_X_dust_disks} for the gas and dust sizes respectively. 
    \item We estimate gas disk masses from thermochemical models and find higher disk masses compared with the earlier \texttt{DALI} grid of \citet{miotello_lupus_2017}. Our $\mathrm{C^{18}O}\,(2-1)$ derived mass estimates based on the \citet{ruaud_C18O_2022} model tracks are compared with an updated \texttt{DALI} grid using a combination of CO isotopologue and $\mathrm{N_2H^+}\,(3-2)$ fluxes in accompanying work by \citet{Trapman_AGEPRO_V_gas_masses}. Masses estimated from both approaches are very similar for large and more massive disks. For smaller and fainter disks, discrepancies are larger but all within the quoted uncertainties. More analysis and/or deeper observations are needed to better constrain the mass estimates for small disks. 
\end{enumerate}

\paragraph{Acknowledgment.}

This paper makes use of the following ALMA data: ADS/JAO.ALMA\#2021.1.00128.L. ALMA is a partnership of ESO (representing its member states), NSF (USA) and NINS (Japan), together with NRC (Canada), MOST and ASIAA (Taiwan), and KASI (Republic of Korea), in cooperation with the Republic of Chile. The Joint ALMA Observatory is operated by ESO, AUI/NRAO and NAOJ. The National Radio Astronomy Observatory is a facility of the National Science Foundation operated under cooperative agreement by Associated Universities, Inc.

D.D. and I.P. acknowledge support from Collaborative NSF Astronomy \& Astrophysics Research grant (ID: 2205870).
L.P. acknowledges support from ANID BASAL project FB210003 and ANID FONDECYT Regular \#1221442.
L.P. and C.A.G. acknowledge support from "Comité Mixto ESO-Gobierno de Chile 2023", under grant 072-2023.
C.A.G. also acknowledges support from FONDECYT de Postdoctorado 2021 \#3210520.
K.Z. and L.T. acknowledge the support of the NSF AAG grant \#2205617.
N.T.K., M.G., and P.P. acknowledge the support provided by the Alexander von Humboldt Foundation in the framework of the Sofja Kovalevskaja Award endowed by the Federal Ministry of Education and Research. 
P.P. and A.S. acknowledge the support from the UK Research and Innovation (UKRI) under the UK government’s Horizon Europe funding guarantee from ERC (under grant agreement No. 101076489).
A.S. also acknowledges support from FONDECYT de Postdoctorado 2022 $\#$3220495.
B.T. acknowledges support from the Programme National “Physique et Chimie du Milieu Interstellaire” (PCMI) of CNRS/INSU with INC/INP and co-funded by CNES.
G.R. and R.A. acknowledge funding from the Fondazione Cariplo, grant no. 2022-1217, and the European Research Council (ERC) under the European Union’s Horizon Europe Research \& Innovation Programme under grant agreement no. 101039651 (DiscEvol). Views and opinions expressed are however those of the author(s) only, and do not necessarily reflect those of the European Union or the European Research Council Executive Agency. Neither the European Union nor the granting authority can be held responsible for them.
L.A.C., C.G.R. and J.M. acknowledge support from the Millennium Nucleus on Young Exoplanets and their Moons (YEMS), ANID - Center Code NCN2021\_080. 
L.A.C. also acknowledges support from the FONDECYT grant \#1241056.
J.M. also acknowledges support from FONDECYT de Postdoctorado 2024 \#3240612.
K.S. acknowledges support from the European Research Council under the Horizon 2020 Framework Program via the ERC Advanced Grant Origins 83 24 28.

All figures were generated with the \texttt{PYTHON}-based package \texttt{MATPLOTLIB} \citep{Hunter2007}. This research made use of Astropy\footnote{\url{http://www.astropy.org}}, a community-developed core Python package for Astronomy \citep{astropy:2013, astropy:2018}, and Scipy \citep{2020SciPy-NMeth}.

\clearpage

\appendix

\restartappendixnumbering 




\onecolumngrid  

\section{Observation log for AGE-PRO Lupus sample}
\label{appendix:observation_log}

The observational log for AGE-PRO observations and the Archival data are summarized in Table~\ref{table:obs_log}.


\startlongtable              
\begin{deluxetable}{lllllllll}
\tabletypesize{\scriptsize}
\tablecaption{ALMA Observational log for AGE-PRO Lupus sample\label{table:obs_log}}
\tablewidth{0pt}
\tablehead{
\colhead{Setup} & \colhead{Source Name} & \colhead{UTC Date} & \colhead{Config} &
\colhead{Baselines (m)} & \colhead{$N_{\rm ant}$} & \colhead{Elev (deg)} &
\colhead{PWV (mm)} & \colhead{Calibrators}
}
\startdata
 Lupus CO Band~6        &  Lupus 1                  &  2022-03-31 06:11   &  C43-2  &  14.6 - 330.6   &     45&   69.8&    1.1&  J1454-3747               \\
  &  Lupus 1                  &  2022-04-09 07:14   &  C43-2  &  15.0 - 313.7   &     44&   77.2&    0.9&  J1454-3747               \\
  &  Lupus 1                  &  2018-09-24 20:09\tablenotemark{a}
  &  C43-5  &  15.1 - 1397.8  &     44&   74.1&    1.5&  J1610-3958               \\
  &  Lupus 1                  &  2018-09-29 21:33\tablenotemark{a}   &  C43-5  &  15.1 - 1397.8  &     44&   55.4&    1.2&  J1610-3958               \\
  &  Lupus 2-5, 7-9           &  2022-03-25 06:16   &  C43-2  &  14.6 - 313.7   &     46&   77.1&    0.2&  J1610-3958               \\
  &  Lupus 2-5, 7-9           &  2022-03-25 07:22   &  C43-2  &  14.6 - 313.7   &     46&   77.1&    0.2&  J1610-3958               \\
  &  Lupus 2-5, 7-9           &  2022-03-25 07:22   &  C43-2  &  14.6 - 313.7   &     46&   77.1&    0.2&  J1610-3958               \\
  &  Lupus 2-5, 7-9           &  2021-12-15 12:58   &  C43-5  &  15.1 - 1917.1  &     19&   71.7&    1.3&  J1610-3958 and J1604-4441\\
  &  Lupus 2-5, 7-9           &  2022-06-21 00:58   &  C43-5  &  15.1 - 1301.6  &     39&   74.1&    1.5&  J1610-3958 and J1604-4441\\
  &  Lupus 2-5, 7-9           &  2022-06-21 02:54   &  C43-5  &  15.1 - 1301.6  &     37&   68.3&    1.1&  J1610-3958 and J1604-4441\\
  &  Lupus 2-5, 7-9           &  2022-06-29 00:56   &  C43-5  &  15.1 - 1301.6  &     47&   75.9&    0.4&  J1610-3958 and J1604-4441\\
  &  Lupus 10                 &  2022-03-28 06:29   &  C43-2  &  14.6 - 313.7   &     46&   63.6&    1.8&  J1610-3958               \\
  &  Lupus 10                 &  2021-12-01 14:05   &  C43-5  &  15.1 - 2617.6  &     43&   66.0&    1.2&  J1610-3958 and J1604-4228\\
  \hline
 Lupus N$_2$H$^{+}$ Band~7&  Lupus 1, Lupus 6         &  2019-10-10 21:24\tablenotemark{b}    &  C43-3  &  15.1 - 783.5   &     45&   49.8&    1.5&  J1610-3958 and J1924-2914\\
  &  Lupus 1, Lupus 6         &  2019-10-30 17:04\tablenotemark{b}    &  C43-3  &  15.1 - 696.9   &     45&   74.3&    1.5&  J1427-4206 and J1610-3958\\
  &  Lupus 1, Lupus 6         &  2019-10-30 20:30\tablenotemark{b}    &  C43-3  &  15.1 - 696.9   &     45&   44.9&    1.5&  J1610-3958 and J1924-2914\\
  &  Lupus 1, Lupus 6         &  2019-11-07 14:50\tablenotemark{b}    &  C43-3  &  15.1 - 500.2   &     45&   62.1&    1.5&  J1427-4206 and J1610-3958\\
  &  Lupus 1, Lupus 6         &  2019-11-07 16:18\tablenotemark{b}    &  C43-3  &  15.1 - 500.2   &     45&   73.4&    1.5&  J1427-4206 and J1610-3958\\
  &  Lupus 2-5, 7-10          &  2022-04-13 9:01     &  C43-3  &  15.1 - 500.2   &     46&   66.7&    0.3&  J1517-2422 and J1610-3958\\
  &  Lupus 2-5, 7-10          &  2022-04-15 7:39     &  C43-3  &  15.1 - 500.2   &     43&   76.2&    0.5&  J1427-4206 and J1610-3958\\
  &  Lupus 2-5, 7-10          &  2022-04-15 9:27     &  C43-3  &  15.1 - 479.1   &     46&   60.5&    0.5&  J1517-2422 and J1610-3958\\
  &  Lupus 2-5, 7-10          &  2022-04-16 5:49     &  C43-3  &  15.1 - 500.2   &     48&   66.2&    0.5&  J1517-2422 and J1610-3958\\
  &  Lupus 2-5, 7-10          &  2022-04-16 7:07     &  C43-3  &  15.1 - 500.2   &     48&   76.2&    0.5&  J1427-4206 and J1610-3958\\
  &  Lupus 2-5, 7-10          &  2022-04-16 8:26     &  C43-3  &  15.1 - 500.2   &     48&   70.6&    0.5&  J1517-2422 and J1610-3958\\
\enddata
\tablenotetext{a}{Archival observations from 2017.1.00569.S, PI: Hsi-Wei Yen}
\tablenotetext{b}{Archival observations from 2019.1.01135.S, PI: Dana Anderson}
\end{deluxetable}         

\clearpage
\twocolumngrid                        

\section{The first-moment maps for CO.}
\label{appendix:m1maps}

\begin{figure*}
   \gridline{\fig{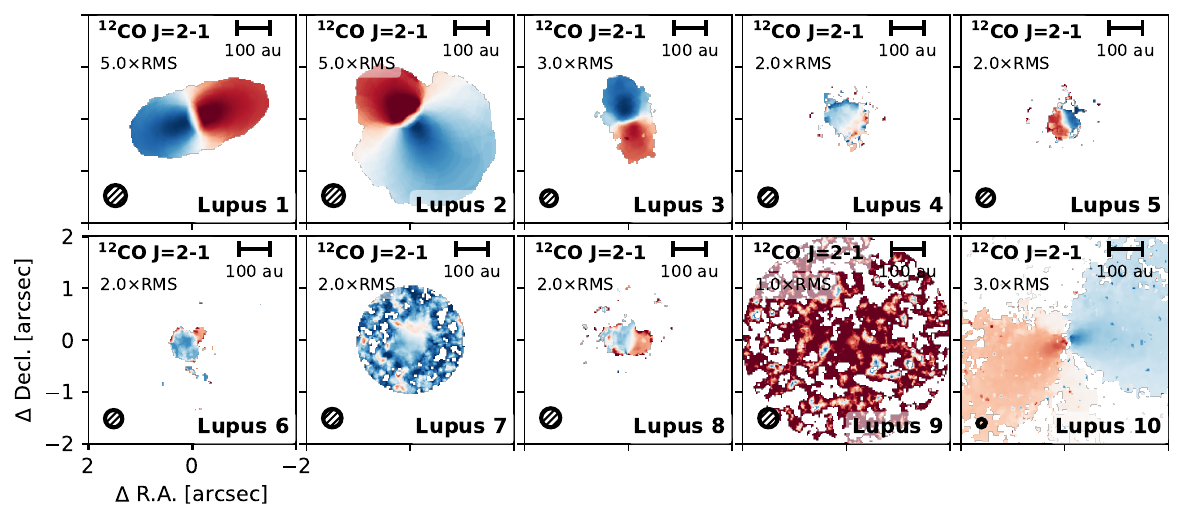}{0.97\textwidth}{}
            }
  \caption{
  The moment 1 maps for the $\mathrm{{}^{12}CO}$\,$J$=2-1 images for AGE-PRO Lupus targets (created with \texttt{bettermoments}). We only show the emission above certain SNR in each image, and the selected SNR is shown in the upper left corners. All other notations follow the same as Figure~\ref{fig:gasline-lineprofile-1}.
  }\label{fig:gasline-gallery-CO_M1maps}
\end{figure*}

In the Band~6 $\mathrm{{}^{12}CO}$ images and spectra, we notice that there are asymmetric structures that due to cloud contamination.
Here we use the \texttt{bettermoments}\footnote{\url{https://github.com/richteague/bettermoments}} to create first-moment maps (Figure~\ref{fig:gasline-gallery-CO_M1maps}) to reassure the reported asymmetric structures and check the potential non-Keplerian components.
We use the same Keplerian masks that used for $\mathrm{C^{18}O}$\,$J$=2-1 line images for zeroth-moment maps and measuring fluxes.

It is clear that the emission from Lupus~2 and 6 are asymmetric due to the cloud contamination, which are also identified from their line spectra (Figure~\ref{fig:gasline-gallery-1} in Section~\ref{subsec:band6observation}). 
The emission from Lupus~3, 4, 7 and 8 are also seem to be asymmetric, but their emission from both sides are not very different and the SNR there was relatively low ($< 3\sigma$ at the edge of the disks), therefore the asymmetry could be due to minor cloud contamination or other disk dynamics such as disk wind or outflow.

The most interesting target is Lupus~10, where we can see two clear red lobes perpendicular to the disk plane in the first-moment map.
This could suggest potential disk outflows from this system. 
But because this first-moment map is made with a pre-assumed disk Keplerian mask on a top of the disk image which also adopted the Keplerian mask during line imaging process, thus part of the disk outflow could have been masked out, and we only see a portion of this large structure.
The analysis of the details of this large disk is out of scope of this work, and will be discussed in more detail in the upcoming accompanying works.



\section{Other lines covered in AGE-PRO observations}
\label{appendix:other_weak_lines}



\begin{figure*}[ht!]
\gridline{
\fig{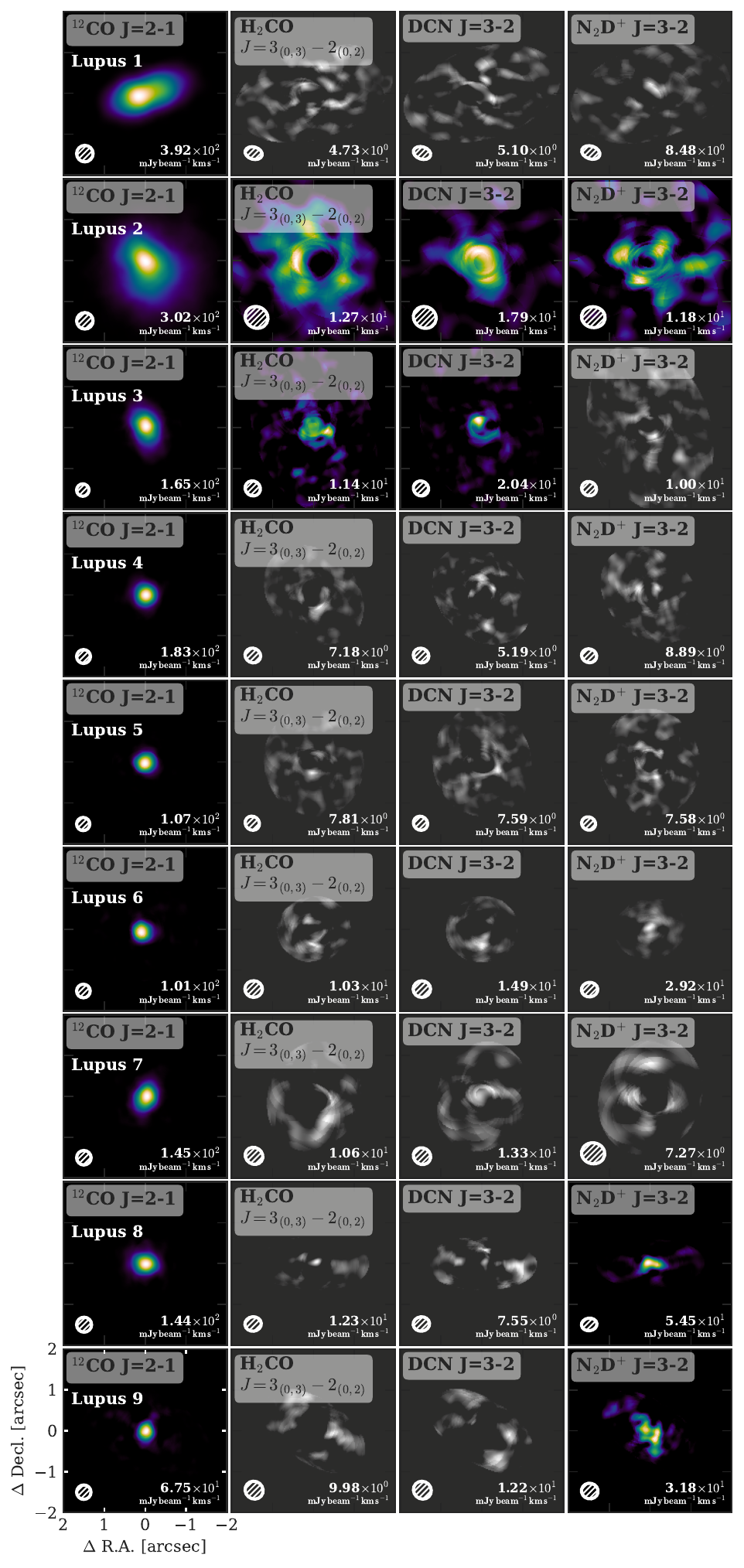}{0.49\textwidth}{}
\fig{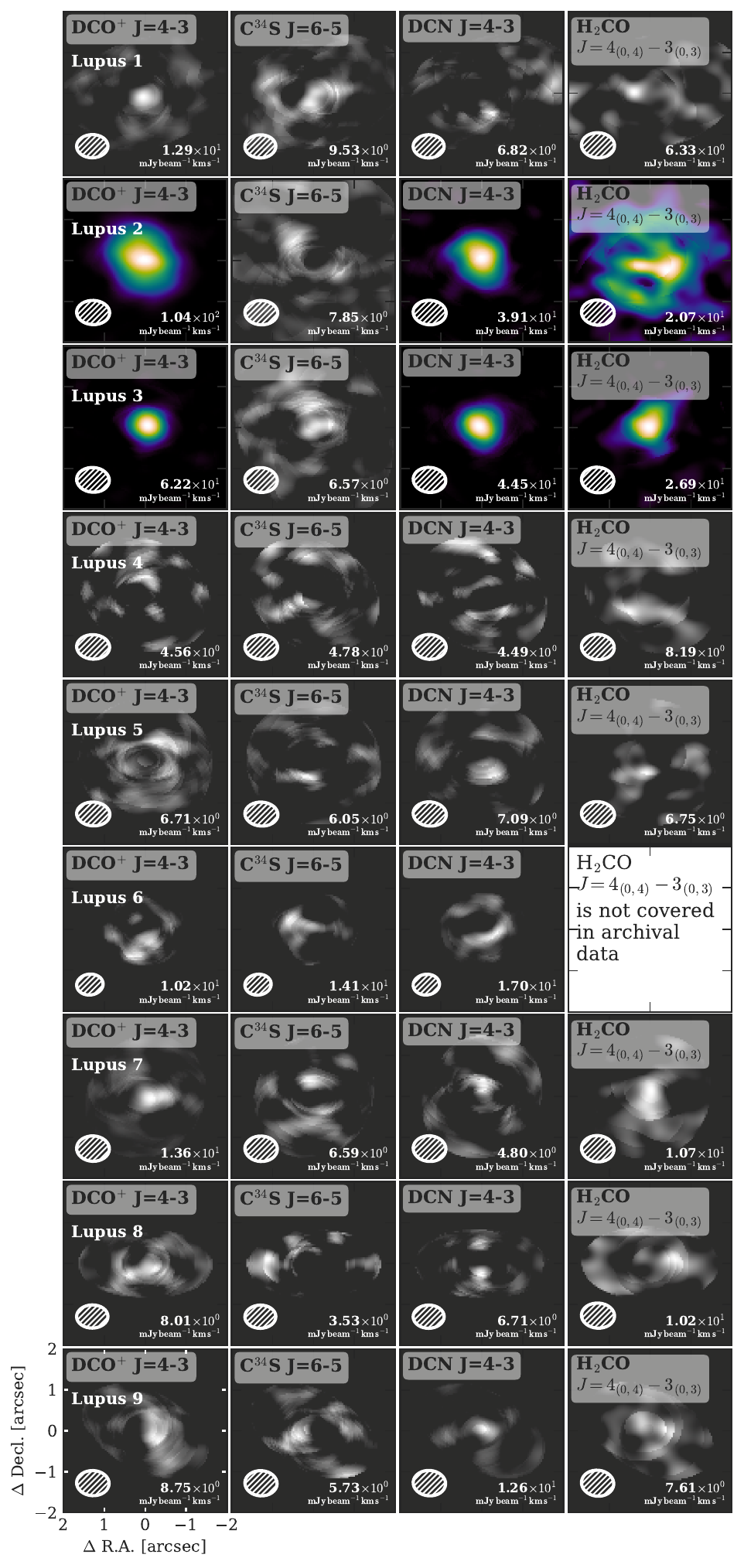}{0.49\textwidth}{}
}
\caption{The image galleries for $\mathrm{{}^{12}CO}$ and other lines. The left columns are $\mathrm{H_2CO}$ J=$\mathrm{3_{03}}$-$\mathrm{2_{02}}$ (218.2\,GHz), $\mathrm{DCN}$\,$J$=3-2 (217.2\,GHz), and $\mathrm{N_2D^+}$\,$J$=3-2 (231.3\,GHz) from Band~6 setup, and the right columns are $\mathrm{DCO^+}$\,$J$=4-3 (288.1\,GHz), $\mathrm{C^{34}S}$\,$J$=6-5 (289.2\,GHz), $\mathrm{DCN}$\,$J$=4-3 (289.6\,GHz), and $\mathrm{H_2CO}$\,$J$=$\mathrm{4_{04}}$-$\mathrm{3_{03}}$ (290.6\,GHz) in Band~7. The notation follows the same as Figure~\ref{fig:gasline-gallery-1} in Section~\ref{subsec:band6observation}.}
\label{fig:allinone_M0_gallery}
\end{figure*}

\begin{figure*}[ht!]
\gridline{
\fig{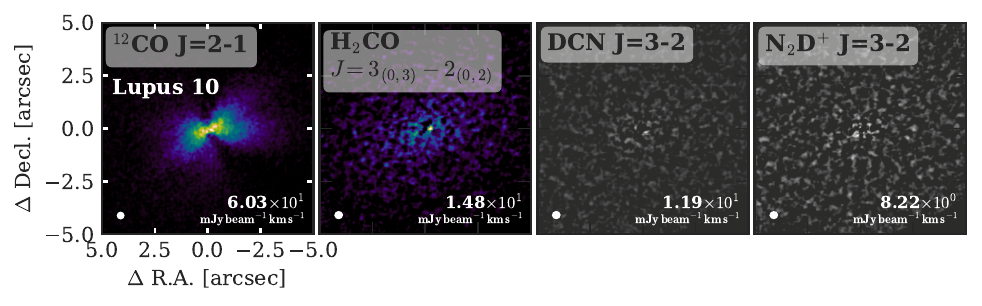}{0.49\textwidth}{}
\fig{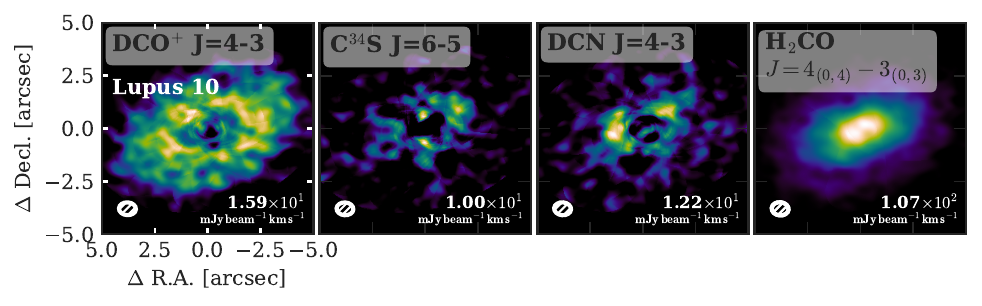}{0.49\textwidth}{}
}
\caption{(continuum) The image galleries for Lupus~10, where the image sizes is extended to 5\arcsec to show its full extended disk. The notation follows Figure~\ref{fig:allinone_M0_gallery}.}
\label{fig:allinone_M0_gallery_Lupus10}
\end{figure*}

Besides the main observational results on continuum, CO isotopologues and $\mathrm{N_2H^+}$ line images presented in Section~\ref{sec:Results}, our observation setups also include some weaker lines, including $\mathrm{H_2CO}$ J=$\mathrm{3_{03}}$-$\mathrm{2_{02}}$ (218.2\,GHz), $\mathrm{DCN}$\,$J$=3-2 (217.2\,GHz), and $\mathrm{N_2D^+}$\,$J$=3-2 (231.3\,GHz) in Band~6, and $\mathrm{DCO^+}$\,$J$=4-3 (288.1\,GHz), $\mathrm{C^{34}S}$\,$J$=6-5 (289.2\,GHz), $\mathrm{DCN}$\,$J$=4-3 (289.6\,GHz), and $\mathrm{H_2CO}$\,$J$=$\mathrm{4_{04}}$-$\mathrm{3_{03}}$ (290.6\,GHz) in Band~7.
We present their zeroth-moment maps in Figure~\ref{fig:allinone_M0_gallery} and~\ref{fig:allinone_M0_gallery_Lupus10}, where we used the same Keplerian masks adopted for $\mathrm{C^{18}O}$\,$J$=2-1 line images.
All the lines that with $> 3\,\sigma$ detections are shown in \texttt{rainforest} color while other lines are gray scales.
We also use the same curve-of-growth method as adopted in Section~\ref{subsec:measure_flux_n_radii} to measure the total flux for these lines, and the results are summarized in Table~\ref{Tab:COG-flux-B6weakerlines}.



\begin{deluxetable*}{lrrrrrrr}
\tablecaption{Measured Fluxes from curve-of-growth method for other lines \label{Tab:COG-flux-B6weakerlines}}
\tablehead{
  \colhead{ID} &  
  \colhead{$F_{\rm H_2CO\,(3_{(0,3)}-2_{(0,2)})}$} & 
  \colhead{$F_{\rm DCN\,(3-2)}$} &  
  \colhead{$F_{\rm N_2D^+\,(3-2)}$} &
  \colhead{$F_{\rm DCO^+\,(4-3)}$} &
  \colhead{$F_{\rm C^{34}S\,(6-5)}$} &
  \colhead{$F_{\rm DCN\,(4-3)}$} &
  \colhead{$F_{\rm H_2CO\,(4_{(0,4)}-3_{(0,3)})}$} \\
  \colhead{} &  \colhead{[${\rm mJy\,km\,s^{-1}}$]} &  \colhead{[${\rm mJy\,km\,s^{-1}}$]}  & \colhead{[${\rm mJy\,km\,s^{-1}}$]} & \colhead{[${\rm mJy\,km\,s^{-1}}$]}& \colhead{[${\rm mJy\,km\,s^{-1}}$]}& \colhead{[${\rm mJy\,km\,s^{-1}}$]}& \colhead{[${\rm mJy\,km\,s^{-1}}$]}
}
\startdata
       Lupus~1 &             11.0$\pm$3.6 &             -- &            -- &            5.4$\pm$3.3 &           17.0$\pm$5.7 &             -- &              -- \\
       Lupus~2 &             96.2$\pm$7.6 &            61.2$\pm$6.7 &           40.8$\pm$8.0 &         341.9$\pm$10.0 &            8.5$\pm$6.3 &            84.3$\pm$9.1 &           145.8$\pm$12.5 \\
       Lupus~3 &             21.5$\pm$3.6 &            29.0$\pm$4.8 &            3.1$\pm$4.6 &           64.6$\pm$6.2 &            9.1$\pm$5.5 &            60.4$\pm$6.0 &             44.1$\pm$8.5 \\
       Lupus~4 &              0.3$\pm$3.0 &             -- &           10.5$\pm$5.7 &            -- &            -- &             -- &              6.0$\pm$6.0 \\
       Lupus~5 &              3.8$\pm$3.2 &             1.9$\pm$4.0 &            3.5$\pm$4.9 &            8.9$\pm$7.2 &            -- &             -- &              -- \\
       Lupus~6 &             10.8$\pm$8.6 &             5.4$\pm$9.6 &          26.2$\pm$10.5 &            -- &            -- &             -- &              -- \\
       Lupus~7 &              -- &             8.2$\pm$4.6 &            -- &           10.3$\pm$5.5 &            4.7$\pm$4.9 &             -- &             12.9$\pm$5.4 \\
       Lupus~8 &              -- &             -- &           53.8$\pm$5.3 &            -- &            3.5$\pm$2.0 &             -- &              5.6$\pm$2.1 \\
       Lupus~9 &              -- &             -- &           42.4$\pm$7.5 &            6.3$\pm$3.2 &            -- &             3.9$\pm$2.2 &              7.1$\pm$3.0 \\
      Lupus~10 &          1838.0$\pm$29.5 &             -- &          11.5$\pm$10.1 &         614.2$\pm$14.8 &          66.2$\pm$11.6 &          165.0$\pm$13.0 &          2269.4$\pm$26.9 \\
\enddata
\end{deluxetable*}


\section{New companion(s) around Lupus 6 (J16085324-3914401)}
\label{appendix:lupus6companions}

\begin{figure*}[ht!]
\gridline{\fig{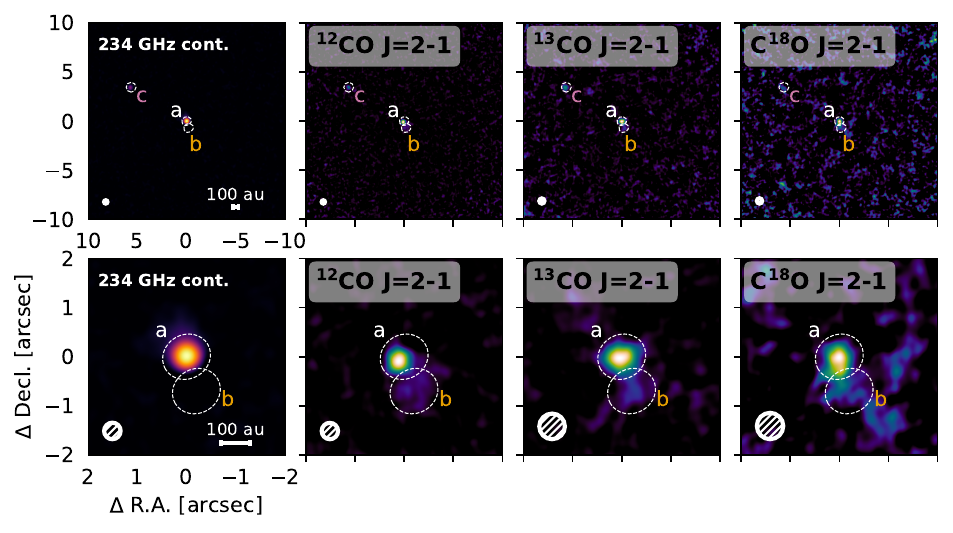}{0.90\textwidth}{}}
\caption{Band~6 continuum image and zeroth-moment maps for Lupus~6. Three targets in this system are marked with white circle and named as `a', `b' and `c'. Images in the two rows are the same with the bottom panels zoomed in to Lupus~6a and b.}
\label{fig:B6_M0_Lupus6_companion}
\end{figure*}

\begin{figure}[ht]
\gridline{\fig{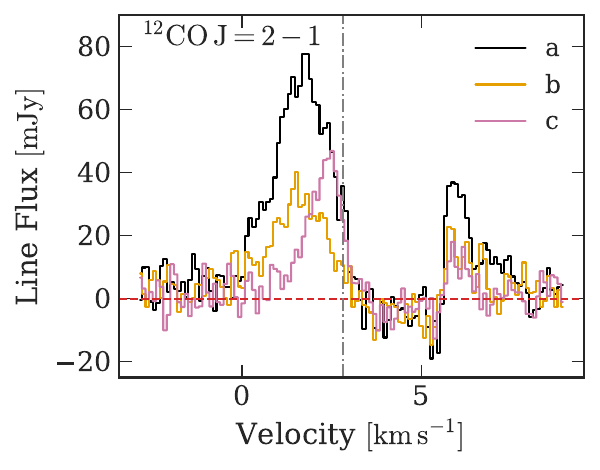}{0.45\textwidth}{}}
\caption{${\rm {}^{12}CO}\,$J=2-1 line spectra for Lupus~6 and its companions. Lupus~6a, b, and c are shown in black, yellow and pink lines, respectively. The red dashed line shows the line flux at 0 mJy, and the dot-dashed gray line shows the systemic velocity measured from Lupus~6a. Cloud contamination appears in all three targets from $\sim 3-6\,\mathrm{km\,s^{-1}}$.}
\label{fig:B6_M0_Lupus6_linespectra}
\end{figure}

In the $\mathrm{{}^{12}CO}$\,$J$=2-1 image, we find there are two potential companions around Lupus~6, and they are marked as target `b' and `c' around the main target `a' in Figure~\ref{fig:B6_M0_Lupus6_companion}, with the white dashed ellipses to circle their emission area.
The Lupus~6b is only $\sim 0.7\arcsec$ away, located at $(\Delta {\rm RA}, \Delta {\rm DEC}) = (-0.20\arcsec, -0.70\arcsec)$, and the Lupus~6c is $\sim 6.6\arcsec$ away, located at $(\Delta {\rm RA}, \Delta {\rm DEC}) = (5.65\arcsec, 3.45\arcsec)$.
All three targets can be seen from three CO isotopologue lines, and their $\mathrm{{}^{12}CO}$\,$J$=2-1 line spectra are shown in Figure~\ref{fig:B6_M0_Lupus6_linespectra}.

We check these three targets in the continuum image with the masks around all three targets identified from the $\mathrm{{}^{12}CO}$\,$J$=2-1 image.
However, only Lupus~6a and c appears in the continuum image while Lupus~6b is missing.
In accompanying paper, \citet{Vioque_AGEPRO_X_dust_disks} reported a 3$\sigma$ residual at the location of Lupus~6b after fitting the visibilities of the Band~6 continuum data.
If the continuum emission appears in the visibility plane is real, then the Lupus~6b is another disk very close to Lupus~6a; but if there is no continuum emission, the CO isotopologue emission at the location of Lupus~6b could also be outflow from or infall into it.
Observations with higher sensitivities are needed to distinguish between the two scenarios and to better understand this system.

We measure the line fluxes for all three CO isotopologue fluxes and also the Band~6 continuum flux densities on all these three targets.
The elliptical masks were adopted based on the morphology we see from the images, and the same curve-of-growth method is used to measure their fluxes.
For $\mathrm{{}^{12}CO}$\,$J$=2-1 images, because we see similar cloud contamination in all three targets, we use the same method as adopted in Lupus~6a (see Section~\ref{subsec:measure_flux_n_radii}) to derive the cloud contamination corrected line fluxes for $\mathrm{{}^{12}CO}$\,$J$=2-1.
All measured fluxes and flux densities are summarized in Table~\ref{Tab:Lupus_6_companion}, and the values for Lupus~6a is the same as the ones reported for Lupus~6 in the main AGE-PRO flux Table~\ref{Tab:COG-flux}.

\begin{deluxetable*}{ccccccc}
\tablecaption{Fluxes of Lupus~6 triple system \label{Tab:Lupus_6_companion}}
\tablewidth{0.90\textwidth}
\tablehead{
\colhead{Target}  &  \colhead{$\Delta {\rm RA}$} &  \colhead{$\Delta {\rm DEC}$} &  \colhead{$F_{\mathrm{234\, GHz\,cont.}}$} &  \colhead{$F_{\mathrm{{}^{12}CO}}$} &  \colhead{$F_{\mathrm{{}^{13}CO}}$} &  \colhead{$F_{\mathrm{C^{18}O}}$}\\
\colhead{} & \colhead{[\arcsec]} & \colhead{[\arcsec]} & \colhead{[${\rm mJy}$]} &  \colhead{[${\rm mJy\,km\,s^{-1}}$]} &  \colhead{[${\rm mJy\,km\,s^{-1}}$]} & \colhead{[${\rm mJy\,km\,s^{-1}}$]}
}
\startdata
Lupus~6a  &  0.00 &  0.00 & 7.8 $\pm$ 0.1 & 205.4 $\pm$ 5.4 & 46.7 $\pm$ 5.3 & 16.3 $\pm$ 3.6 \\
Lupus~6b  & -0.20 & -0.70 & --            & 72.8  $\pm$ 5.1 & 15.7 $\pm$ 5.1 & 8.1 $\pm$ 3.5 \\
Lupus~6c  &  5.63 &  3.45 & 1.1 $\pm$ 0.1 & 96.8  $\pm$ 3.9 & 23.1 $\pm$ 3.9 & 5.6 $\pm$ 2.6 \\
\enddata
\tablecomments{$F_{\mathrm{{}^{12}CO}}$ are cloud contamination corrected. We used the same method described in Section~\ref{subsec:measure_flux_n_radii} to carry out the correction.}
\end{deluxetable*}

\section{Flux uncertainties measured through bootstrapping}
\label{appendix:bootstrap}

To compute uncertainties on the total fluxes, we adopt the following approach, often called {\it RMS method}.
First, we measure ${\rm RMS/beam}$ in an annulus outside the disk emission in the moment~0 maps (Figures~\ref{fig:gasline-gallery-1} and~\ref{fig:gasline-gallery-B7-0} in Section~\ref{sec:Results}).
Then, we multiply this value by the square root of the number of beams covering the disk emission itself. 
A different but more time-consuming approach is to adopt a
bootstrapping method \citep[e.g.,][]{Bergner_uncertainties_2019, Harikane_statistical_uncertainties_2020, anderson_N2Hp_2022, Tsukui_statistical_uncertainties_2023}.
This method can be carried out by placing the same aperture as used to measure the total flux at random locations on an emission-free part of the cube.
The uncertainty of the total flux is taken as the standard deviation of the fluxes within the sampled apertures.
Here we outline how the method is implemented:

\begin{enumerate}
    \item We select an emission-free part of the data cube, which are all channels with velocities outside a user-specified velocity range that contains line emission from the source. Moving forward, we refer to the emission-free part as the ``line-free data cube'' and the part inside the user-specified velocity range as the ``line-containing data cube''. Note that the noise bootstrapping should be done with a so-called dirty image, including the channels outside of the masks that are used to CLEAN the data. 
    \item A subset of equal velocity width as the line-containing data cube is selected from the line-free data cube, and we refer to this subset as the ``noise cube''. By default, the channels in the noise cube are selected at random, but it is possible to select them continuously in velocity if inter-channel noise correlations are suspected to be relevant.
    \item The noise cube is collapsed across the spectral axis to create an integrated intensity map using \texttt{bettermoments} python package. 
    Any (Keplerian) masks used to create the integrated intensity map from the line-containing cube are also used during this collapse.
    \item The integrated flux is measured from the collapsed noise cube using the same aperture used to measure the line flux from the source.
    \item Steps 2, 3, and 4 are repeated $N$ times to create a distribution of flux measurements from empty parts of the cube. Each time, the position of the flux-measuring aperture is varied randomly, with at least one aperture radius away from the edge of the image. Any (Keplerian) masks used in step 3 are similarly shifted to the new aperture location. 
    \item The flux measurement uncertainty is computed as the standard deviation of the fluxes measured during Step 5.
\end{enumerate}

We provide the \texttt{Python} script \footnote{\url{https://zenodo.org/records/14925816}}\citep{trapman_2025_ageproobs_pythoncode} to the community so that this method can be easily implemented on other targets.

\begin{figure*}
    \gridline{\fig{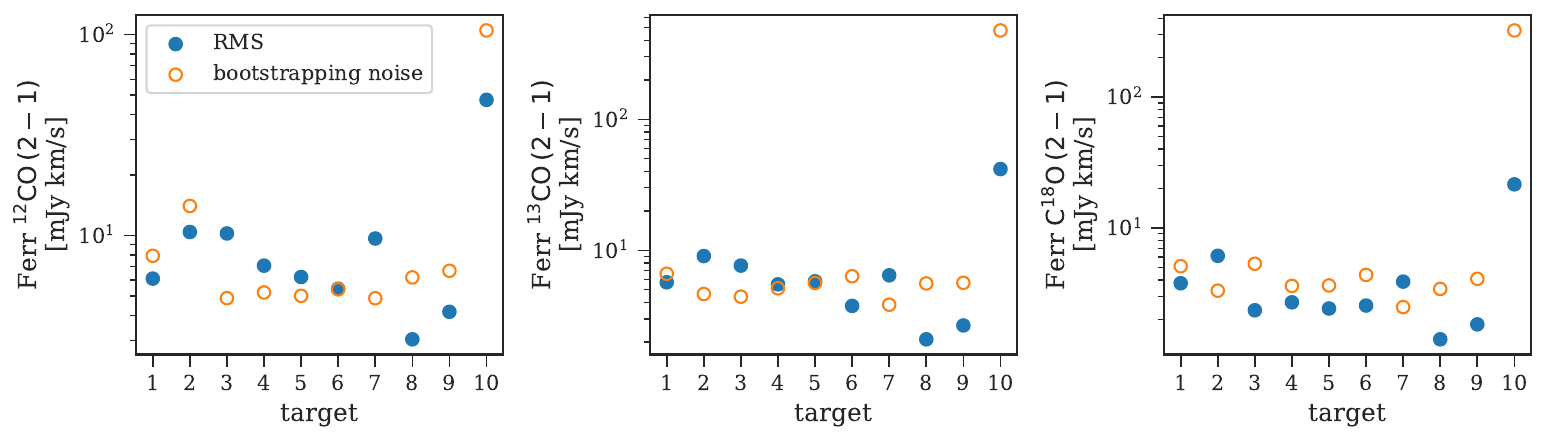}{0.95\textwidth}{}
            }
  \caption{
  The comparisons between the flux uncertainties estimated from the ${\rm RMS}$ in the moment~0 maps and from the bootstrapping method for three CO isotopologue lines. The uncertainties measured from two methods are similar within a factor of few except for Lupus~10, where the bootstrapping method provides more conservative estimates.
  }
  \label{fig:flux_comp_bootstrap}
\end{figure*}

We apply the time-consuming bootstrapping method only on the ALMA Band~6 CO isotopologue line cubes. 
We bootstrap the emission-free part of the datacube for $N=10^{4}$ times to estimate the uncertainties in the total fluxes and show that for all sources, except Lupus~10, the two methods give similar uncertainties (Figure~\ref{fig:flux_comp_bootstrap}).

Lupus~10 (V1094~Sco) has the largest disk in our sample, therefore the largest flux uncertainty in both methods. 
For this large disk, the {\it RMS method} samples an annulus that is so large that only a few channels contribute to the moment~0 map, hence we believe that it is less reliable than the bootstrapping method. 
However, the bootstrapping method is too time-consuming to apply it to all sources in all transitions and, as shown in Figure~\ref{fig:gasline-gallery-1} and Table~\ref{Tab:COG-radius}, all other disks are small and the two methods give similar uncertainties.

\section{Literature comparison of fluxes and disk radii}
\label{appendix:compare_literature}

While our selection criteria for the Lupus AGE-PRO disks were based on the shallow and moderate resolution Band~7 survey by \citet{Ansdell_2016ApJ_Lupus}, the Lupus region has been also observed in Band~6 \citep{Andrews_DSHARP_2018ApJ...869L..41A, Ansdell_Lupus_2018ApJ}.
Hence, we compare our results with those obtained from the previous shallower surveys respect to AGE-PRO observations.

\begin{figure*}
   \gridline{\fig{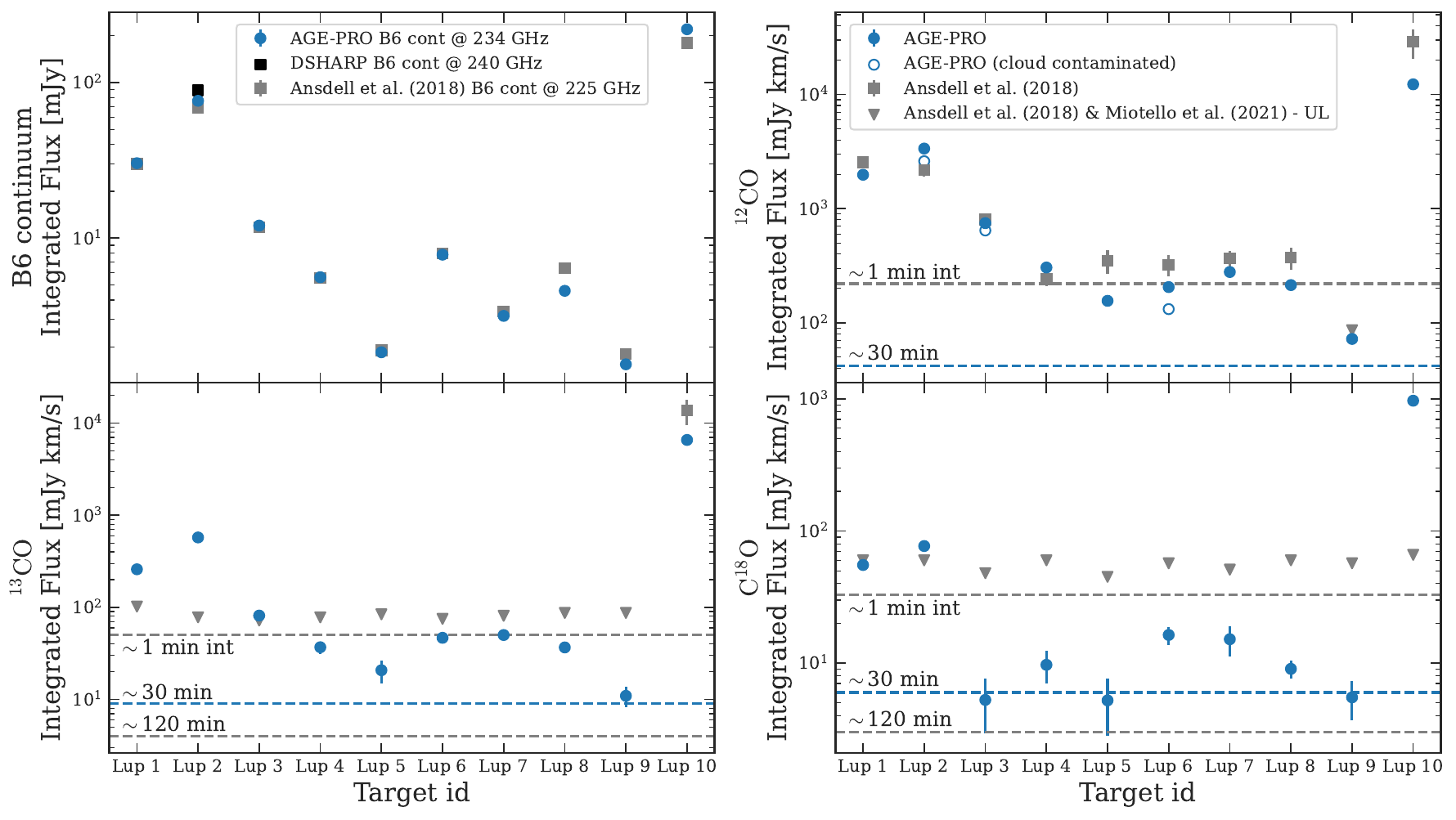}{0.95\textwidth}{}
            }
  \caption{
   The total flux compared with literature values.
   The AGE-PRO results are shown in blue points, and the literature measurements are represented as gray and black squares.
   The upper limits are shown as triangles.
   The horizontal dashed lines in three panels of CO isotopologue line fluxes show the approximate integrated flux that can be achieved with different observational integration time of $\sim 1$ min (past observation), $\sim 30$ min (AGE-PRO observation), and $\sim 120$ min (required to achieve 3$\sigma$ detections on $\mathrm{C^{18}O}$ for all targets).
  }
  \label{fig:comp-with-literature-flux}
\end{figure*}

\paragraph{Continuum flux densities and line fluxes.}
Figure~\ref{fig:comp-with-literature-flux} shows the comparisons between our results with the ones reported in the literature \citep{Andrews_DSHARP_2018ApJ...869L..41A, Ansdell_Lupus_2018ApJ, Miotello_Compat_Disks_2021A&A...651A..48M}.
We also plot three dashed lines of the expected integrated fluxes for three CO isotopologues with different observational integration time of $\sim 1$ min (past observation), $\sim 30$ min (AGE-PRO observation), and $\sim 120$ min. 
The values of the dashed lines are the sensitivities ($\sim 1\,\sigma$) calculated based on the ALMA sensitivity calculator \footnote{\url{https://almascience.eso.org/proposing/sensitivity-calculator}} and the disk sizes ($\sim 0.5$\arcsec) and line width (full-width at half maxima of $\sim 3\,{\rm km\,s^{-1}}$) for small disks (see Section~\ref{sec:Results}).
It shows that our AGE-PRO observation is $\sim 5 \times$ deeper than the past observations and indeed we have $> 3\sigma$ detections on all of the CO isotopologue lines except $F_\mathrm{C^{18}O}$ in Lupus~9.
Moreover, based on the flux, line width and emission size from AGE-PRO observation, if keep the same observational setup but further push the integration time to $\sim 120$ min, we would be able have $> 3\sigma$ detection on $F_\mathrm{C^{18}O}$ in Lupus~9, and also higher SNR for other disks.

The AGE-PRO measured continuum fluxes (blue points) are consistent with the literature values (gray and black squares) within their uncertainties, except the Lupus~10 where appears to have a discrepancy of $<20\%$ of the flux density, which could due to its extended emission and previous survey did not have enough sensitivity to cover its faint emission at the edge of its disk.
We have high SNR measurements of $\mathrm{{}^{12}CO}$ line flux, and all of our measured $F_\mathrm{{}^{12}CO}$ agree with the previous detected fluxes within the uncertainty for Lupus 1, 2, 3, 4 and 7, and 
within the $3\times$ the uncertainties for other targets.
For two other rarer CO isotopologues, only $F_\mathrm{{}^{13}CO}$ on Lupus~10 has been detected in the past and our new measurement agrees within $2\times$ the uncertainty with that.
We have $> 3\sigma$ detections of $F_\mathrm{{}^{13}CO}$ and $F_\mathrm{C^{18}O}$ for other targets for the first time, and our measured fluxes are lower than the estimated upper limits in the literature as expected in most cases.

However, there are obvious discrepancies on the $F_\mathrm{{}^{13}CO}$ for Lupus 1 and 2, and $F_\mathrm{C^{18}O}$ for Lupus 2 and 10, where our measurements are higher than the upper limits reported in the literature with more than an order of magnitude.
That is because in the past shallow surveys by \citet{Ansdell_Lupus_2018ApJ}, the upper limits are calculated from an area of the size of the beam ($\sim 0.25\,\arcsec$) and the area was not scaled to the source, i.e., those upper limits could be underestimated for large disks. 
The three disks here (Lupus~1, 2 and 10) are indeed the largest three in our sample, so that it is not surprise that those upper limits are not appropriate.
Besides, the earlier shallow surveys are not deep enough to pick up the faint extended emission inside each beam for large disks.
The $\mathrm{C^{18}O}$ line emission from Lupus 10 is a good example: based on its radial profile (Figure~\ref{fig:gasline-radialprofile-1}), the peak ($\sim\,18\,\mathrm{mJy\,beam^{-1}\,km\,s^{-1}}$) would indeed not have been detected by the shallow observations ($\mathrm{RMS}$ per beam $=\,22\,\mathrm{mJy\,beam^{-1}\,km\,s^{-1}}$ and they have similar beam size as ours).
Therefore, the AGE-PRO observations, with its higher sensitivity, provide more robust flux measurements.

\paragraph{Disk radii.}

\begin{figure*}
   \gridline{\fig{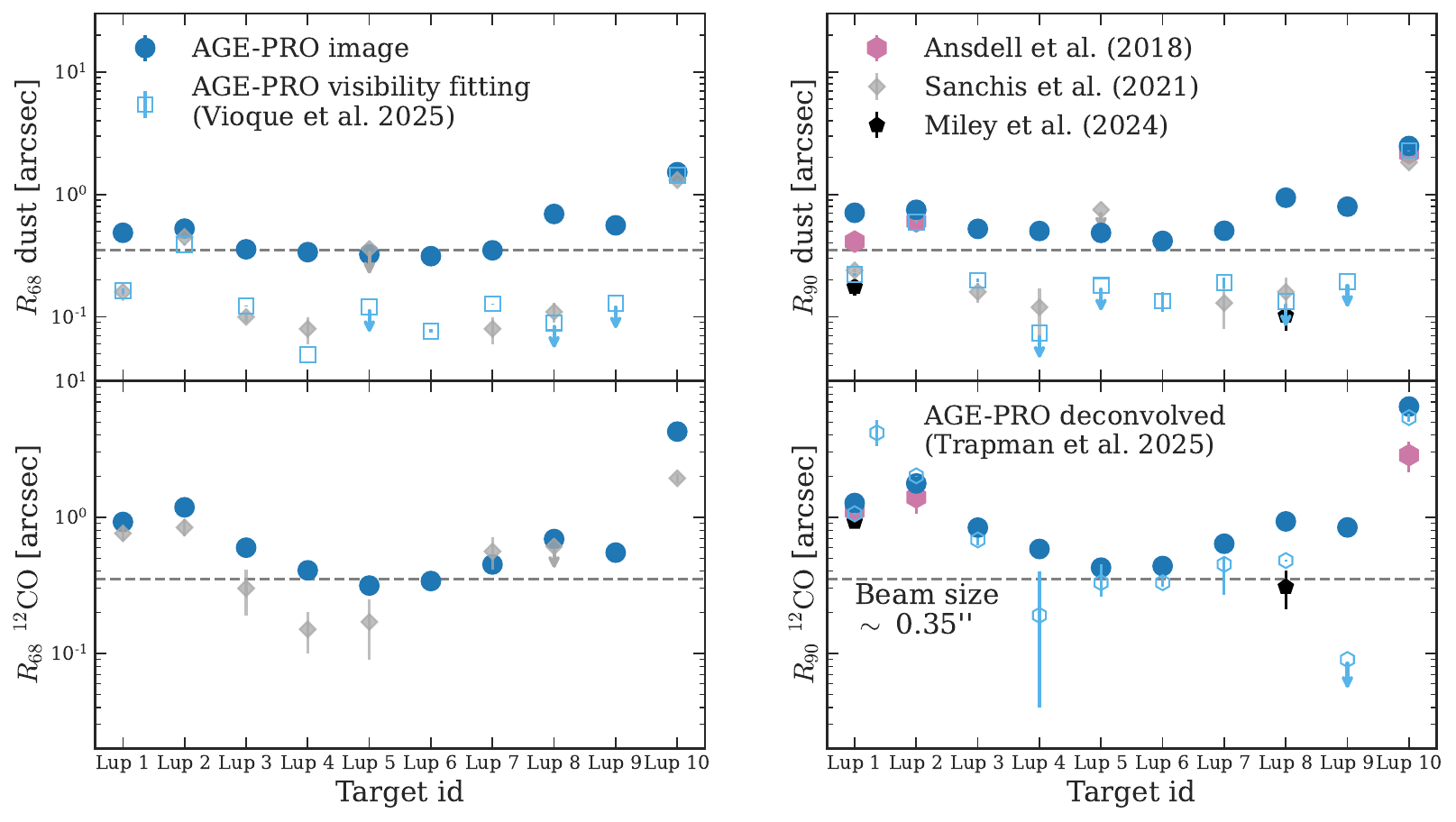}{0.95\textwidth}{}
            }
  \caption{
   The radius compared with literature values. The 68\% and 90\% dust radii are on the top left and right panels, while the two panels below are those for gas radii measured from $\mathrm{{}^{12}CO}$.
   The blue points are the AGE-PRO results measured from the images and they are not de-convolved with the beam.
   The empty squares are the AGE-PRO results from visibility fitting \citep{Vioque_AGEPRO_X_dust_disks}, and empty hexagans are the AGE-PRO results from beam de-convolved image fitting \citep{Trapman_AGEPRO_XI_gas_disk_sizes}.
   The black pentagons are the results from \citet{Miley_Sz65_Sz66_2024} from observations with higher spatial resolution.
   The gray points are the results from \citet{Sanchis_Lupus_radius_2021A&A...649A..19S}, where they carried out visibility fitting to derive dust radii, and their gas radii are also from beam de-convolved image fitting.
   The upper limits are represented with downward pointing arrows.
   For the large disks (Lupus 2 and 10) that are well-resolved in the image, all measurements are consistent; and for other targets, results from \citet{Vioque_AGEPRO_X_dust_disks} and \citet{Trapman_AGEPRO_XI_gas_disk_sizes} are consistent with other measurements in the literature.
  }
  \label{fig:comp-with-literature-radius}
\end{figure*}

Figure~\ref{fig:comp-with-literature-radius} compares our dust disk radii measured in the image plane (blue points) with literature, including the radii measured also from curve-of-growth method in \citet{Ansdell_Lupus_2018ApJ}, the radii obtained from visibility fittings reported in the \citet{Sanchis_Lupus_radius_2021A&A...649A..19S} (gray diamonds) and from the accompanying work (\citealt{Vioque_AGEPRO_X_dust_disks}; empty squares), and the radii measured from images with higher angular resolution \citep{Miley_Sz65_Sz66_2024} (black pentagons).
In the Band 6 continuum, our observation clearly resolved Lupus 2 and 10, and therefore the derived $R_{90}$ and $R_{68}$ for the Band~6 continuum is close to that from the visibility fitting, and also close to the visibility fitting results by \citet{Sanchis_Lupus_radius_2021A&A...649A..19S}.
However, all other disks are not well-resolved and therefore the radii measured in the image plane are strongly affected by the beam smearing, hence overestimated.
The visibility fitting from the AGE-PRO are still close to previous ones when literature values they are available for comparison.
Fig.~\ref{fig:comp-with-literature-radius} also compares the $R_{90}$ and $R_{68}$ for $\mathrm{{}^{12}CO}$ from our work (blue points) with the radii reported in the literature, where both the radii reported in \citet{Sanchis_Lupus_radius_2021A&A...649A..19S} and accompanying work by \citet{Trapman_AGEPRO_XI_gas_disk_sizes} are from beam de-convolved images.
Because gas disks are generally larger, more targets are resolved in the image plane.
The results for Lupus 1, 2, and 8 are all close to the results from both \citet{Sanchis_Lupus_radius_2021A&A...649A..19S} and \citet{Trapman_AGEPRO_XI_gas_disk_sizes}.
For the smaller targets -- Lupus 3, 4, 5 and 8 -- the $R_{68}$ measured from our images are overestimated by a factor of $\sim 2$, and their $R_{90}$ are also overestimated compared to others, suggesting they are unresolved.
Lupus 10 is the largest target in our sample and our image gives nearly a factor of $\sim 2$ larger $R_{68}$ than \citet{Sanchis_Lupus_radius_2021A&A...649A..19S}.
That is because we have deeper observations with higher sensitivity in each pixel, therefore our observation is better on this extended target.

\section{Relationships between fluxes, flux densities, and radii.}
\label{appendix:compare_f_n_f_n_r}

\begin{figure*}
    \gridline{\fig{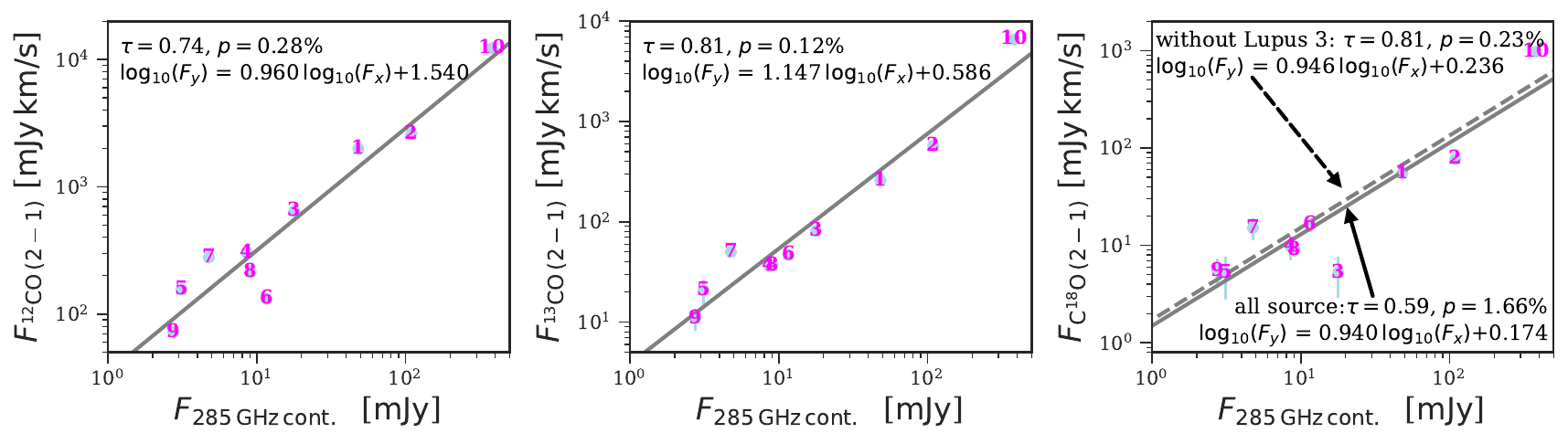}{0.95\textwidth}{}
            }
  \caption{
  The comparisons between the CO isotopologue fluxes and the continuum flux densities from ALMA Band 7 at 285 GHz.
  The notations follow Figure~\ref{fig:FF_comp} in Section~\ref{subsec:compare_f_n_f}.
  The results of \texttt{pymccorrelation} Kendall's $\tau$ tests for the Lupus sample are reported in each panel: There are positive correlations between the CO isotopologue fluxes with the Band 7 continuum fluxes at 285\,GHz.
  The best-fit results from linear-regression in log-log scale are plotted in black solid lines (for all source) and dashed lines (without Lupus~3) together with the functions at the corners of each panel.
  }
  \label{fig:FF_comp_Band7}
\end{figure*}

In Section~\ref{subsec:compare_f_n_f}, we find positive correlations between different CO isotopologue lines and the Band~6 continuum flux densities at 234\,GHz; and we also find correlation between the optically thin  $F_{\mathrm{C^{18}O\,(2-1)}}$ with $F_{\mathrm{N_2H^+\,(3-2)}}$ for the Lupus disks in our sample.
Here, we also present the comparisons between the three CO isotopologue lines in Band~6 with the Band~7 continuum flux densities at 285\,GHz (Figure~\ref{fig:FF_comp_Band7}).
Following the same tests that carried out in Section~\ref{subsec:compare_f_n_f}, we find strong positive correlations between the $F_{\rm 285\,GHz\,cont.}$ and $F_{\mathrm{{}^{12}CO\,(2-1)}}$ as well as $F_{\mathrm{{}^{13}CO\,(2-1)}}$, where we find positive $\tau$ and small $p < 1\%$ from Kendall's $\tau$ test.
Lupus~3 appears to be an outlier again in the correlation between $F_{\rm 285\,GHz\,cont.}$ and $F_{\mathrm{C^{18}O\,(2-1)}}$ and because of its very small $F_{\mathrm{C^{18}O\,(2-1)}}$ compared to its other CO isotopologue fluxes.
Once we exclude Lupus~3, the $F_{\rm 285\,GHz\,cont.}$ is also strongly correlated with the $F_{\mathrm{C^{18}O\,(2-1)}}$ with $p < 1\%$.

\begin{figure*}
    \gridline{\fig{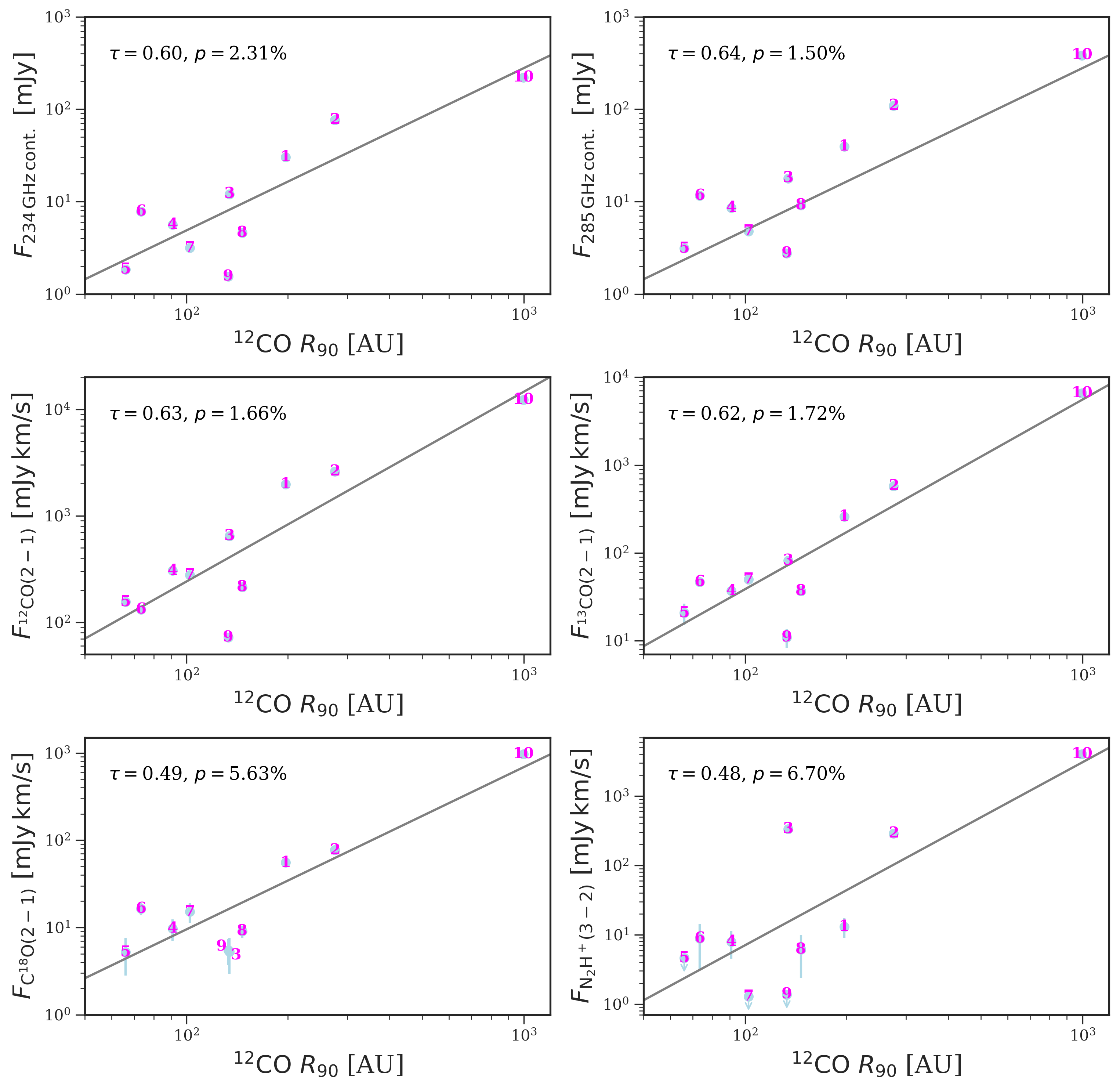}{0.9\textwidth}{}
            }
  \caption{
  The comparisons between disk sizes measured from images with the continuum flux densities, CO isotopologue fluxes and ${\rm N_2H^+\,(3-2)}$ flux.
  The notations follow Figure~\ref{fig:FF_comp} in Section~\ref{subsec:compare_f_n_f}.
  The results of \texttt{pymccorrelation} Kendall's $\tau$ tests for the Lupus sample are reported in each panel.
  }
  \label{fig:FR_comp}
\end{figure*}

We also test the correlations between the continuum flux densities, CO isotopologue fluxes, and $F_{\mathrm{N_2H^+\,(3-2)}}$ with the measured 90\% radius from the $\mathrm{{}^{12}CO\,(2-1)}$ images (Figure~\ref{fig:FR_comp}).
We find positive correlations for all of them, and the continuum flux densities from both bands, $F_{\mathrm{{}^{12}CO\,(2-1)}}$ and $F_{\mathrm{{}^{13}CO\,(2-1)}}$ are with $p < 3\%$ from the Kendall's $\tau$ tests, suggesting they are all likely to be optically thick, while $F_{\mathrm{C^{18}O\,(2-1)}}$ and $F_{\mathrm{N_2H^+\,(3-2)}}$ are optically thin as they are less likely to be correlated with the disk sizes.
But none of these tests show significant correlations (with $p < 1\%$), which could be due to the fact that more than half of the disks in our sample are not well-resolved in the images and their measured disk sizes are the beam sizes.
Therefore, observations with higher angular-resolution that can resolve these small disks in our sample could help further conduct detailed analysis on the fluxes and the disk sizes, and confirm the correlations found here.

\bibliography{agepro_Lupus}{}
\bibliographystyle{aasjournal}

\end{document}